\documentclass[twocolumn,preprintnumbers,amsmath,amssymb,superscriptaddress,nofootinbib,longbibliography]{revtex4-1}

\usepackage{graphicx}
\usepackage{bm}
\usepackage{dsfont}
\usepackage[usenames,dvipsnames]{xcolor}
\usepackage{pstricks}
\usepackage[tight]{subfigure}
\usepackage{verbatim}
\usepackage{units}
\usepackage{multirow}
\usepackage{enumitem}
\usepackage{mathrsfs}
\usepackage{leftidx}
\usepackage{xspace}
\usepackage{braket}
\usepackage{bbm}
\usepackage{overpic}
\usepackage{amsfonts,amssymb,amsmath}

\usepackage[a4paper]{hyperref}
\hypersetup{colorlinks=true,linktoc=all,linkcolor=blue,breaklinks=true,citecolor=blue,urlcolor=blue}
\usepackage[english]{babel}
\newcommand{\ta}{T_\mathrm{A}}
\newcommand{\ra}{R_\mathrm{A}}
\newcommand{\tb}{T_\mathrm{B}}
\newcommand{\rb}{R_\mathrm{B}}
\newcommand{\da}{d_\mathrm{A}}
\newcommand{\db}{d_\mathrm{B}}

\usepackage[normalem]{ulem}

\usepackage{hyperref}

\AtBeginDocument{%
    \newwrite\bibnotes
    \def\bibnotesext{Notes.bib}
    \immediate\openout\bibnotes=\jobname\bibnotesext
    \immediate\write\bibnotes{@CONTROL{REVTEX41Control}}
    \immediate\write\bibnotes{@CONTROL{%
    apsrev41Control,author="08",editor="1",pages="1",title="0",year="1"}}
     \if@filesw
     \immediate\write\@auxout{\string\citation{apsrev41Control}}%
    \fi
}%

\begin{document}
		
	\title{Influence of channel mixing in fermionic Hong-Ou-Mandel experiments}
	
	\author{Matteo Acciai}
	\affiliation{Department of Microtechnology and Nanoscience (MC2), Chalmers University of Technology, S-412 96 G\"oteborg, Sweden}

	\author{Preden Roulleau}
	\affiliation{Universit\'e Paris-Saclay, CEA, CNRS, SPEC, 91191, Gif-sur-Yvette, France}
	
	\author{Imen Taktak}
	\affiliation{Universit\'e Paris-Saclay, CEA, CNRS, SPEC, 91191, Gif-sur-Yvette, France}
	
	\author{D. Christian Glattli}
	\affiliation{Universit\'e Paris-Saclay, CEA, CNRS, SPEC, 91191, Gif-sur-Yvette, France}

	\author{Janine Splettstoesser}
	\affiliation{Department of Microtechnology and Nanoscience (MC2), Chalmers University of Technology, S-412 96 G\"oteborg, Sweden}
	
	\date{\today}
	
	\begin{abstract}
We consider an electronic Hong-Ou-Mandel interferometer in the integer quantum Hall regime, where the colliding electronic states are generated by applying voltage pulses (creating for instance Levitons) to ohmic contacts. The aim of this work is to investigate possible mechanisms leading to a reduced {visibility of the} Pauli dip, i.e., the noise suppression expected for synchronized sources.
{It is known that electron-electron interactions cannot account for this effect and always lead to a full suppression of the Hong-Ou-Mandel noise.} Focusing on the case of filling factor $\nu=2$, we show {instead} that a reduced visibility of the {Pauli dip} can result from mixing of the copropagating edge channels, {arising from tunneling events between them}.
	\end{abstract}

	\maketitle

\section{Introduction}

The advent of time-dependently driven, on-demand single-electron sources~\cite{Feve2007,Hermelin2011,Dubois2013,Fletcher2013} has been pivotal for the development of quantum optics with electrons~\cite{Grenier2011,Bocquillon2013review} and in particular for the arrival of quantum information applications using electron flying qubits based on Levitons~\cite{Bauerle2018}. It is therefore of fundamental importance to improve our understanding of the ac transport regime and of effects that can possibly be detrimental for  the operation of these single-electron devices.
Electron quantum optics deals with highly controllable single-electron excitations that propagate in solid-state systems and, ideally, can be coherently manipulated. A natural platform where this can be implemented is represented by quantum Hall systems, where chiral edge channels play the role of waveguides and quantum point contacts can be used as beamsplitters. Several theoretical works~\cite{Olkhovskaya2008,Jonckheere2012,Dubois2013PRB,Haack2013,Moskalets2013,Grenier2013,Ferraro2013,Wahl2014,Ferraro2014,Moskalets2015,Moskalets2016,Moskalets2016b,Hofer2017,Glattli2018,Cabart2018,Misiorny2018,Dashti2019,Yin_2019,Yue2021} have investigated various properties of single-electron sources in this regime, and many experimental results~\cite{Bocquillon2012,Parmentier2012,Bocquillon2013,Bocuillon2013Science,Jullien2014,Battista2014,Freulon2015,Ubbelohde2015,Vanevic2016,Kataoka2016,Glattli2016,Bisognin2019} have shown that a high degree of control in the manipulation of single-electron excitations can be achieved. Moreover, extensions to interacting systems~\cite{Ferraro2015,Ferraro2016,Slobodeniuk2016,Glattli2017,Rech2017,Litinski2017,Vannucci2017,Ferraro2018,Ronetti2018a,Acciai2019,Acciai_2019b,Ronetti2019,Ronetti2020} have also been considered. In particular, the fractional quantum Hall regime plays a special role, due to the presence of fractionally charged quasiparticles, whose anyonic statistics can in principle be probed by electron quantum optics setups~\cite{Carrega2021}. Very recently, first experiments in this regime have been reported~\cite{Kapfer2019,Bartolomei2020}.

\begin{figure}[b]
\begin{overpic}[percent=true,width=\columnwidth,grid=false]{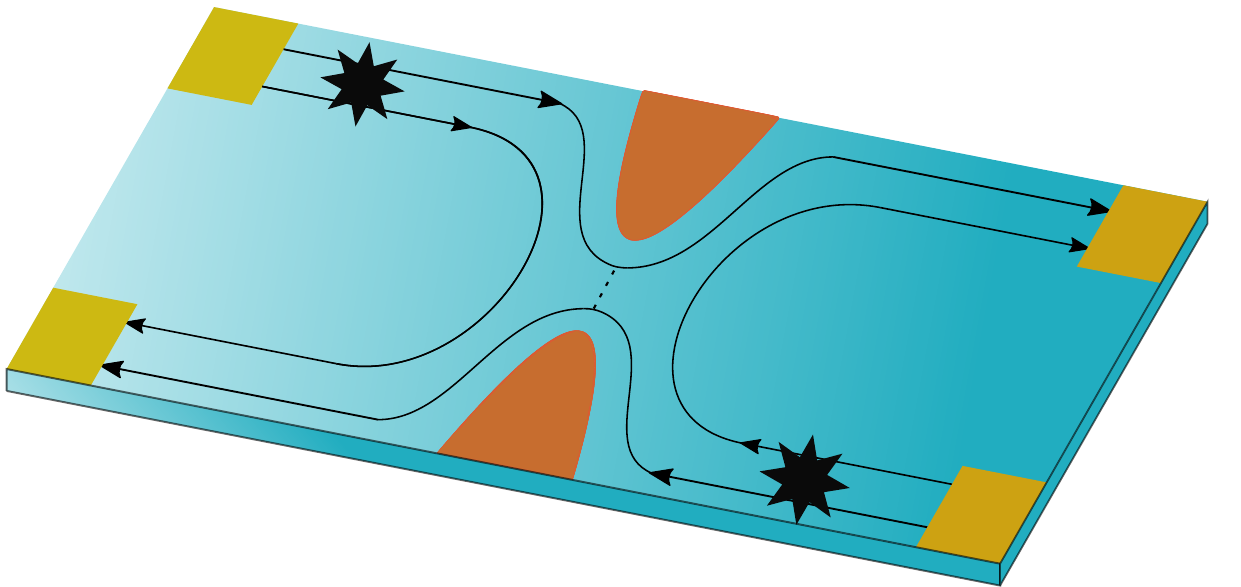}
\put(16.3,42){$V_\text{A}$}
\put(77,5.2){$V_\text{B}$}
\put(32,34){\small $i=2$}
\put(37,41.2){\small $i=1$}
\put(74,9){\small 4}
\put(72,5.7){\small 3}
\put(12,22.2){\small $I_2^\text{out}$}
\put(8,13.8){\small $I_3^\text{out}$}
\put(79.2,24.8){\small $I_4^\text{out}$}
\put(83,33){\small $I_1^\text{out}$}
\put(90.5,27.2){R}
\put(4.6,19){L}
\end{overpic}
\caption{Sketch of a Hong-Ou-Mandel (HOM) interferometer in the integer quantum Hall regime at filling factor $\nu=2$. Chiral edge states serve as waveguides in which electrons can propagate and are used to send excitations, generated by voltages $V_\text{A}$ and $V_\text{B}$, towards a quantum point contact (QPC) implementing an electronic beamsplitter. The QPC is set in such a way that the inner edge channels are fully reflected, so that the HOM interference only occurs on the outer channels. Inner and outer channels are labelled by $i=1,2$ (upper edge) and $i=3,4$ (lower edge).
Mixing points (black stars) may be present between the sources and the QPC, inducing inter-channel tunneling.
}
\label{fig:hom_sketch}
\end{figure}
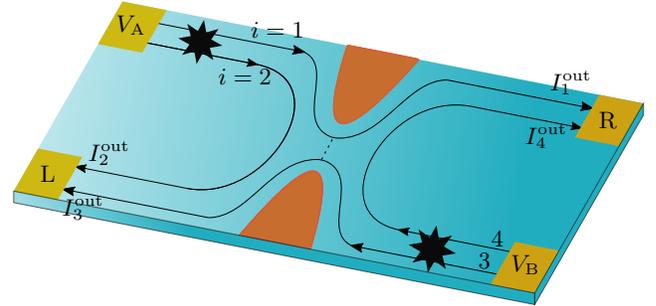

In the present paper, we deal with an electronic Hong-Ou-Mandel (HOM) interferometer~\cite{Hong1987}, which is a quantum-optics setup in a quantum Hall device, where electrons are brought to ``collide" at a quantum point contact (QPC), see Fig.~\ref{fig:hom_sketch}. {Such a setup is a milestone in electron quantum optics and has been used to directly demonstrate fermionic antibunching, by observing a suppression of the current fluctuations at the output of the interferomenter. However, in realistic experiments, this suppression can be incomplete~\cite{Bocuillon2013Science,Freulon2015,Taktak2021}, calling for a mechanism able to explain this effect. Remarkably, when the electronic states colliding at the QPC are generated by voltage drives, Coulomb interactions between edge channels are found to preserve a full suppression of the HOM noise~\cite{Ferraro2014,safi2014,Safi2019,Rebora2020}. In contrast,} we identify \emph{mixing}---namely, tunneling of quasiparticles between edge channels---as a possible source of the partial reduction of the characteristic HOM dip in the detected current correlation functions.

Edge state mixing has been considered since long in the stationary transport regime: for co-propagating edge channels in the integer quantum Hall regime, it determines the equilibration length of edge channels. The first works, done at moderately low temperature (around 1K) addressed spin-degenerate edge channels, where elastic tunneling of electrons occurs due  to random impurities~\cite{Alphenaar_1990,Hirai_1995,Palacios_1992}. Also, the equilibration length of spin-resolved channels was studied, at lower temperature, giving equilibration lengths ranging from few tens of $\mu$m~\cite{Takagaki_1994,Acremann_1999} to few hundreds of $\mu$m~\cite{Muller_1990}. Here, the combination of spin-orbit coupling and a scattering potential is necessary to allow for the mixing of opposite spin edge channels~\cite{Khaetskii_1992,Polyakov_1996,Pala_2005}. Mixing has a dramatic effect in the case of counter-propagating edges, a situation occurring for fractional quantum Hall edge states at filling factor $1/2<\nu<1$, like the $\nu=2/3$ hole-conjugated fractional quantum Hall state \cite{Johnson1991,Kane1994,Meir1994,Spaanslaett2019}. In this regime,  mixing has been found to be  essential to the explanation of the 2/3 conductance quantization of edge channels and the appearance of neutral counter propagating modes. However, as the mixing of equilibrated co-propagating edge channels does not affect the dc conductance, see below, the importance of mixing for other transport regimes has been overlooked.  This paper is aimed at filling this gap in ac-driven integer quantum Hall devices.

Here, we investigate the effect of mixing of co-propagating edge channels in the integer quantum Hall regime, which is likely to occur in 2D electron systems due to the native disorder arising from random ionization of donor impurities in the host material. 
Concretely, we consider a HOM interferometer, as shown in Fig.~\ref{fig:hom_sketch}, in the integer quantum Hall regime at filling factor $\nu=2$. Currents are injected from contacts A and B due to ac voltage drivings and particles are brought to collide at the central QPC. When the drivings are appropriately synchronized, this leads in the ideal case to a full suppression of the shot noise. Here, we show that mixing significantly affects the ac conductance and noise properties of co-propagating edge channels and, in particular, the visibility of two-particle interference effects leading to noise suppression in electronic HOM experiments. We consider one or two mixing points in any of the interferometer arms. Tunneling between edge channels, {together with different propagation velocities}, results in {a reduced visibility of the HOM dip.} In the case of multiple mixing points, additional interference effects {arise,}
similar to those occurring in Mach-Zehnder interferometers fed by single-electron sources~\cite{Splettstoesser2009,Haack2011,Juergens_2011,Hofer2014,Rossello2015,Kotilahti2021}. We show this both for sine-wave driving as well as for Lorentzian voltage pulses carrying single electrons (Levitons). We analyze in detail how the position of tunneling points affects the noise properties.
 
Our results are shown for the special case of two edge channels, but generalizations to more co-propagating edge channels is straightforward. We also anticipate that, regarding the qualitative effects of channel mixing, our results can be applicable to the fractional quantum Hall case, like for filling factor $\nu=2/5$, which corresponds to $\nu=2$ in the fractional quantum Hall Jain's hierarchy based on composite fermions~\cite{Jain1989}. 
 Furthermore, as HOM shot noise experiments measure the interference between two-particle braided and non-braided paths, they are expected to give a measure of the statistical angle characterising anyonic particles. The present work shows that, in the future,  the effect of mixing should be addressed in order to get sensible statistical angle measurements from shot noise.

The paper is organized as follows. First, in Sec.~\ref{sec:basic} we recall some basic results on HOM noise in the absence of mixing, that will be needed throughout the rest of the paper. Next, in Sec.~\ref{sec:int} we discuss the role of interactions, showing that they do not modify the HOM visibility. Then, we present a qualitative picture of the influence of channel mixing on photo-assisted and HOM noise in Sec.~\ref{sec:mix1}. Sec.~\ref{sec:mix_full} shows instead the result of a full Floquet calculation of the HOM noise in the presence of mixing in various conditions. Finally, in Sec.~\ref{sec:conclusion} we draw our conclusions and outline some directions for further investigations.

\section{Basic results on HOM noise without mixing}
\label{sec:basic}
We consider the HOM interferometer sketched in Fig.~\ref{fig:hom_sketch}, in the absence of mixing, namely when no inter-channel tunneling processes can occur. In addition, the QPC is tuned in such a way that the inner channels are fully reflected\footnote{{Note that the opposite regime, with fully transmitted outer channels and partially reflected inner ones, is equivalent and leads to similar results.}}. This choice is motivated both by the experimental relevance of such a configuration~\cite{Bocuillon2013Science,Freulon2015,Marguerite2016} and by the easier interpretation of the results arising when only one pair of channels is partitioned at the beamsplitter. Furthermore, we assume that the injecting contacts are driven by periodic voltages, with period $\mathcal{T}=2\pi\Omega^{-1}$. In the HOM configuration, these voltages are typically identical, apart from a controllable time delay $\delta$:
\begin{equation}
    V_\text{A}(t)=V(t)\,,\qquad V_\text{B}(t)=V(t+\delta)\,.
    \label{eq:voltage_HOM}
\end{equation}
It is also useful to split them into dc and ac contributions, namely $V(t)=V_\text{dc}+V_\text{ac}(t)$, with $\int_0^\mathcal{T}V_\text{ac}(t)dt=0$.
A HOM experiment deals with the zero-frequency noise of currents $\hat I_{i}^\text{out}(t)$ that are collected at the contacts at the output of the interferometer, see Fig.~\ref{fig:hom_sketch}. The zero-frequency noise is defined as
\begin{equation}
    \mathcal{S}_{ij}=2\int_{-\mathcal{T}/2}^{\mathcal{T}/2}\!\frac{dt}{\mathcal{T}}\int_{-\infty}^{+\infty}\!\!dt'\Braket{\delta\hat{I}_{i}^\text{out}\!\!\left(t+\frac{t'}{2}\right)\delta\hat{I}_{j}^\text{out}\!\!\left(t-\frac{t'}{2}\right)}
    \label{eq:noise_general_definition}
\end{equation}
where $\delta\hat{I}_i^\text{out}(t)=\hat{I}_i^\text{out}(t)-\braket{\hat{I}_i^\text{out}(t)}$ are the fluctuations in the output currents. These output currents are given by $\hat{I}_i^\text{out}(t)=-ev_i\hat\psi_{i,\text{out}}^\dagger(t)\hat\psi_{i,\text{out}}(t)$, where $e>0$ is the elementary charge and $\hat\psi_{i,\text{out}}(t)$
are the fermionic annihilation operators at the output of the interferometer. They can be expressed in terms of the input operators $\hat\psi_{i,\text{in}}(t)$, describing the electron states exiting the injecting contacts, by using a Floquet scattering approach~\cite{moskalets-book}, in the following way. First, due to the periodic voltage drives, the fermionic operators before the QPC can be written as (a similar expression holds for $\hat{\psi}_{i,\text{out}}$)
\begin{equation}\label{free}
\hat{\psi}_{i,\text{in}}(x,t)=\int_{-\infty}^{+\infty}\frac{d E}{\sqrt{h v_{i}}} \sum_{l\in\mathbb{Z}}p_{l}  \hat{a}_{i,\text{in}}(E-l\hbar\Omega)e^{-\frac{i E}{\hbar}\left(t - \frac{x}{v_{i}}\right)},
\end{equation}
where $\hat a_{i,\text{in}}(E)$ annihilates an electron at energy $E$ on channel $i$ and $v_i$ denotes the propagation velocity along that channel. We keep the setup as general as possible, allowing all velocities $v_i$ to be different. In Eq.~\eqref{free}, we have introduced
\begin{equation}
    p_l=\frac{1}{\mathcal{T}}\int_{-\mathcal{T}/2}^{\mathcal{T}/2}dt\,e^{il\Omega t}e^{\frac{ie}{\hbar}\int_{-\infty}^t dt' V_\text{ac}(t')},
    \label{eq:pl}
\end{equation}
that are the photoassisted Floquet probability amplitudes to absorb ($l>0$) or emit ($l<0$) $|l|$ energy quanta.
Next, the QPC is described via the scattering matrix\footnote{Here, as typically done in the description of HOM interferometry in quantum Hall setups, we have assumed an energy-independent scattering matrix. Considering an energy-dependent transmission does affect the HOM dip, leading to a loss of visibility. However, in standard experimental conditions~\cite{Taktak2021}, such an effect only plays a minor role from a quantitative point of view.}
\begin{equation}
    S_\text{QPC}=\begin{pmatrix}
    \sqrt{T} & i\sqrt{R}\\
    i\sqrt{R} & \sqrt{T}
    \end{pmatrix},
    \label{eq:S_QPC}
\end{equation}
where $T$ and $R$ are, respectively, the transmission and reflection probabilities. 
{Note that, for simplicity, the phase factors describing the propagation between input and output contacts have been disregarded in the scattering matrix. They will be included in the next part when edge mixing will be introduced}. The scattering matrix relates the ingoing and outgoing fermionic operator via
\begin{equation}
    \begin{pmatrix}
    \hat{a}_{1,\text{out}}\\ \hat{a}_{3,\text{out}}
    \end{pmatrix}
    =S_\text{QPC}
    \begin{pmatrix}
    \hat{a}_{1,\text{in}} \\ \hat{a}_{3,\text{in}}
    \end{pmatrix},
\end{equation}
while $\hat a_{2,\text{out}}=\hat a_{2,\text{in}}$ and $\hat a_{4,\text{out}}=\hat a_{4,\text{in}}$ as the inner channels are fully reflected.

The total currents collected at the output of the interferometer are (see Fig.~\ref{fig:hom_sketch})
\begin{equation}
\begin{split}
    I_\text{R}^\text{out}(t)&=I_1^\text{out}(t)+I_4^\text{out}(t),\\
    I_\text{L}^\text{out}(t)&=I_2^\text{out}(t)+I_3^\text{out}(t).
\end{split}
\label{eq:out-currents}
\end{equation}
In a HOM experiment, one usually considers the cross correlations of output currents, namely $\mathcal{S}_\text{RL}$. In the limit of zero frequency fluctuations, due to current conservation, the following relation holds: $\mathcal{S}_\text{RR}=\mathcal{S}_\text{LL}=-\mathcal{S}_\text{RL}=-\mathcal{S}_\text{LR}$. Moreover, in order to discard purely thermal contributions to the noise, which are independent of the applied drive, it is a standard procedure to define the HOM noise as the following difference
\begin{equation}
    \mathcal{S}_\text{HOM}=-\Delta\mathcal{S}_\text{RL}=-(\mathcal{S}_\text{RL}^\text{on,on}-\mathcal{S}_\text{RL}^\text{off,off})\,.
\end{equation}
Here ``on, on'' (``off, off'') indicate that both sources $V_\text{A}(t)$ and $V_\text{B}(t)$ are switched on (off). Another standard convention, is to normalize the HOM noise with respect to the Hanbury Brown-Twiss (HBT) noise, obtained when only one of the two sources is active. This defines the ratio
\begin{equation}
    \mathcal{R}=\frac{\mathcal{S}_\text{HOM}}{\mathcal{S}_\text{HBT}}=\frac{\mathcal{S}_\text{RL}^\text{on,on}-\mathcal{S}_\text{RL}^\text{off,off}}{\mathcal{S}_\text{RL}^\text{on,off}+\mathcal{S}_\text{RL}^\text{off,on}}.
    \label{eq:ratio}
\end{equation}

By combining the above Eqs.~(2)-(8), the HOM noise can be calculated. In the mixing- and interaction-free scenario we are considering in this Section, the excitations injected in channels 1 and 3 freely propagate until they are partitioned at the QPC, eventually generating the following noise:
\begin{equation}
    \mathcal{S}_\text{HOM}=RT\mathcal{S}_0(\Delta_{13}),
    \label{eq:hom_nomix}
\end{equation}
where~\cite{Dubois2013PRB}
\begin{equation}
\mathcal{S}_0(\Delta_{ij})=\frac{e^2}{\pi}\Omega\sum_{l\in\mathbb{Z}} \Pi_l(\Delta_{ij})\left[l\coth\left(\frac{l\hbar\Omega}{2k_\text{B}\theta}\right)-\frac{2k_\text{B}\theta}{\hbar\Omega}\right]
\label{eq:base-HOM}
\end{equation}
and $\theta$ is the temperature. The argument of this function, that we have generically indicated with $\Delta_{ij}$, represents the time delay with which the excitations injected into channels $i$ and $j$ arrive at the QPC. In this Section, the only relevant channels for the HOM noise are 1 and 3, so $\Delta_{13}$ appears in~\eqref{eq:hom_nomix}. Other values of $i$ and $j$ become relevant in the presence of mixing, see Sec.~\ref{sec:mix_full}. If the two identical injecting sources are shifted by $\delta$, see Eq.~\eqref{eq:voltage_HOM}, then the delay between the arrival times of the excitations at the QPC is $\Delta_{13}=\delta+d_\text{A}/v_1-d_\text{B}/v_3$, where $d_\text{A}$ and $d_\text{B}$ are, respectively, the lengths of the interferometer arms (i.e., the distances between the injecting sources and the QPC). This clearly shows that any asymmetry in the interferometer ($v_1\neq v_3$ and/or $d_\text{A}\neq d_\text{B}$) influences the effective time delay $\Delta_{13}$ at the QPC.
This time delay appears in Eq.~\eqref{eq:base-HOM} via the photoassisted probabilities $\Pi_l(\Delta_{13})$, given by
\begin{equation}
\Pi_l(\Delta_{ij})=\Big|\sum_{m\in\mathbb{Z}} p_m p_{m-l}^* e^{im\Omega\Delta_{ij}}\Big|^2\equiv|\tilde{p}_l(\Delta_{ij})|^2\,,
\label{eq:ptilde}
\end{equation}
where $p_l$ are defined in Eq.~\eqref{eq:pl}. When the arrival times are synchronized ($\Delta_{13}=0$), one finds $\mathcal{S}_0(0)=0$, by exploiting the property $\sum_m p_mp_{m-l}^*=\delta_{l,0}$. Thus, by measuring the noise as a function of the time delay, a full suppression of the signal is expected at $\Delta_{13}=0$. This is known as HOM dip or Pauli dip~\cite{Bocquillon2013review}.

The HOM noise takes a simple form when the injecting sources are driven by a periodic train of Lorentzian voltage pulses, each carrying a single electron, namely
\begin{equation}
V(t)=\sum_{k\in\mathbb{Z}}\mathscr{L}(t-k\mathcal{T})\,,\quad \mathscr{L}(t)=\frac{2\hbar}{ e w}\frac{1}{1+(t/w)^2}\,,
\label{eq:lor}
\end{equation}
where $w$ is the width of the pulses. The Lorentzian drive $\mathscr{L}(t)$ has a special role in the domain of electron quantum optics, as it is known to generate pure single-electron states on top of the Fermi sea, without any spurious neutral particle-hole pairs~\cite{Levitov1996,Ivanov1997,Keeling2006}. Such single-electron states are known as Levitons~\cite{Dubois2013}.
With the voltage~\eqref{eq:lor}, and in the limit of zero temperature, $\mathcal{S}_0$ reduces to~\cite{Dubois2013PRB,Glattli2016}
\begin{equation}
    \mathcal{S}_0(\Delta_{13})\big|_{\theta=0}=\frac{4e^2}{\mathcal{T}}\frac{\sin^2(\pi\Delta_{13}/\mathcal{T})}{\sin^2(\pi\Delta_{13}/\mathcal{T})+\sinh^2(2\pi w/\mathcal{T})}\,.
    \label{eq:ratio-lev}
\end{equation}
Moreover, the noise ratio $\mathcal{R}$ for Levitons was predicted~\cite{Dubois2013PRB} and confirmed~\cite{Glattli2016} to maintain the same form as in Eq.~\eqref{eq:ratio-lev}, independently of temperature.
Another interesting feature of the single-particle nature of Levitons is that they allow one to make a direct comparison with photonic HOM experiments. Indeed, in the same way as HOM interference of two single-particle bosonic states is related to the corresponding wavefunction overlap~\cite{Hong1987}, the HOM noise resulting from the interference of two single-particle fermionic states $\ket{\psi_\text{A}}$ and $\ket{\psi_\text{B}}$ is given by
\begin{equation}
\mathcal{S}_\text{HOM}\propto 1-|\langle \psi_\text{A}(t)|\psi_\text{B}(t)\rangle |^{2},
\end{equation}
where the overlap $J=|\langle \psi_\text{A}(t)|\psi_\text{B}(t)\rangle |^{2}$ is the coincidence probability to find the two particles at different outputs of the interferometer. For indistinguishable states, $J=1$ and the noise vanishes, as a consequence of the Pauli principle forcing the two particles to exit the interferometer on different outputs. Such an expression can be found from~\eqref{eq:ratio-lev} in the limit $w\ll \mathcal{T}$, where just a single pulse can be considered, instead of a periodic train. One finds~\cite{Olkhovskaya2008,Dubois2013PRB}
\begin{subequations}
\begin{align}
    \mathcal{S}_\text{HOM}&\approx 4e^2RT[1-J(\Delta_{13})]/\mathcal{T}\,,\label{eq:single_lev_noise}\\
    J(\Delta_{13})&=|\langle \psi_\text{A}(t)|\psi_\text{B}(t)\rangle |^{2}=\frac{4w^{2}}{4w^{2}+\Delta_{13}^{2}}\,.\label{eq:lev_overlap}
    \end{align}
\end{subequations}
Here, $J(\Delta_{13})$ is interpreted as the overlap of two Leviton wavefunctions $\psi_\text{A}(t)=\psi_\text{Lev}(t-\da/v_1)$ and $\psi_\text{B}(t)=\psi_\text{Lev}(t-\db/v_3+\delta)$, with
\begin{equation}
    \psi_\text{Lev}(t)=\sqrt{\frac{w}{\pi}}\frac{1}{t-iw}\,.
    \label{eq:lev_wf}
\end{equation}
For $\Delta_{13}=0$ (achieved e.g. with $\delta=0$ and a symmetric interferometer) the right Leviton wavefunction $\psi_\text{B}(t)$ matches the left Leviton wavefunction $\psi_\text{A}(t)$ at all times. In this situation, the incoming states are perfectly indistinguishable, so that the coincidence probability is exactly one and the HOM noise vanishes. For $\Delta_{13} \gg w$, keeping $\Delta_{13}<\mathcal{T}$, the right and left Leviton wavefunctions no longer overlap and we are left with the independent partitioning of left and right Levitons, each contributing to the current noise by the amount $2e^{2}RT/\mathcal{T}$.

The $100\%$ visibility of the HOM dip as a function of $\Delta_{13}$ is characteristic of electron anti-bunching as opposed to photon bunching and is common to all voltage-generated fermionic excitations, as Eq.~\eqref{eq:base-HOM} shows.
However, previous experimental results in the quantum Hall regime at $\nu=2$~\cite{Bocuillon2013Science,Freulon2015}, as well as more recent ones using voltage pulses~\cite{Taktak2021}, observed imperfect dips and a reduced visibility of the HOM signal. {The central result of this paper is the discussion of a mechanism -- channel mixing -- explaining this effect. Before coming to this point, we recall some known results concerning interactions.}

\section{Interacting copropagating edge channels}
\label{sec:int}

Electron-electron interactions often play a relevant role in the description of copropagating edge channels in the integer quantum Hall effect. In particular, they are known to affect the propagation of an injected electronic wavepacket, leading to its decomposition into charge and neutral modes~\cite{Bocquillon2013,Ferraro2014,Inoue2014,Freulon2015,Hashisaka2017,Acciai2018}. This is an instance of charge fractionalization, a well-known phenomenon in various interacting one-dimensional systems~\cite{Safi1995,Pham2000,Steinberg2008,Kamata2014,Acciai2017,Lin2021}. Fractionalization was found to be a possible explanation~\cite{Wahl2014} for the reduced visibility of HOM experiments performed with single electrons injected from a driven quantum dot into quantum Hall edge channels~\cite{Bocuillon2013Science,Freulon2015}. However, it was noted that if the injected states are obtained by applying voltage pulses, the HOM noise signal remains unaffected when the sources are synchronized, even in the presence of interactions and fractionalization~\cite{Rebora2020}. In this Section, we review these results and generalize them, including the effect of possible further dissipation channels. By this, we show that interaction is not expected to be at the origin of the reduced visibility even in this case. Note the later Sections (\ref{sec:mix1} and~\ref{sec:mix_full}) on mixing effects do not require the material contained in the present one, and can therefore be read independently of it.

\subsection{Model of electron-electron interactions}\label{sec:int_a}
We consider the pair of copropagating channels on the upper edge in Fig.~\ref{fig:hom_sketch} $(i=1,2)$. Including density-density inter-channel interactions, the two channels are modeled by the following Hamiltonian density~\cite{Levkivskyi2008,Slobodeniuk2016}
\begin{equation}
\mathcal{H}_0=\sum_{i=1,2}\frac{\hbar v_i}{4\pi}(\partial_x{\hat{\phi}}_i)^2+\frac{\hbar u_{12}}{2\pi}(\partial_x{\hat{\phi}}_1)(\partial_x{\hat{\phi}}_2)\,.
\label{eq:hamiltonian_TLL}
\end{equation}
Here, $v_i$ is the propagation velocity on channel $i$ and ${\hat{\phi}}_i$ are bosonic operators, satisfying the commutation relations $[{\hat{\phi}}_i(x),{\hat{\phi}}_j(y)]=i\pi\text{sign}(x-y)\delta_{ij}$. They are connected to the fermionic operators $\hat\psi_i(x)$, annihilating an electron on channel $i$ at position $x$, via the bosonization identity
\begin{equation}
\hat\psi_i(x)=\frac{\mathcal{F}_i}{\sqrt{2\pi a}}e^{-i\hat{\phi}_i(x)}\,,
\end{equation}
where $\mathcal{F}_i$ are Klein factors and $a$ is a short-distance cutoff. Furthermore, the charge density on channel $i$ can be conveniently expressed as
\begin{equation}
\hat\rho_i=-e:\hat\psi_i^\dagger\hat\psi_i:=-\frac{e}{2\pi}\partial_x\hat{\phi}_i\,.
\end{equation}
Finally,
the coupling strength between the channels is denoted by the parameter $u_{12}$. Note that we have neglected intra-channel interactions, as they can be readily included in a redefinition of the velocities $v_i$.
The Hamiltonian~\eqref{eq:hamiltonian_TLL} can be diagonalized by introducing a new basis via the rotation
\begin{equation}
\begin{pmatrix}
\hat{\phi}_\rho\\
\hat{\phi}_\sigma
\end{pmatrix}=
\begin{pmatrix}
\cos\chi & \sin\chi\\
-\sin\chi & \cos\chi
\end{pmatrix}
\begin{pmatrix}
\hat{\phi}_1\\
\hat{\phi}_2
\end{pmatrix}.
\end{equation}
Here, $\chi$ is called mixing angle and can be expressed as
\begin{equation}
\tan(2\chi)=\frac{2u_{12}}{v_1-v_2}\,.
\end{equation}
It ranges in the interval $0\le\chi\le\pi/4$, where $\chi=0$ means no interactions $(u_{12}=0)$, while $\chi=\pi/4$ is achieved at maximal coupling between the channels.
The new fields $\hat{\phi}_\rho$ and $\hat{\phi}_\sigma$ are the eigenfields of the problem and describe charge density wave excitations propagating with velocities
\begin{equation}
v_{\rho,\sigma}=\frac{v_1+v_2}{2}\pm\frac{1}{\cos(2\chi)}\frac{v_1-v_2}{2}\,.
\end{equation}

Suppose now that the fields associated with the physical channels have initial amplitudes $\tilde{\phi}_{1,2}(0,\omega)$. Here, we are using the frequency representation
\begin{equation}
    \tilde{{\phi}}_i(x,\omega)=\int dt\,e^{i\omega t}\hat{\phi}_i(x,t)
\end{equation}
and the argument $x=0$ denotes the initial position where the fields are evaluated. How do these amplitudes change after a propagation length $\da$? To answer this question, one has first to consider the equations of motion for the eigenfields $(\eta=\rho,\sigma)$
\begin{equation}
    (-i\omega+v_\eta\partial_x)\tilde{\phi}_\eta(x,\omega)=0\implies\tilde{\phi}_\eta(x,\omega)=e^{i\frac{\omega x}{v_\eta}}\tilde{\phi}_\eta(0,\omega)
    \label{eq:eom1}
\end{equation}
and then perform the change of basis to relate $\tilde{\phi}_{\rho,\sigma}$ to $\tilde{\phi}_{1,2}$. The result is the following expression
\begin{equation}
\begin{pmatrix}
\tilde{\phi}_1(\da,\omega)\\
\tilde{\phi}_2(\da,\omega)
\end{pmatrix}
=S(\da,\omega)
\begin{pmatrix}
\tilde{\phi}_1(0,\omega)\\
\tilde{\phi}_2(0,\omega)
\end{pmatrix},
\end{equation}
where $S(\da,\omega)$ is called edge-magnetoplasmon scattering matrix. It reads~\cite{Degiovanni2010}
\begin{equation}
S(\da,\omega)=
\begin{pmatrix}
c^2\,e^{i\omega\tau_\rho^\text{A}}+s^2\,e^{i\omega\tau_\sigma^\text{A}} & cs(e^{i\omega\tau_\rho^\text{A}}-e^{i\omega\tau_\sigma^\text{A}}) \\
cs(e^{i\omega\tau_\rho^\text{A}}-e^{i\omega\tau_\sigma^\text{A}}) & s^2\,e^{i\omega\tau_\rho^\text{A}}+c^2\,e^{i\omega\tau_\sigma^\text{A}}
 \end{pmatrix},
 \label{eq:smatrix}
\end{equation}
where $\tau_{\rho,\sigma}^\text{A}=\da/v_{\rho,\sigma}$ are the times of flight associated with the two different eigenmodes and we have introduced the shorthand notation $c\equiv\cos\chi$, $s\equiv\sin\chi$.

Next, we consider the effect of an applied voltage acting on the edge channels. This can be described by the Hamiltonian
\begin{equation}
    \mathcal{H}_\mathcal{U}=[\hat\rho_1(x)+\hat\rho_2(x)]\mathcal{U}(x,t)
\end{equation}
where $\mathcal{U}(x,t)=\Theta(-x)V_\text{A}(t)$ and $V_\text{A}(t)$ is the voltage applied to the upper-edge channels. In the absence of interactions, the above term would result in the following time evolution of the fermionic operators:
\begin{equation}
    \hat\psi_i(x,t)=\hat\psi_i^{(0)}(x,t)e^{\frac{ie}{\hbar}\int_{-\infty}^{t}dt'V_\text{A}(t'-x/v_i)}\,,
\end{equation}
where the superscript $(0)$ denotes the time evolution in the absence of $V_\text{A}(t)$. Thus, we see that the effect of a voltage drive modifies the accumulated phase of the operators. {Remarkably, the same is true in the presence of electron-electron interactions, which only lead to a modification of the voltage appearing in the previous equation.} Indeed, as discussed in Ref.~\cite{Grenier2013}, a classical voltage generates a bosonic coherent state, onto which interactions act as a frequency-dependent beamsplitter via the edge-magnetoplasmon scattering matrix~\eqref{eq:smatrix}. As a result, the outgoing state on channel $i$ after a propagation length $\da$ is equivalent to the state generated by a distorted voltage $U_{i}(t)$, such that its Fourier transform satisfies $\tilde{U}_{i}(\omega)=\sum_{j=1,2}S_{ij}(\da,\omega)\tilde{V}_\text{A}(\omega)$. By using Eq.~\eqref{eq:smatrix}, one then finds
\begin{equation}
    \begin{split}
        U_{1}(t)&=c^2V_\text{A}(t-\tau_\rho^\text{A})+s^2V_\text{A}(t-\tau_\sigma^\text{A})\\
        &\quad+cs[V_\text{A}(t-\tau_\rho^\text{A})-V_\text{A}(t-\tau_\sigma^\text{A})],\\
        U_{2}(t)&=s^2V_\text{A}(t-\tau_\rho^\text{A})+c^2V_\text{A}(t-\tau_\sigma^\text{A})\\
        &\quad+cs[V_\text{A}(t-\tau_\rho^\text{A})-V_\text{A}(t-\tau_\sigma^\text{A})].
    \end{split}
    \label{eq:voltage_out}
\end{equation}
Notice that in the absence of interactions, one correctly recovers $U_{i}(t)=V_\text{A}(t-\da/v_{i})$.

The above model can be modified to take into account (unwanted) energy losses. Such a possibility is suggested by some experimental results~\cite{Sueur2010,Bocquillon2013,Krahenmann2019,Rodriguez2020}, where energy losses after a given propagation distance were reported and had to be taken into account to explain the observations. {Here}, we follow Refs.~\cite{Bocquillon2013,Rodriguez2020,Rebora2021} and phenomenologically introduce a damping factor in the equations of motion~\eqref{eq:eom1}, which become
\begin{equation}
    (-i\Gamma(\omega)+v_\eta\partial_x)\tilde{\phi}_\eta(x,\omega)=0.
\end{equation}
Here, $\Gamma(\omega)=\omega+i\gamma(\omega)$ and $\gamma(\omega)$ is the damping factor. In Ref.~\cite{Bocquillon2013}, the form $\gamma(\omega)=\gamma_2\omega^2$ was assumed, whereas a linear dependence $\gamma(\omega)=\gamma_1\omega$ was shown to be the best choice to explain the experimental data in Ref.~\cite{Rodriguez2020}. It is also worth mentioning that a different and more refined approach to include dissipation mechanisms in the context of copropagating quantum Hall states was introduced in~\cite{Goremykina2018}. Independently of the specific form of $\gamma(\omega)$, it is clear that in the presence of this damping factor the scattering matrix~\eqref{eq:smatrix} has to be modified by replacing $\omega\to\omega+i\gamma(\omega)$. Consequently, the voltages in Eq.~\eqref{eq:voltage_out} are modified by replacing $V_\text{A}(t)\to W_\text{A}(t)$, where
\begin{equation}
    W_\text{A}(t)=\int dt'\int\frac{d\omega}{2\pi}V_\text{A}(t')\,e^{-i\omega(t-t')}e^{-\gamma(|\omega|)(t-t')}\,.
\end{equation}
Note that in the absence of dissipation $[\gamma(\omega)=0]$ one recovers $W_\text{A}(t)=V_\text{A}(t)$.
We observe from the last equation that the inclusion of a dissipative term in the model only leads to a further modification of the outgoing voltages, compared to Eq.~\eqref{eq:voltage_out}. As we will see in the following, this modification does not lead to a reduction of the visibility of the HOM dip.

\subsection{HOM noise}
We now consider the HOM interferometer sketched in Fig.~\ref{fig:hom_sketch}. It is clear that the discussion in Sec.~\ref{sec:int_a} can be repeated identically for the lower-edge channels $(i=3,4)$, leading to voltages $U_3$ and $U_4$ which can be obtained, respectively, from $U_1$ and $U_2$ in Eq.~\eqref{eq:voltage_out} by replacing $\text{A}\to\text{B}$.
This allows one to evaluate the HOM noise, which can be written as~\cite{Ferraro2013}
\begin{equation}
    \mathcal{S}_\mathrm{HOM}=(e v_1)^2 RT\int  \Delta Q(t,t')dt dt'\,,
    \label{eq:noise}
\end{equation}
where the integrand is given by
\begin{equation}
\begin{split}
    \Delta Q(t,t')&=2\mathcal{G}^{(e)}_{\text{F}}(t'-t)\mathcal{G}^{(h)}_{\text{F}}(t-t')\\
    &\quad\times\big[1-\cos(\varphi_{\text{A}}(t,t')-\varphi_\text{B}(t,t'))\big]\,.
\end{split}
 \label{eq:new}
\end{equation}
In this equation, we have introduced the equilibrium correlation functions $\mathcal{G}^{(e)}_{\text{F}}(t)=\braket{\hat\psi^\dagger_1(0)\hat\psi_1(t)}_\text{F}$ and $\mathcal{G}^{(h)}_{\text{F}}(t)=\braket{\hat\psi_1(t)\hat\psi^\dagger_1(0)}_\text{F}$ and the phases $\varphi_\text{A/B}$ are due to the modified voltages taking interactions into account. Explicitly, they read:
\begin{equation}
\varphi_{\text{A/B}}(t,t')=\frac{e}{\hbar}\int_{t'}^{t}U_{1/3}(\tau)d\tau\,.
\label{varphi}
\end{equation} 
By using Eq.~\eqref{eq:voltage_out} to calculate these phases and recalling Eq.~\eqref{eq:voltage_HOM}, one immediately realizes that, in the case of a synchronized emission $(\delta=0)$ and if no asymmetry is present in the interferometer ($\da=\db$, $v_1=v_3$, $v_2=v_4$), then $\varphi_\text{A}(t,t')=\varphi_\text{B}(t,t')$ and the HOM noise~\eqref{eq:noise} vanishes. Note that the mentioned potential asymmetries are typically absent due to experimental design.
The vanishing of the HOM noise at $\delta=0$ is not only valid for the simplest interaction model in Eq.~\eqref{eq:hamiltonian_TLL}, but also for the extended version including dissipation. It also applies to modified models with long range interactions~\cite{Grenier2013}, because, in that case, interaction effects certainly give a more complicated voltage modification at the end of the interacting region, but this does not change the finding $\varphi_\text{A}(t,t')=\varphi_\text{B}(t,t')$ for a synchronized emission.
We hence conclude that Coulomb interactions between the channels do not explain a reduced visibility of the HOM dip. For these reasons, we neglect interactions in the rest of the paper and we focus on a different mechanism that instead is able {(by itself)} to explain such an effect, namely channel mixing induced by inter-channel tunneling. {As a further reason to neglect interactions, we also emphasize that the description presented in the rest of this paper can already explain recent experimental results where a reduced HOM dip is reported~\cite{Taktak2021}.}

We conclude this Section by mentioning that the impact of interactions on the HOM noise strongly depends on how the injected states are generated. As discussed e.g. in Refs.~\cite{Grenier2013,Ferraro2014}, voltage-generated states are coherent states for edge magnetoplasmons and hence they do not suffer from decoherence, leading to a full HOM dip. On the contrary, single electrons injected from driven quantum dots quickly relax at energies close to the Fermi level before undergoing fractionalization~\cite{Ferraro2014}. In this case, interactions are important and can explain a reduced HOM dip~\cite{Wahl2014}.

\section{Mixing induced by interchannel tunneling}
\label{sec:mix1}
This Section is dedicated to a qualitative understanding of mixing and is divided into two parts. First, in Sec.~\ref{sec:mix1_a}, we discuss the effect of mixing two co-propagating chiral edge channels in the Quantum Hall regime when a dc or an ac voltage is applied to the contact feeding the two channels with electrons (injecting contact). For simplicity, we will consider a single point-like scatterer mixing the two edge channels. We show that for an ac voltage applied to the contact, mixing generates a photo-assisted current noise in each channel which vanishes for a pure dc voltage bias. 
\begin{figure}[t]
\centering
\includegraphics[width=\columnwidth]{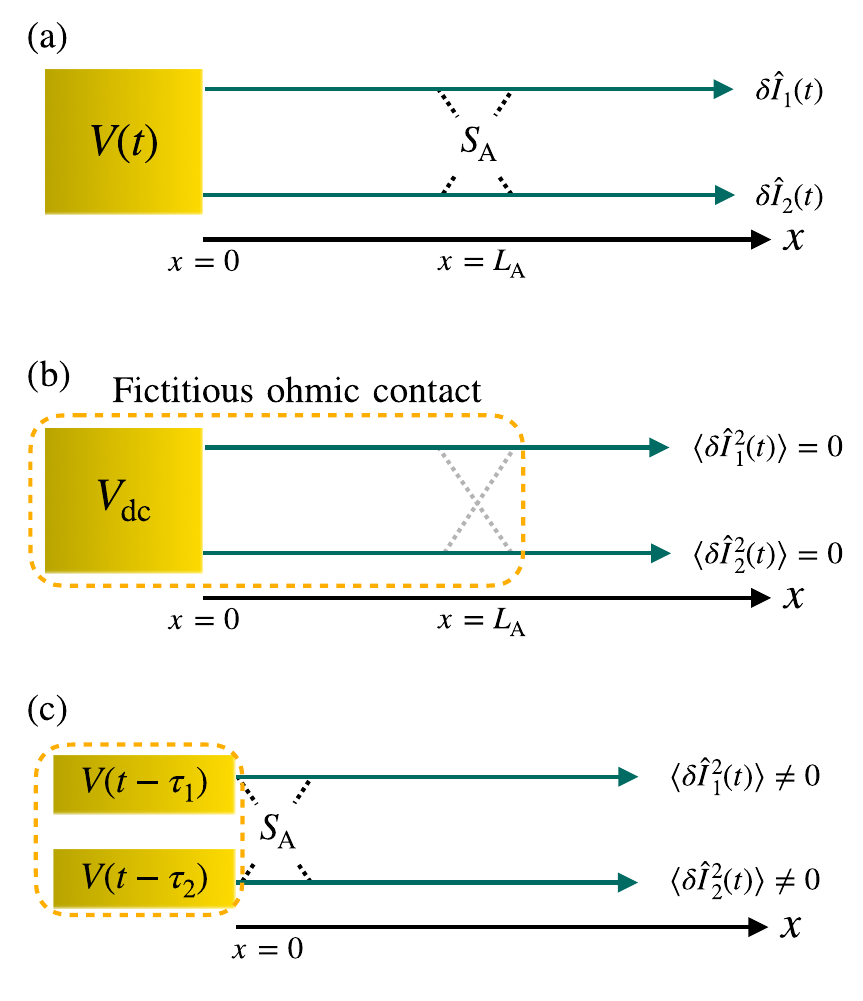}
\caption{(a) Two edge channels connected to the same ohmic contact. At a distance $x=L_\text{A}$ from the contact, a mixing point is located. It is characterized by a mixing strength ${\ra}=1-{\ta}$ {yielding the probability of inter-channel tunneling.} (b) {Special case of (a):} when a dc voltage is applied, one can {ignore the mixing point and include it in an equivalent} ohmic contact, without changing the transport and current noise properties. (c) When an ac bias is applied, the ohmic contact can be replaced by two separate ohmic contacts, one connecting the outer edge with a voltage shifted by the time delay $\tau_{1}=L_\text{A}/v_{1}$, and the second connecting the inner edge including the time delay $\tau_{2}=L_\text{A}/v_{2}$. A current noise results from the two fictitious voltages not being equal when $\tau_{1}\ne \tau_{2}$. All panels are sketches of the upper-edge channels in Fig.~\ref{fig:hom_sketch}, before the QPC; the mixing point in this figure corresponds to the black star in Fig.~\ref{fig:hom_sketch}.}
\label{fig:mixing1}
\end{figure}
Then, in Sec.~\ref{sec:mix_hom_qualitative}, we consider the injection of single-charge Levitons in the input arms of the HOM interferometer, in order to understand how mixing may affect the two-particle HOM interference of single-electron quantum states. 

\subsection{{Transport properties of two mixed copropagating chiral channels}}\label{sec:mix1_a}
Consider two edge channels $(i=1,2)$ emitted by the same contact where a voltage is applied, see Fig.~\ref{fig:mixing1}(a). At a distance $L_\text{A}$ from the contact, a pointlike elastic scatterer mixes the two edge channels, which then freely propagate. We describe the effect of mixing by using again a scattering approach, as discussed in Sec.~\ref{sec:basic}.
We therefore introduce the following unitary matrix
\begin{equation}
S_\text{A}=
\begin{pmatrix}
\sqrt{\ta} & i\sqrt{\ra}\\
i\sqrt{\ra} & \sqrt{\ta}
\end{pmatrix},
\label{eq:sA}
\end{equation}
which relates the fermionic operators before and after the mixing point $x=L_\text{A}$, in the same way as $S_\text{QPC}$ in Eq.~\eqref{eq:S_QPC} relates the fermionic operators before and after the QPC.

\subsubsection{Constant voltage bias}
Let us first consider applying a dc voltage $V_\text{dc}$ to the contact. In this case, the photoassisted amplitudes~\eqref{eq:pl} reduce to $p_l=\delta_{l,0}$, which simplifies the expression~\eqref{free} for the incoming fermionic operators. Then, the energy distributions of electrons injected by the contact at $x=0$ in channel $i$ are given by  $f_{1}(E)=f_{2}(E)=f(E-eV_\text{dc})$ where $f(E)=[1+\exp(E/k_{B}\theta)]^{-1}$ is the equilibrium Fermi distribution at temperature $\theta$, taking the Fermi energy as the energy reference. As we anticipate that the temperature does not play an important qualitative role, in the following we will choose $\theta=0$ for simplicity. After mixing, the new energy distributions are: $f_{1,+}={\ta} f_{1}+{\ra}f_{2}$ and $f_{2,+}={\ra}f_{1}+{\ta} f_{2}$. They remain equal to $f(E-eV_\text{dc})$ thanks to the unitarity of $S_\text{A}$. From this, we {see that mixing has} no consequence {on} dc transport properties. In particular, mixing does not generate current noise as equally occupied electronic states lead, by antibunching, to the Pauli suppression of quantum partition noise. We could {then completely disregard the presence of the mixing point and include it in an equivalent injecting contact always biased by $V_\text{dc}$}, see Fig.~\ref{fig:mixing1}(b). However, this picture is no longer appropriate when driving the injecting contact with a time-dependent voltage.

\subsubsection{Ac voltage bias: {time-resolved current}}
Let us assume that an ac drive is applied to the contact in Fig.~\ref{fig:mixing1}(a). Even if the following discussion is valid for an arbitrary periodic signal, one can take for concreteness $V(t)=V_\text{ac}\cos(\Omega t)$, in which case one has $p_l=J_l(eV_\text{ac}/\hbar\Omega)$, with $J_l$ the Bessel functions of the first kind, and $\Pi_{l}(\Delta_{ij})=J_{l}^{2}(2eV_\text{ac}\sin(\Omega \Delta_{ij}/2)/\hbar\Omega)$.

As electrons propagate freely between the contact and the scatterer, one sees from Eq.~\eqref{free} that by a redefinition of the Floquet amplitudes $p_{l}\rightarrow p_{l}e^{-iE(t - x/v_{i})/\hbar}$, the problem becomes equivalent to having two separate ohmic contacts placed immediately before the scatterer. These artificial contacts, separately acting on channels 1 and 2, are respectively biased by voltages $V(t-\tau_{1})$ and $V(t-\tau_{2})$, where $\tau_{1}=L_\text{A}/v_{1}$ and $\tau_{2}=L_\text{A}/v_{2}$, see Fig.~\ref{fig:mixing1}(c).

{As a first ac transport property,} it is interesting to see how mixing affects the time-resolved current generated in the outer and inner edge due to the cosine-wave potential $V(t)=V_\text{ac}\cos(\Omega t)$ applied to the contact. In the absence of mixing, the outer and inner ac currents are equal in amplitude while their phase at a given point reflects the propagation velocity of each channel. One has $I_{i}(x,t)=\frac{e^{2}}{h}V(t-x/v_{i})$. In the presence of mixing one finds, for $x \geq L_\text{A}$:
\begin{equation}
    \begin{split}
    I_{1}(x,t)&=\frac{e^{2}}{h}\left[{\ta}V\!\left(t-\frac{x}{v_{1}}\right)+{\ra}V\!\left(t-\tau_{2}+\frac{x-L_\text{A}}{v_{2}}\right)\right]\\
    I_{2}(x,t)&=\frac{e^{2}}{h}\left[{\ra}V\!\left(t-\tau_{1}-\frac{x-L_\text{A}}{v_{1}}\right)+{\ta}V\!\left(t-\frac{x}{v_{2}}\right)\right]
    \end{split}
\end{equation}
Right at the mixing point output, the two currents share equal amplitude: $V_{\text{ac}}\sqrt{1-2{\ra}{\ta}(1-\cos(\Omega\Delta \tau))}$, a value smaller than in absence of mixing, and they present a relative phase shift which not only depends on $\tau_{1}-\tau_{2}$ but also on the mixing strength.

{\subsubsection{Noise after the mixing point}}
For $v_{1}\neq v_{2}$, the two fictitious ac voltages {in Fig.~\ref{fig:mixing1}(c)} are not equal and we expect the generation of a finite photoassisted shot noise (PASN) in the outer and inner currents, $I_1$ and $I_2$. Indeed, the corresponding zero-temperature current noise spectral densities $\mathcal{S}_{I_1}$ and $\mathcal{S}_{I_2}$ after the mixing point are given by \begin{equation}
    \mathcal{S}_{I_1}=\mathcal{S}_{I_2}=\frac{e^{2}}{\pi}\Omega {\ra}{\ta}\sum_{l\in\mathbb{Z}}|l|\Pi_l(\Delta_{12})\,,
    \label{eq:Si1-Si2}
\end{equation}
where $\Delta_{12}=\tau_{1}-\tau_{2}$. Notice that the noise in the previous equation is equivalent to $\mathcal{S}_{22}$, as defined in Eq.~\eqref{eq:noise_general_definition}. One thus observes that a finite-frequency excitation on the original contact is responsible for a finite partition noise due to mixing, in contrast with the dc case ($\Omega=0$), where channel mixing does not affect the dc transport and noise.

\subsubsection{{Energy density matrix}}

An{other} important quantity fully characterizing fermionic states generated by an ac voltage bias is the energy density matrix $D(E',E)$. Its measurement has been proved possible in recent electronic quantum tomography experiments~\cite{Jullien2014,Bisognin2019}. A calculation of this quantity requires the fermionic operators in energy representation. 
Considering the input fermionic operators $\hat\psi_{i,\text{in}}(L_\text{A},t)$ before the scatterer, their energy representation reads, using Eq.~\eqref{free}:
\begin{equation}
\begin{split}
   \hat{\psi}_{i,\text{in}}(E) &=\sqrt{\frac{v_i}{2\pi\hbar}}\int_{-\infty}^{+\infty}dt\,e^{iEt/\hbar}\hat{\psi}_{i,\text{in}}(L_\text{A},t)\\
   &=\sum_{l\in\mathbb{Z}}\hat{a}_{i}(E-l\hbar\Omega) p_{l}e^{iE\tau_{i}/\hbar}.
\end{split}
\end{equation}
The quantum statistical average of the energy density matrix for inner and outer states incoming on the scatterer, writes:
\begin{equation}\label{inputdensitymatrix}
\begin{split}
   &D_{i,\text{in}}(E',E) =\langle \hat{\psi}_{i,\text{in}}^\dagger(E') \hat{\psi}_{i,\text{in}}(E) \rangle-\langle \hat{\psi}_{i,\text{in}}^\dagger(E') \hat{\psi}_{i,\text{in}}(E) \rangle_{\text{off}} \\
   &=\sum_{k,l\in\mathbb{Z}}\    p_{l-k}^{*}p_{l} e^{ik\Omega \tau_{i}}f(E-l\hbar\Omega)\delta(E'-E-k\hbar\Omega)\\
   &\quad -f(E)\delta(E-E').
\end{split}
\end{equation}
The term with the subscript ``off'' is the equilibrium Fermi sea contribution with no ac voltage applied (i.e., obtained by setting $p_{l}=\delta_{l,0}$). In Eq.~\eqref{inputdensitymatrix}, the nonzero diagonal terms connecting energies separated by a multiple $k$ of the photon energy $\hbar\Omega$ are typical of a photo-assisted excitation of the Fermi sea. The term $k=0$, yielding the diagonal part of $D_{i,\text{in}}(E',E)$, is the energy distribution and does not contain any information on the propagation times. Therefore, the inner and outer input states share the same photo-assisted energy distribution:
\begin{equation}
\tilde{f}_{\text{in}}(E)=D_{i,\text{in}}(E,E)=\sum_{l\in\mathbb{Z}}|p_{l}|^{2}[ f(E-l\hbar\Omega)-f(E)].
\end{equation}

Now, let us consider the output operators after the mixing point. Using the scattering matrix~\eqref{eq:smatrix}, they are given by\footnote{Here, we use the notation $\hat\psi_i^+$ for the output operators after the mixing point, instead of $\hat\psi_{i,\text{out}}$, which we keep for the operators after the QPC.}:
\begin{equation}\label{operatormixing}
\begin{split}
\hat{\psi}_{1}^+(E) &=\sqrt{{\ta}} \hat{\psi}_{1,\text{in}}(E)+i\sqrt{{\ra}}\hat{\psi}_{2,\text{in}}(E)\\
\hat{\psi}_{2}^+(E) &=i\sqrt{{\ra}} \hat{\psi}_{1,\text{in}}(E)+\sqrt{{\ra}}\hat{\psi}_{2,\text{in}}(E)
\end{split}
\end{equation}
 This allows one to compute the output density matrix $D_{i}^+(E',E)$. We find, using ${\ra}+{\ta}=1$, that the new energy distribution $\tilde{f}_{i}^+(E)=D_{i}^+(E,E)$ is identical to that of input states (and thus channel-independent):
\begin{equation}
    \tilde{f}_{i}^+(E)=\tilde{f}_{\text{in}}(E)\,.
\end{equation}
The new energy distribution therefore displays, at zero temperature, the same step-wise variation at energies $E_{l}=l\hbar\Omega$, as the one expected if mixing was not present.
\begin{figure}[t]
\centering
\includegraphics[width=\columnwidth]{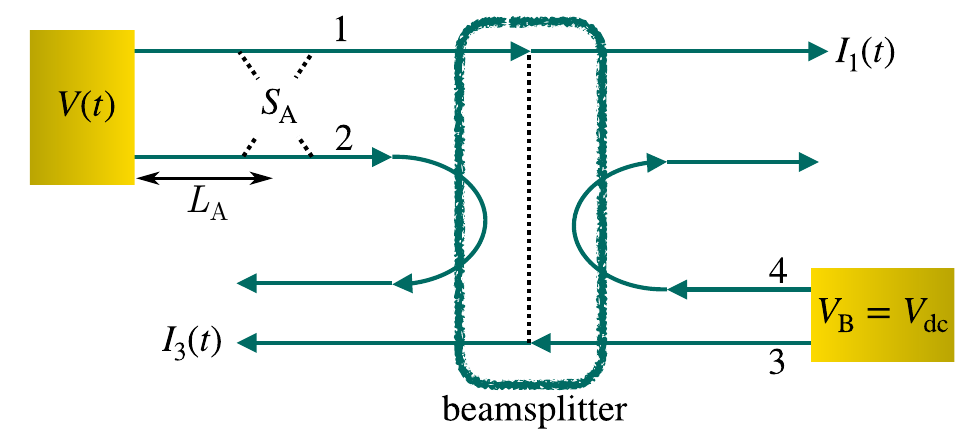}
\caption{A beamsplitter is used to partition the excitations {generated by a time-dependent drive $V(t)$} on channel 1, while fully reflecting those on channel 2. A dc voltage applied on the right contact is used to probe the photo-assisted shot noise in the presence of mixing. The cross-correlation $\mathcal{S}_{13}$ between the outer edge currents $I_{1}(t)$ and $I_{3}(t)$ exiting the beamsplitter is considered in the main text.
At the beamsplitter, outer channels 1 and 3 are transmitted with probability $T$ and reflected with probability $R=1-T$.}
\label{fig:PASN}
\end{figure}
{This has a direct consequence when considering the setup in Fig.~\ref{fig:PASN}, where the mixed edge channels meet at a beamsplitter with another channel incoming from the opposite side and biased by a dc voltage $V_\text{dc}$. Namely, the PASN due to partitioning by the beamsplitter will show singularities when $V_\text{dc}$ is a multiple of $\hbar\Omega/e$.}
This means that the PASN Josephson relation~\cite{Lesovik1994,Pedersen1998,Crepieux2004,Kapfer2019,Safi2019}
\begin{equation}
eV_\text{dc}=k\hbar\Omega\,,\quad k\in\mathbb{Z},
\label{eq:josephsonPASN}
\end{equation}
is not affected by channel mixing. Clearly, this strong result reflects the elastic scattering property of the scatterer mixing the outer and inner edge channels.

\subsubsection{Photoassisted shot noise {after a QPC}}
The last statement can be checked by a direct calculation of the PASN that includes both the mixing in the input copropagating edge channels and the subsequent partitioning at a beamsplitter, described by the scattering matrix~\eqref{eq:S_QPC}. {Notice that here we are not yet considering a HOM configuration, as one of the input sources is simply biased by a constant voltage, used to probe the PASN generated by the other source, see Fig.~\ref{fig:PASN}.} We find the following cross-correlated noise:
\begin{equation}\label{mixedPASN}
\begin{split}
    \mathcal{S}_{13}&=\frac{e^{2}\Omega}{\pi}RT\left[{\ta}{\ra}\sum_{l\in\mathbb{Z}} |l|\Pi_{l}(\Delta_{12})\right.\\
    &\left.-{\ta}\sum_{l\in\mathbb{Z}}P_{1,l} \left|l+\frac{eV_\text{dc}}{\hbar\Omega}\right|-{\ra}\sum_{l\in\mathbb{Z}}P_{2,l}\left|l+\frac{eV_\text{dc}}{\hbar\Omega}\right|\right].
\end{split}
\end{equation}
Here, $P_{1,l}=P_{2,l}=|p_l|^2$ are the photoassisted probabilities induced by the fictitious contacts acting on the outer and inner channels, respectively biased by $V(t-\tau_1)$ and $V(t-\tau_2)$. Moreover, $\Pi_{l}(\Delta_{12})$ is the photoassisted probability stemming from an ac bias $V(t-\tau_1)-V(t-\tau_2)$. Notice that $\Delta_{12}=\tau_1-\tau_2$ is precisely the time delay between $V(t-\tau_1)$ and $V(t-\tau_2)$.
Eq.~\eqref{mixedPASN} shows that the singularities in the shot noise occurring for dc voltages obeying the Josephson relation~\eqref{eq:josephsonPASN} are not affected by mixing before the beamsplitter. Let us now comment on the three terms appearing in Eq.~\eqref{mixedPASN}. The first one leads to positive correlations. It is due to mixing by the scatterer which makes the outer edge incoming at the beamsplitter noisy. Like a thermal noise, it has a bosonic character, leading to a positive contribution to the cross-correlation. The second and the third terms give instead negative correlations, which result from the partitioning of electron-hole pairs photo-excited by the fictitious voltages $V(t-\tau_{1})$ and $V(t-\tau_{2})$ of Fig.~\ref{fig:mixing1}(c) and respectively transmitted with probability ${\ta}$ and ${\ra}$. As $P_{1,l}=P_{2,l}=|p_l|^2$, Eq.~\eqref{mixedPASN} simplifies to:
\begin{equation}\label{mixedPASN2}
\begin{split}
    \mathcal{S}_{13}&=-\frac{e^{2}\Omega}{\pi}RT\left[\sum_{l\in\mathbb{Z}}|p_{l}|^2 \left|l+\frac{eV_\text{dc}}{\hbar\Omega}\right|-{\ta}{\ra}\sum_{l\in\mathbb{Z}}\Pi_l |l|\right].
\end{split}
\end{equation}
We observe from this result that mixing does not affect the excess cross-correlated PASN $\Delta \mathcal{S}_{13}= \mathcal{S}_{13}-\mathcal{S}_{13}|_{V_\text{dc}=0}$, as it just adds a finite, bias-independent offset to the total cross-correlated noise. {Finally, note that $\mathcal{S}_{13}$ is the current noise observed at the outer output channels only, while in Sec.~\ref{sec:mix_full} we will consider the total noise of the inner and outer edge channels $\mathcal{S}_{\text{RL}}$.}

\subsection{Mixing and Hong Ou Mandel fermionic noise: qualitative approach with Levitons}
\label{sec:mix_hom_qualitative}

\begin{figure}[t]
\centering
\includegraphics[width=\columnwidth]{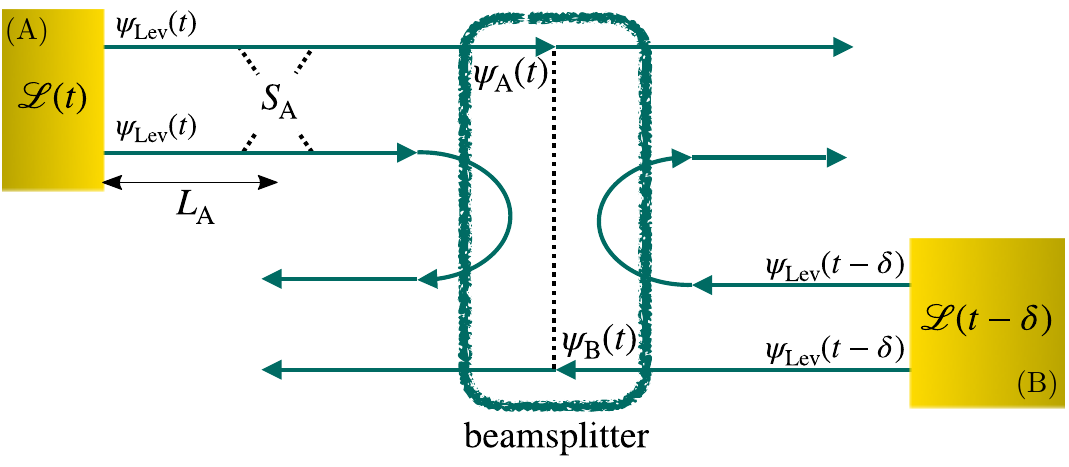}
\caption{A HOM interferometer where both sources inject Levitons and a mixing point is present on the left input arm. Contact A is driven by a Lorentzian pulse $\mathscr{L}(t)$, while a shifted pulse $\mathscr{L}(t-\delta)$ is applied to contact B. The resulting injected Levitons are denoted by the wavefunctions $\psi_\text{Lev}(t)$ and $\psi_\text{Lev}(t-\delta)$. The right- and left-moving Levitons propagate towards the electronic beamsplitter, where they interfere. The two-particle interference manifests in the cross-correlation between the outer edge currents exiting the beam-splitter. }
\label{fig:mixing}
\end{figure}

We now move to discussing how mixing can affect the HOM shot noise. To this aim, we consider a qualitative picture based on Levitons. As sketched in Fig.~\ref{fig:mixing}, we assume that single-pulse Lorentzian voltages $\mathscr{L}(t)$ and $\mathscr{L}(t-\delta)$, see Eq.~\eqref{eq:lor}, are applied to the left and right contacts, thereby injecting identical Levitons on both the inner and outer channels. The current noise at the outputs of the outer edge channels provides a measure of the two-particle HOM interference.

The mixing-free scenario has been discussed in Sec~\ref{sec:basic} and leads to the noise in Eq.~\eqref{eq:single_lev_noise}.
Let us now consider a finite mixing probability ${\ra}=1-{\ta}$ due to the scatterer located at $x=L_\text{A}$. The left and right wavefunctions incoming at the HOM beamsplitter are now given by 
\begin{equation}\label{innerouter1}
\begin{split}
\psi_\text{A}(t) &=\sqrt{{\ta}} \psi_\text{Lev}(t-\tau_{1}-\tau'_{1})+i\sqrt{{\ra}}\psi_{\text{Lev}}(t-\tau_{2}-\tau'_{1})\\
\psi_\text{B}(t) &=\psi_\text{Lev}(t-\delta-\tau_\text{B}).
\end{split}
\end{equation}
Here, $\psi_\text{Lev}(t)$ is the time-domain representation of the Leviton wavefunction, see Eq.~\eqref{eq:lev_wf}.
Moreover, we have introduced the propagation time $\tau'_{1}$ between the mixing scatterer and the left input of the QPC and the propagation time $\tau_\text{B}$ between the right contact and the right input of the QPC. These two times do not play a particular physical role and will be absorbed in a redefinition of the time $t \rightarrow t+\tau'_{1}$ and of the HOM time delay: $\delta \rightarrow \delta-(\tau_\text{B}-\tau'_{1})$. With this redefinition we get:
\begin{equation}\label{innerouter2}
\begin{split}
\psi_\text{A}(t) &=\sqrt{{\ta}} \psi_\text{Lev}(t-\tau_{1})+i\sqrt{{\ra}}\psi_\text{Lev}(t-\tau_{2})\\
\psi_\text{B}(t) &=\psi_\text{Lev}(t-\delta).
\end{split}
\end{equation}
From this equation, we can qualitatively observe that $\psi_\text{A}(t)$ can never match $\psi_\text{B}(t)$ unless $\tau_{1}=\tau_{2}$, namely when the inner and outer propagation velocities are equal. In this case, for $\delta=\tau_1=\tau_{2}$, perfect indistinguishability occurs and the joint probability of finding two Levitons in different outputs is $|\langle \psi_\text{A}(t)|\psi_\text{B}(t)\rangle |^{2}=1$, so the HOM noise vanishes. For a general situation with different propagation velocities ($\tau_{1}\neq \tau_{2}$), the smallest finite mixing ${\ra}\ne 0$ lifts the HOM dip, meaning that the HOM noise is never fully vanishing. For $|\tau_{1}-\tau_{2}|>w$ we can even expect two HOM minima located in the vicinity of $\tau=\tau_{1}$ and $\tau=\tau_{2}$, still having a nonzero noise at each minimum. Indeed, using the representation of fictitious contacts discussed above, one can use a gauge transformation by subtracting from all contacts the same ac voltage $V_\text{B}(t)=\mathscr{L}(t-\delta)$. For $\delta=\tau_{1}$, the HOM noise is obtained by calculating the PASN where only the inner fictitious contact is biased by a finite ac voltage equal to $\mathscr{L}(t-\tau_2)-\mathscr{L}(t-\tau_{1})$ while the two remaining contacts are grounded. Similarly, one finds the HOM noise close to the second minimum at $\delta=\tau_{2}$ by evaluating the PASN while grounding both the fictitious outer contact and the right contact and applying the voltage $\mathscr{L}(t-\tau_{1})-\mathscr{L}(t-\tau_{2})$ on the fictitious inner contact. 

Finally, the qualitative picture just discussed is confirmed by the full calculation of section~\ref{sec:mix_full}, from which one gets the following noise observed at the output contacts:
\begin{equation}
\begin{split}
\mathcal{S}_\text{HOM}(\delta)&\propto RT\{{\ta}[1-J(\delta-\tau_{1})]+{\ra}[1-J(\delta-\tau_{2})]\}\\
&\quad+T^2{\ra}{\ta}[1-J(\tau_{1}-\tau_{2})].
\end{split}
\label{eq:HOM-Lev-1mix}
\end{equation}
Here, $J(\delta)$ is again the coincidence probability given by Eq.~\eqref{eq:lev_overlap}, which measures the overlap of the wavefunctions of Levitons incoming onto the beamsplitter in the mixing-free scenario.
For ${\ra}=0$ and ${\ta}=1$, only the first term in~\eqref{eq:HOM-Lev-1mix} contributes and a full HOM dip is reached at $\delta=\tau_1$, corresponding to the process in which the Leviton injected on channel 1 is transmitted at the mixing point and collides with the other incoming from channels 3. Likewise, for perfect reflection at the mixing point, ${\ra}=1$ and ${\ta}=0$, the only contribution, i.e.~the second term in~\eqref{eq:HOM-Lev-1mix}, comes from the process where the Leviton injected into channel 2 is transferred to channel 1 at the mixing point and then collides with the incoming state injected into channel 3 (lower edge, outer channel). Then a full dip occurs when $\delta=\tau_2$. At finite mixing, $\ra\neq 0$ and $\ta\neq 0$, both terms contribute and the HOM noise does not vanish.

\section{Full HOM noise in the presence of mixing}
\label{sec:mix_full}
\begin{figure}[t]
\centering
\begin{overpic}[percent=true,width=\columnwidth,grid=false]{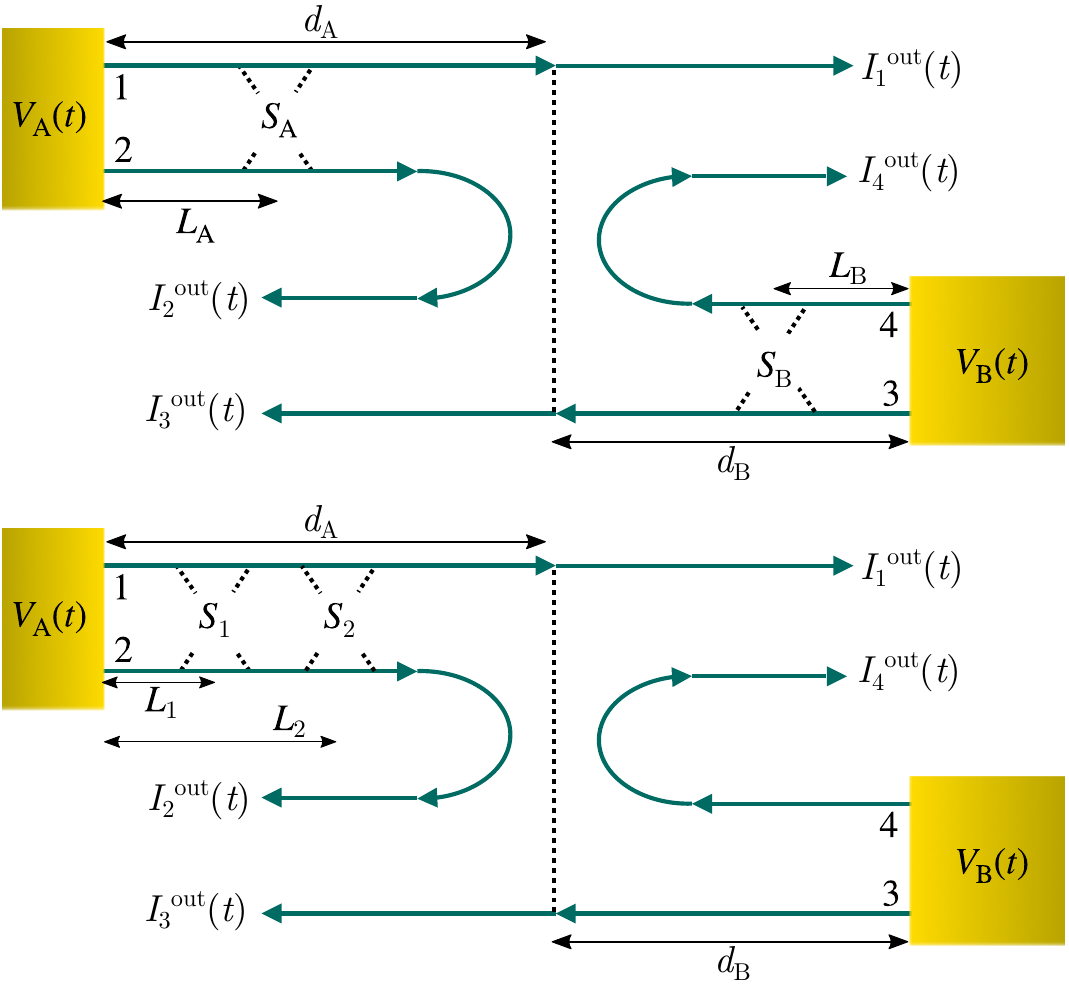}
\put(0,6){(b)}
\put(0,52.5){(a)}
\end{overpic}
\caption{Setup for the evaluation of the full HOM noise in the presence of mixing. (a) One mixing point is present on each edge. Mixing is described by the scattering matrices $S_\text{A}$ and $S_\text{B}$. (b) Configuration with two mixing points (described by $S_1$ and $S_2$) on the same edge. This second setting allows us to investigate the effect of additional interferences.}
\label{fig:HOM-full}
\end{figure}
We have shown in the previous Section that when two copropagating channels are mixed, their ac transport properties are modified: in particular, unlike in the dc case, current noise is generated. We have then argued that a reduced visibility of the HOM dip is expected. However, a precise expression for the HOM noise in the presence of mixing, as provided in~\eqref{eq:HOM-Lev-1mix} cannot be obtained by a qualitative reasoning and requires a more careful evaluation.

In this Section, we therefore address the full calculation of the HOM noise in the presence of mixing and present explicit results in various configurations. Specifically, we consider the settings sketched in Fig.~\ref{fig:HOM-full}.
For the sake of generality, we allow the HOM interferometer to be asymmetric, namely we consider different distances ($d_\text{A}$ and $d_\text{B}$) from the injecting contacts to the central QPC. Finally, as far as mixing points are concerned, we investigate three possibilities.
The first one, considered in Sec.~\ref{sec:hom_1vs0}, was already sketched in Fig.~\ref{fig:mixing}, where a single mixing point is present in one edge only. Here, the difference with Fig.~\ref{fig:mixing} is that we consider generic periodic sources instead of the injection of Levitons.
The second case, described in Sec.~\ref{sec:hom_1vs1}, is shown in Fig.~\ref{fig:HOM-full}(a), where one mixing points is present on both edges, at distances $L_\text{A}$ and $L_\text{B}$ from the corresponding injecting contact. These mixing points are characterized by scattering matrices $S_\text{A}$, given in Eq.~\eqref{eq:sA}, and $S_\text{B}$ having an identical expression, with $\ta\to\tb$ and $\ra\to\rb$.
In the third case, drawn in Fig.~\ref{fig:HOM-full}(b) and addressed in Sec.~\ref{sec:mix2}, two mixing points are present in the same edge, respectively located at positions $L_1$ and $L_2$ from the injecting source. As before, the mixing points are described by scattering matrices $S_1$ and $S_2$, parametrized as in~\eqref{eq:sA}, but with transmissions $T_1$ and $T_2$, respectively.

{The reasons why we choose these three configurations are the following. First, the case of a single mixing point in one input arm only (Sec.~\ref{sec:hom_1vs0}) provides the minimal model able to explain a reduced HOM dip. Then, in a realistic setup mixing is expected in both input arms, motivating us to consider the configuration discussed in Sec.~\ref{sec:hom_1vs1}. Finally, in Sec.~\ref{sec:mix2} we provide the simplest example that illustrates the effect of consecutive mixing points in the same input arm, showing that this leads to additional interference terms. More complicated scenarios are beyond the scope of this works and, importantly, our results already provide a sensible explanation of recent experimental data~\cite{Taktak2021}.}

\subsection{Single mixing point}
\label{sec:hom_1vs0}
If a single mixing point is present, say in the left incoming arm of the interferometer, $S_\text{B}=\mathbbm{1}$. In this case, we find
\begin{equation}
\begin{split}
\mathcal{S}_\text{HOM}&=T^2\ta\ra\mathcal{S}_0(\Delta_{12})+RT\ta\mathcal{S}_0(\Delta_{13})\\
&+RT\ra\mathcal{S}_0(\Delta_{23}),
\end{split}
\label{eq:HOM-full_1vs0}
\end{equation}
where the effective time delays appearing in the $\mathcal{S}_0$ functions are
\begin{subequations}
\begin{align}
    \Delta_{12}&=\frac{L_\text{A}}{v_1}-\frac{L_\text{A}}{v_2},\\
    \Delta_{13}&=\delta+\frac{d_\text{A}}{v_1}-\frac{d_\text{B}}{v_3},\\
    \Delta_{23}&=\delta+\frac{L_\text{A}}{v_2}+\frac{d_\text{A}-L_\text{A}}{v_1}-\frac{d_\text{B}}{v_3}\,.
\end{align}
\label{eq:delays_1}
\end{subequations}
In these expressions, the subscripts in $\Delta_{ij}$ indicate that the excitations responsible for the contribution to the noise associated with $\mathcal{S}_0(\Delta_{ij})$ are injected into channels $i$ and $j$.

Some comments about the result~\eqref{eq:HOM-full_1vs0} are appropriate. Firstly, it reduces to Eq.~\eqref{eq:hom_nomix} in the absence of mixing ($\ta=1$ and $\ra=0$). Secondly, we note that when $d_\text{A}=d_\text{B}$, $v_3=v_1$, and for the injection of Levitons, Eq.~\eqref{eq:HOM-full_1vs0} reproduces (after a suitable redefinition of the HOM delay $\delta$) what we anticipated in Eq.~\eqref{eq:HOM-Lev-1mix}.
Thirdly, the structure of Eq.~\eqref{eq:HOM-full_1vs0} makes the interpretation of the result straightforward. Indeed, we observe that the total HOM noise is composed of a combination of independent HOM noises of the form~\eqref{eq:base-HOM}, each with its own time delay, depending on the different paths that the colliding electrons took to arrive at the QPC. For instance, the term involving the delay $\Delta_{13}$ is the simplest and originates from electrons being injected in channels 1 and 3 and remaining in the same channels when passing through the respective mixing point. Similarly, the term associated with $\Delta_{23}$ comes from the collision of an excitation injected into channel 3 and propagating directly to the QPC and an excitation injected into channel 2 and transferred to channel 1 at the mixing point at distance $L_\text{A}$ from the source. The difference in the times of flight of these two paths is exactly given by $\Delta_{23}$. A sketch illustrating such paths is provided in Fig.~\ref{fig:sketch_paths_direct}(a). Finally, the coexistence of three different terms in Eq.~\eqref{eq:HOM-full_1vs0} shows that the presence of mixing is able to introduce a ``which-path'' information (encoded in the different time delays), thus leading to an overall breaking of indistinguishability and a reduction of the HOM dip.

Such a reduction can indeed be easily understood from Eq.~\eqref{eq:HOM-full_1vs0}: it is sufficient to recall that the function $\mathcal{S}_0$ vanishes when its argument does. Then, it is clear that in the presence of mixing and if channels propagate with different velocities $v_1\neq v_2$, it is not possible to synchronize all arrival times at the QPC and this results in a never vanishing noise. Clearly, the amount by which the HOM dip is reduced depends both on the difference in the time delays~\eqref{eq:delays_1}, but also on the mixing strength that modifies the relative weight of the three terms in Eq.~\eqref{eq:HOM-full_1vs0}.

\subsection{Two mixing points in different arms}
\label{sec:hom_1vs1}
As expected from the discussion of Sec.~\ref{sec:hom_1vs0}, the addition of a second mixing point in the other arm of the interferometer, see Fig.~\ref{fig:HOM-full}(a), introduces more independent HOM-like contributions to the total noise, because more possible paths from the sources to the QPC are present. The total HOM noise in the presence of two mixing points as in Fig.~\ref{fig:HOM-full}(a) is found to be
\begin{equation}
    \begin{split}
    \mathcal{S}_\text{HOM}&=RT\ta\tb\mathcal{S}_0(\Delta_{13})+RT\ta\rb\mathcal{S}_0(\Delta_{14})\\
    &\quad+RT\ra\tb\mathcal{S}_0(\Delta_{23})+RT\ra\rb\mathcal{S}_0(\Delta_{24})\\
    &\quad+T^2\ra\ta\mathcal{S}_0(\Delta_{12})+T^2\tb\rb\mathcal{S}_0(\Delta_{34}).
\end{split}\label{eq:HOM-full_1vs1_2}
\end{equation}
\begin{figure}[t]
    \centering
    \begin{overpic}[percent=true,width=0.9\columnwidth,grid=false]{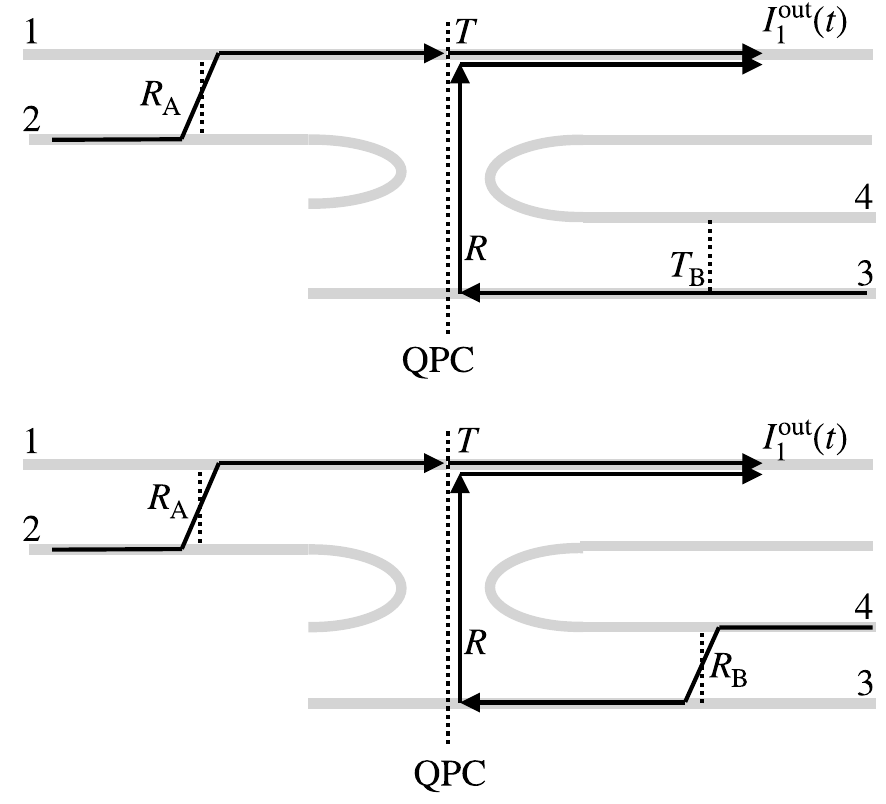}
    \put(7,24){$L_\text{A}/v_1$}
    \put(7,71){$L_\text{A}/v_1$}
    \put(25,41){$(d_\text{A}-L_\text{A})/v_1$}
    \put(25,87){$(d_\text{A}-L_\text{A})/v_1$}
    \put(82,22){$L_\text{B}/v_4$}
    \put(82,53){$L_\text{B}/v_3$}
    \put(53.5,6.5){$(d_\text{B}-L_\text{B})/v_3$}
    \put(53.5,53){$(d_\text{B}-L_\text{B})/v_3$}
    \put(4,5){(b)}
    \put(4,55){(a)}
    \end{overpic}
    \caption{Sketch of two paths contributing to the HOM noise~\eqref{eq:HOM-full_1vs1_2}. In both examples, the times of flight for each part of the paths are indicated, together with the required transmission and reflection probabilties with which excitations are scattered at each mixing point and at the QPC. In (a) the total time delay is $\Delta_{23}$ and the probability of such a collision is $RT\ra\tb$, while in (b) one finds $\Delta_{24}$, with probability $RT\ra\rb$. All terms in~\eqref{eq:HOM-full_1vs1_2} can be understood in this way.}
    \label{fig:sketch_paths_direct}
\end{figure}
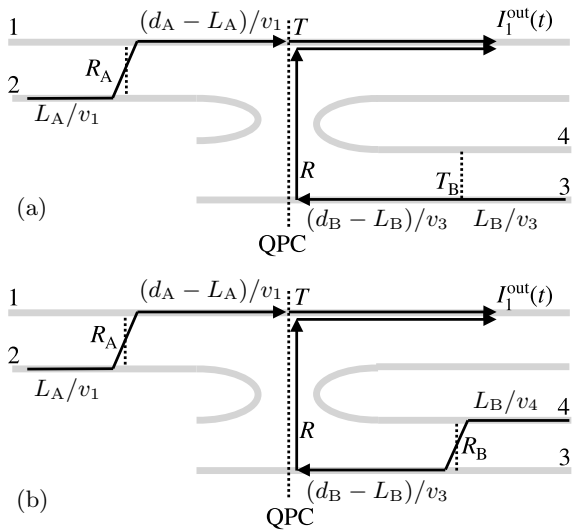
Here, some of the effective time delays $\Delta_{ij}$ are already given in Eq.~\eqref{eq:delays_1}. The remaining ones read
\begin{subequations}
\begin{align}
    \Delta_{14}&=\delta+\frac{d_A}{v_1}-\frac{L_\text{B}}{v_4}-\frac{d_B-L_\text{B}}{v_3},\\
    \Delta_{24}&=\delta+\frac{L_\text{A}}{v_2}+\frac{d_A-L_\text{A}}{v_1}-\frac{L_\text{B}}{v_4}-\frac{d_B-L_\text{B}}{v_3},\\
    \Delta_{34}&=\frac{L_\text{B}}{v_3}-\frac{L_\text{B}}{v_4}\,.
\end{align}
\label{eq:delays}
\end{subequations}
Once again, these time delays do not just contain the bare time shift $\delta$ between the injecting sources, but also other asymmetry factors due to different propagation lengths in the interferometer and velocity mismatches among the different channels.

The origin of the various terms in Eq.~\eqref{eq:HOM-full_1vs1_2} is the following. Since the HOM noise is the cross-correlation of currents $I_\text{R}$ and $I_\text{L}$, given in Eq.~\eqref{eq:out-currents}, one has in general $\mathcal{S}_\text{RL}=\mathcal{S}_{12}+\mathcal{S}_{13}+\mathcal{S}_{24}+\mathcal{S}_{34}$. However, $\mathcal{S}_{24}=0$ because channels 2 and 4 are fully reflected at the QPC and therefore do not contribute to the partition noise. In the remaining terms, $\mathcal{S}_{13}$ leads to the first two lines in Eq.~\eqref{eq:HOM-full_1vs1_2}, plus two additional terms: $-RT\ra\ta\mathcal{S}_0(\Delta_{12})$ and $-RT\tb\rb\mathcal{S}_0(\Delta_{34})$. Instead, $\mathcal{S}_{34}=T\tb\rb\mathcal{S}_0(\Delta_{34})$ and $\mathcal{S}_{12}=T\ra\ta\mathcal{S}_0(\Delta_{12})$, so that they combine with the corresponding terms from $\mathcal{S}_{13}$ to yield the last line of~\eqref{eq:HOM-full_1vs1_2}. {Note that the terms proportional to $T^2$ would be present even in the absence of the central QPC. Indeed, these contributions are uniquely due to mixing, that makes the input channels of the interferometer noisy. In contrast, the terms involving the HOM time delay $\delta$ require the presence of the QPC.}

As noted in Sec.~\ref{sec:hom_1vs0}, we can interpret every term in Eq.~\eqref{eq:HOM-full_1vs1_2} by considering the associated time delays resulting from one of the possible paths that the colliding excitations follow from the source to the QPC. This is shown in Fig.~\ref{fig:sketch_paths_direct}.
Compared to Eq.~\eqref{eq:HOM-full_1vs0}, here more terms appear in the total HOM noise, {due to the presence of the second mixing point.}
\begin{figure}[t]
    \centering
    \begin{overpic}[percent=true,width=0.9\columnwidth,grid=false]{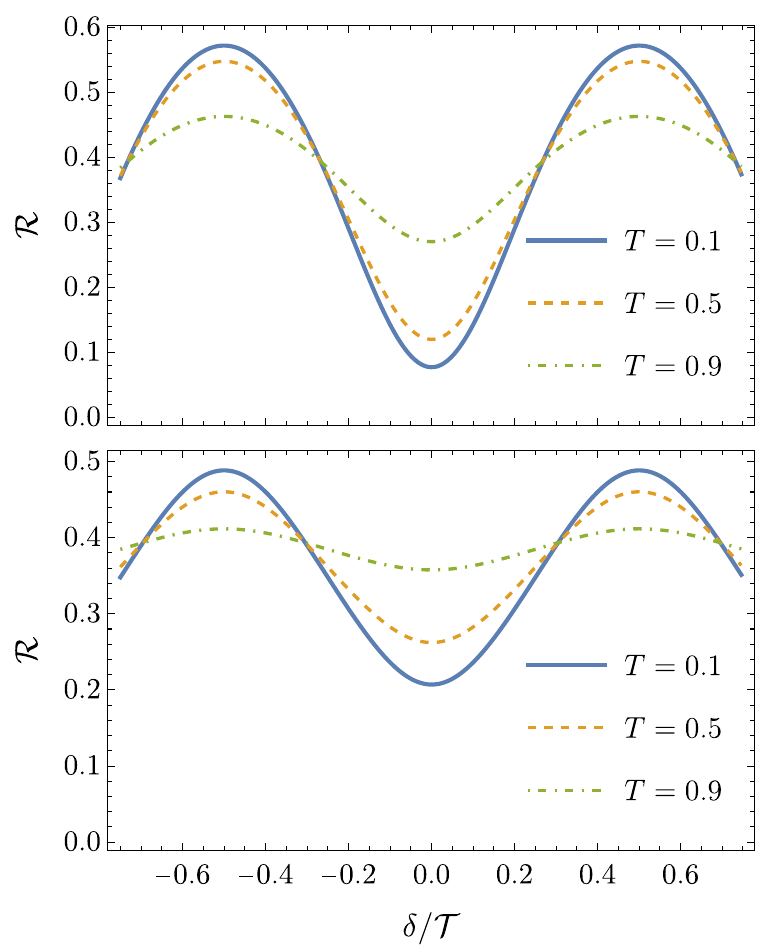}
    \put(12,13){(b) $R_\text{A}=R_\text{B}=0.5$}
    \put(12,57){(a) $R_\text{A}=R_\text{B}=0.1$}
    \end{overpic}
    \caption{Normalized HOM noise as a function of the time delay $\delta$ between the sources $V_\text{A}(t)$ and $V_\text{B}(t)$ and for different values of the QPC transmission $T$. Both sources are sinusoidal and inject on average $q=1$ electron per period, see Eq.~\eqref{eq:q}. (a) Weak mixing scenario, with $\ra=\rb=0.1$. (b) Strong mixing case, with $\ra=\rb=0.5$. In both (a) and (b), we considered $d_\text{A}=d_\text{B}=12\,\mu$m, $\theta=30\,$mK, $L_\text{A}=L_\text{B}=6\,\mu$m. Finally, the velocity mismatch is $v_\text{out}=1.1v_\text{in}$, with $v_\text{in}=3\times 10^4\,$m/s, and the driving frequency is $14\,$GHz.}
    \label{fig:HOM_weak-vs-strong-mixing}
\end{figure}
In the following, we discuss some results for the case of a symmetric interferometer with $d_\text{A}=d_\text{B}$. Furthermore, we assume that the velocities on the outer channels 1 and 3 are equal, $v_1=v_3=v_\text{out}$ and, similarly, $v_\text{2}=v_\text{4}=v_\text{in}$ for the inner channels. In most of the plots, we consider a cosine drive
\begin{equation}
    V(t)=V_0[1-\cos(\Omega t)],
    \label{eq:cosine}
\end{equation}
{injecting on average} $q$ electrons per period into the edge channel, with
\begin{equation}
    q=\frac{eV_0}{\hbar\Omega}\,.
    \label{eq:q}
\end{equation}
When $q=1$, one electron per period is injected. However, a cosine drive does not generate a purely electronic excitation, unlike Lorentzian pulses.

\begin{figure}[t]
    \centering
    \begin{overpic}[percent=true,width=0.95\columnwidth,grid=false]{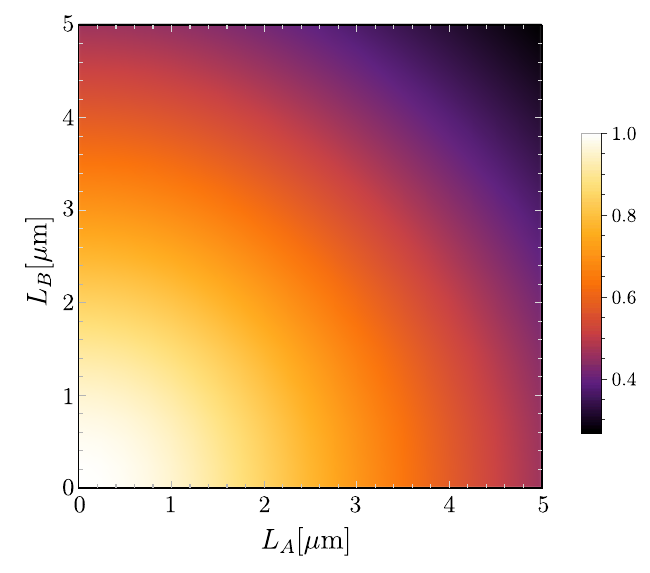}
    \put(90,70){$\mathcal{V}$}
    \end{overpic}
    \caption{Visibility $\mathcal{V}$ of the HOM signal, see Eq.~\eqref{eq:vis}, as a function of the positions $L_\text{A}$ and $L_\text{B}$ of the mixing points along the edges. In this plot, we set $T=0.7$, $v_\text{out}=1.2 v_\text{in}$, and a strong mixing $\ra=\rb=0.5$. All other parameters are as in Fig.~\ref{fig:HOM_weak-vs-strong-mixing}.}
    \label{fig:visibility}
\end{figure}
In Fig.~\ref{fig:HOM_weak-vs-strong-mixing}, we show the normalized HOM noise in a scenario where mixing occurs at the same distance in both edges $(L_\text{A}=L_\text{B})$. Both in (a) and (b), we clearly observe that the HOM noise never vanishes and the contrast of the HOM dip is thus reduced. The difference between the two plots is that in (a) we have considered a weak mixing scenario, with $\ra=\rb=0.1$, while in (b) the channels are more strongly mixed ($\ra=\rb=0.5$). As is intuitively clear, the stronger the mixing, the more the contrast in the HOM dip is reduced. This is because when mixing is strong, all the terms in Eq.~\eqref{eq:HOM-full_1vs1_2} are relevant, each having its own time delay. On top of this, we also notice that by increasing the beamsplitter transmission, the contrast is further reduced. This is because, when $T$ is large, the terms proportional to $T^2$ in Eq.~\eqref{eq:HOM-full_1vs1_2} are dominant. But those terms do not depend on the time delay $\delta$, see Eq.~\eqref{eq:delays}, which makes the modulation of the HOM noise as a function of $\delta$ very weak. As a last comment in Fig.~\ref{fig:HOM_weak-vs-strong-mixing}, we notice that the curves are symmetric with respect to $\delta=0$ and have a minimum at that point. This is a consequence of the specific symmetric choice of parameters ($L_\text{A}=L_\text{B}$, $\ta=\tb$). By relaxing this condition, the minimum of the HOM noise can drift both towards positive and negative values of $\delta$ and asymmetries may appear as well (not shown). 
\begin{figure}[t]
    \centering
    \begin{overpic}[percent=true,width=\columnwidth,grid=false]{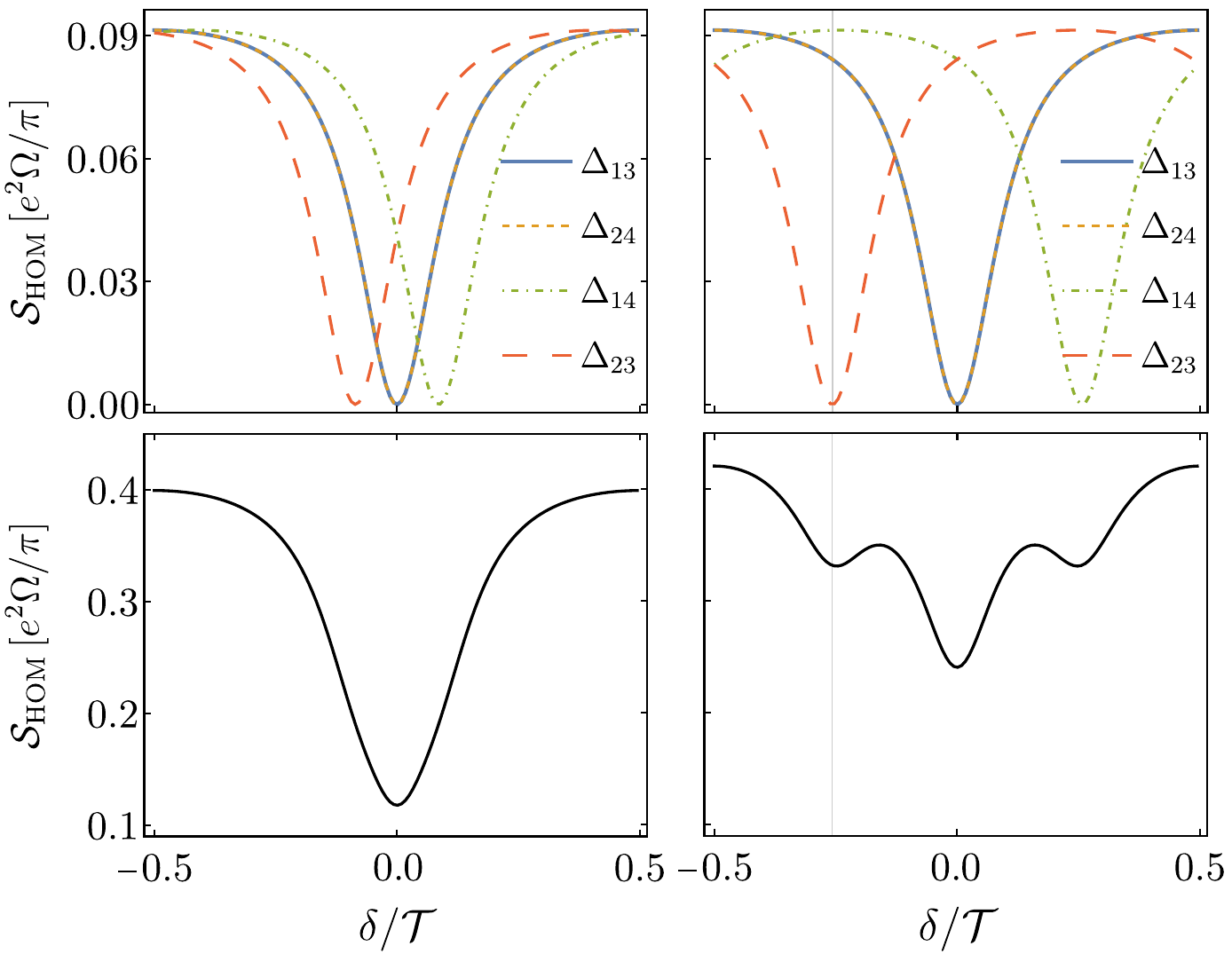}
    \put(12.5,11){(b)}
    \put(12.5,45.5){(a)}
    \put(58,45.5){(c)}
    \put(58,11){(d)}
    \end{overpic}
     \caption{Appearance of side dips in the HOM noise due to mixing. In this figure, we have chosen a periodic train of Lorentzian pulses $\mathscr{L}(t)$, see Eq.~\eqref{eq:lor}, with width $w/\mathcal{T}=0.05$. In the top row, different terms contributing to the HOM noise~\eqref{eq:HOM-full_1vs1_2} are plotted separately. {The delays $\Delta_{ij}$ in the legend are shorthand notations for $\mathcal{S}_0(\Delta_{ij})$.} The bottom row shows instead the total HOM noise. For the left column, we have chosen $L_\text{A}=L_\text{B}=2\,\mu$m, while in the right column $L_\text{A}=L_\text{B}=6\,\mu$m. For both rows, $\ta=\tb=0.5$ and $T=0.3$. All other parameters are the same as in Fig.~\ref{fig:HOM_weak-vs-strong-mixing}. In (a), the minima of the terms involving $\Delta_{24}$ and $\Delta_{23}$ are not separated enough and the corresponding total HOM noise (b) then only shows a single dip. In (c), instead, a greater separation of these minima is responsible for the appearance of side dips in the HOM noise (d).}
    \label{fig:side-dips}
\end{figure}
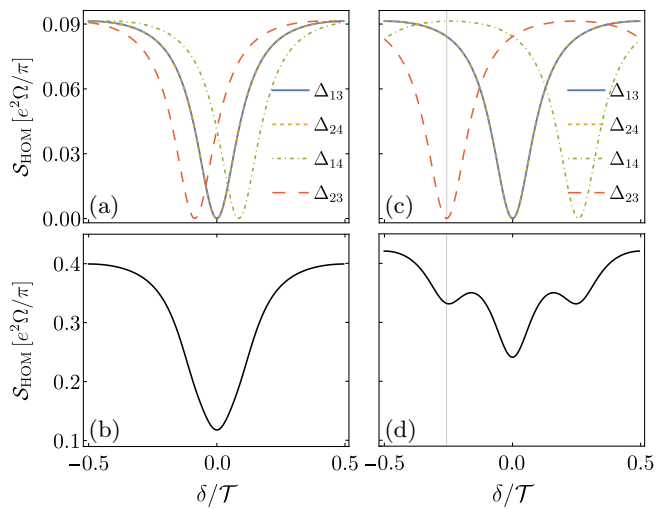

By increasing the distances $L_\text{A,B}$ from the sources to the mixing points, the effective time delays~\eqref{eq:delays} differ more and more, resulting in a decrease of the HOM contrast. This is seen in Fig.~\ref{fig:visibility}, where the visibility of the HOM dip is shown as a function of $L_\text{A}$ and $L_\text{B}$. The visibility is defined as
\begin{equation}
    \mathcal{V}=\frac{\max_\delta(\mathcal{S}_\text{HOM})-\min_\delta(\mathcal{S}_\text{HOM})}{\max_\delta(\mathcal{S}_\text{HOM})+\min_\delta(\mathcal{S}_\text{HOM})}.
    \label{eq:vis}
\end{equation}
In Fig.~\ref{fig:visibility}, we observe that indeed an increase of both $L_\text{A}$ and $L_\text{B}$ leads to a decrease in the visibility.

\subsection{Influence of the pulse width}
Until now, all the results have been shown by considering a cosine drive, see Eq.~\eqref{eq:cosine}. Apart from the frequency $\Omega$, this drive only has one additional parameter, namely the amplitude $V_0$ determining the average injected charge per period. Another important feature that can influence the HOM noise is the width of each pulse with respect to the period. For this, other drives than the cosine voltage~\eqref{eq:cosine} must be considered. Here, we choose a periodic train of Lorentzian pulses $\mathscr{L}(t)$, see Eq.~\eqref{eq:lor}, whose width is parametrized by $w$.
Thanks to this additional parameter, important qualitative differences in the HOM signal can be observed compared to Fig.~\ref{fig:HOM_weak-vs-strong-mixing}, namely the appearance of multiple minima (of different height) in the HOM noise in the presence of mixing. This effect is best observed for the symmetric case $L_\text{A}=L_\text{B}=L$. Under this condition (and recalling that we have assumed $v_1=v_3=v_\text{out}$, $v_2=v_4=v_\text{in}$, and $d_\text{A}=d_\text{B}$), we find from Eq.~\eqref{eq:delays} that $\Delta_{13}=\Delta_{24}=\delta$, while $\Delta_{14}=\delta+L(v_\text{out}^{-1}-v_\text{in}^{-1})$ and $\Delta_{23}=\delta-L(v_\text{out}^{-1}-v_\text{in}^{-1})$. As already mentioned, the remaining time delays $\Delta_{12}$ and $\Delta_{34}$ are independent of $\delta$ and therefore the associated terms do not contribute to the $\delta$-dependent modulation of the HOM noise. Because $\Delta_{13}=\Delta_{24}=\delta$, the terms in~\eqref{eq:HOM-full_1vs1_2} containing these delays share the same minima (for $\delta=k\mathcal{T},k\in\mathbb{Z}$) and maxima (for $\delta=(2k+1)\mathcal{T}/2,k\in\mathbb{Z}$), and may only differ in amplitude. On the other hand, the two terms associated with $\Delta_{14}$ and $\Delta_{23}$ have minima that shift in opposite directions when increasing $L$ and/or the velocity mismatch. If these minima become sufficiently well separated within a given period of the drive, a local minimum in the full HOM signal may occur in correspondence of $\pm L(v_\text{out}^{-1}-v_\text{in}^{-1})$. The other condition for this to happen is that the minima in the terms containing $\Delta_{14}$ and $\Delta_{23}$ are well localized. This happens for narrow excitations, with a small width compared to the driving period. We illustrate this effect in Fig.~\ref{fig:side-dips}.

Our findings suggest that there might be an alternative interpretation of previous results, where the appearance of side dips and a nonvanishing central HOM dip were attributed to fractionalization~\cite{Bocuillon2013Science,Wahl2014,Marguerite2016}. Indeed, we have shown (at least qualitatively) that mixing is able to produce both these features. {This constitutes a starting point for a quantitative comparison with experimental data.} {For completeness, we remind the reader that in the experiments in~\cite{Bocuillon2013Science,Marguerite2016} the excitations were injected into the outer edge channel only.}

\subsection{Two mixing points on the same edge}
\label{sec:mix2}
We now {address the final step of our analysis and} derive an exact expression for the noise in the case of two mixing points on the same edge, for arbitrary mixing strength.
Let us consider the setup sketched in Fig.~\ref{fig:HOM-full}(b). Here, the main difference compared to Sec.~\ref{sec:hom_1vs0} is the larger number of possible paths incoming at the beamsplitter. This difference is reflected in the form of the HOM noise, which is not as simple as Eq.~\eqref{eq:HOM-full_1vs0}. In fact, we find the following decomposition:
\begin{equation}
    \mathcal{S}_\text{HOM}=\mathcal{S}_\text{a}+\mathcal{S}_\text{b}+\mathcal{S}_\text{c}+\mathcal{S}_\text{d}\,,
    \label{eq:HOM-full_2vs0}
\end{equation}
where the different contributions arise from the collision of distinct types of paths involving different time delays, as we now explain.
The first term, $\mathcal{S}_\text{a}$ reads
\begin{equation}
\begin{split}
    &\mathcal{S}_\text{a}=2T^2T_2T_1R_2R_1\mathcal{S}_0(\Delta)+RT[T_2T_1\mathcal{S}_0(\Delta_{13})\\
    &+T_2R_1\mathcal{S}_0(\Delta_{23}^1)+R_2T_1\mathcal{S}_0(\Delta_{23}^2)+R_2R_1\mathcal{S}_0(\Delta+\Delta_{13})]\\
    &+T^2[T_2^2R_1T_1\mathcal{S}_0(\Delta_{12}^1)+T_1^2R_2T_2\mathcal{S}_0(\Delta_{12}^2)\\
    &+R_1^2T_2R_2\mathcal{S}_0(\Delta+\Delta_{12}^1)+R_2^2T_1R_1\mathcal{S}_0(\Delta_{12}^1)],
\end{split}
\label{eq:HOM_2vs0_direct}
\end{equation}
where $\Delta_{13}$ is given in Eq.~\eqref{eq:delays_1} and
\begin{subequations}
\begin{align}
    \Delta&=(L_2-L_1)(v_1^{-1}-v_2^{-1}),\label{eq:delta}\\
    \Delta_{12}^{1,2}&=L_{1,2}(v_1^{-1}-v_2^{-1}),\\
    \Delta_{23}^{1,2}&=\Delta_{13}+L_{1,2}(v_1^{-1}-v_2^{-1}).
\end{align}
\end{subequations}
Each term in Eq.~\eqref{eq:HOM_2vs0_direct} can be interpreted by identifying paths of the same type of those shown in Fig.~\ref{fig:sketch_paths_direct} and calculating their arrival time delay.

Next, $\mathcal{S}_\text{b}$ can be expressed as
\begin{equation}
    \mathcal{S}_\text{b}=T^2\sqrt{T_2T_1R_2R_1}(R_2-T_2)(T_1-R_1)\mathcal{X}(\Delta).
    \label{eq:HOM_2vs0_dir-int}
\end{equation}
\begin{figure}[t]
    \centering
    \begin{overpic}[percent=true,width=0.8\columnwidth,grid=false]{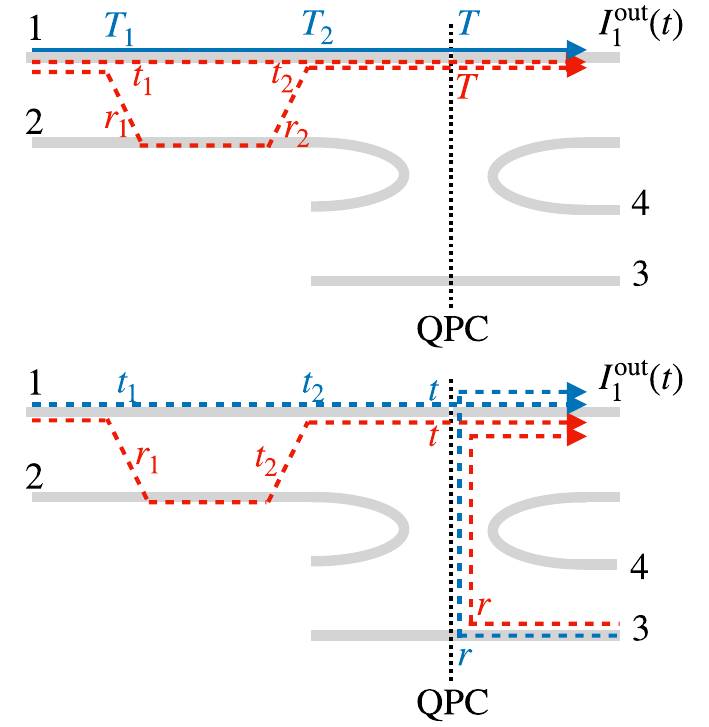}
    \put(4,5){(b)}
    \put(4,60){(a)}
    \end{overpic}
    \caption{(a) Paths contributing to the term $\mathcal{S}_\text{b}$ in~\eqref{eq:HOM_2vs0_dir-int}. The blue line represents a process weighted by probabilities $T_{1,2}$, while the red, dashed lines are weighted by probability amplitudes $r_{1,2}=\sqrt{R_{1,2}}$ and $t_{1,2}=\sqrt{T_{1,2}}$. Notice that the solid blue line results from a combination of two identical dashed lines, each weighted by a probability amplitude. Taken together, the paths in (a) produce one of the terms in Eq.~\eqref{eq:HOM_2vs0_dir-int}. (b) One of the possible pairs of paths contributing to $\mathcal{S}_\text{d}$ in Eq.~\eqref{eq:HOM_2vs0_tpi}. Specifically, the drawn paths generate the term with $\mathcal{Z}(\Delta_{13},\Delta+\Delta_{13})$. Indeed, the two arguments of the function $\mathcal{Z}$ are the difference in times of flight between paths of the same color.}
    \label{fig:sketch_paths_interf}
\end{figure}
It arises from the combination at the output of the interferometer of paths as those sketched in Fig.~\ref{fig:sketch_paths_interf}(a). Here, we have the blue line representing a direct path and the red, dashed ones standing for an interfering path. By calculating the difference in the times of flight between the two red lines, one precisely finds $\Delta$, see Eq.~\eqref{eq:delta}, which is the parameter entering the function $\mathcal{X}$, whose explicit expression is reported in Appendix~\ref{app:2mixing-points}. {It is important to notice that $\mathcal{S}_\text{b}$ is independent of the time delay $\delta$ between the injecting sources. Therefore, it affects the visibility of the HOM signal by modifying the denominator of Eq.~\eqref{eq:vis} only.}

The third term, $\mathcal{S}_\text{c}$ arises from current fluctuations associated with a pair of paths like the dashed ones in Fig.~\ref{fig:sketch_paths_interf}(a). There are only a few terms with this origin, and the final result has the simple expression
\begin{equation}
    \mathcal{S}_\text{c}=2T^2T_2T_1R_2R_1\mathcal{Y}(\Delta)\,.
    \label{eq:HOM_2vs0_1pi}
\end{equation}
The function $\mathcal{Y}$ is explicitly reported in App.~\ref{app:2mixing-points}, but the presence of the time delay $\Delta$ is again intuitively understood by considering the difference in times of flight in the amplitudes of the interfering paths, see Fig.~\ref{fig:sketch_paths_interf}(a). {Since $\mathcal{S}_\text{c}$ is independent of $\delta$, it affects the visibility in Eq.~\eqref{eq:vis} is the same way as $\mathcal{S}_\text{b}$.}

Finally, $\mathcal{S}_\text{d}$ is given by
\begin{equation}
\begin{split}
    \mathcal{S}_\text{d}&=-T^2T_2T_1R_2R_1[\mathcal{Z}(\Delta_{12}^2,\Delta+\Delta_{12}^2)+\mathcal{Z}(\Delta_{12}^1,\Delta_{12}^1)]\\
    &+RT\sqrt{T_2T_1R_2R_1}[\mathcal{Z}(\Delta_{23}^1,\Delta_{23}^2)-\mathcal{Z}(\Delta_{13},\Delta+\Delta_{13})]\\
    &+T^2\sqrt{T_2T_1R_2R_1}[T_1(T_2-R_2)\mathcal{Z}(\Delta_{12}^1,\Delta_{12}^2)\\
    &+R_1(R_2-T_2)\mathcal{Z}(\Delta_{12}^1,\Delta+\Delta_{12}^1)]\,.
\end{split}
\label{eq:HOM_2vs0_tpi}
\end{equation}
As shown in Fig.~\ref{fig:sketch_paths_interf}(b), each of the terms in the previous equation can be described in terms of a pair of interfering paths that combine at the output of the interferometer and produce current fluctuations detected in the HOM noise. While referring the reader to App.~\ref{app:2mixing-points} for the full expression of $\mathcal{Z}$, it is worth mentioning here that the two time delays appearing in this function can be understood by considering the different pairs of interfering paths. For instance, with the help of Fig.~\ref{fig:sketch_paths_interf}(b), one can notice that the difference in times of flight in the red paths is given by $\Delta+\Delta_{13}$, while the blue ones are delayed by $\Delta_{13}$. Moreover, by considering the transmission and reflection amplitudes, one also finds that the process in Fig.~\ref{fig:sketch_paths_interf}(b) contributes with a probability $RT\sqrt{T_2T_1R_2R_1}$, which is precisely the prefactor of the function containing the above mentioned time delays.

\begin{figure}[t]
    \centering
    \begin{overpic}[percent=true,width=0.9\columnwidth,grid=false]{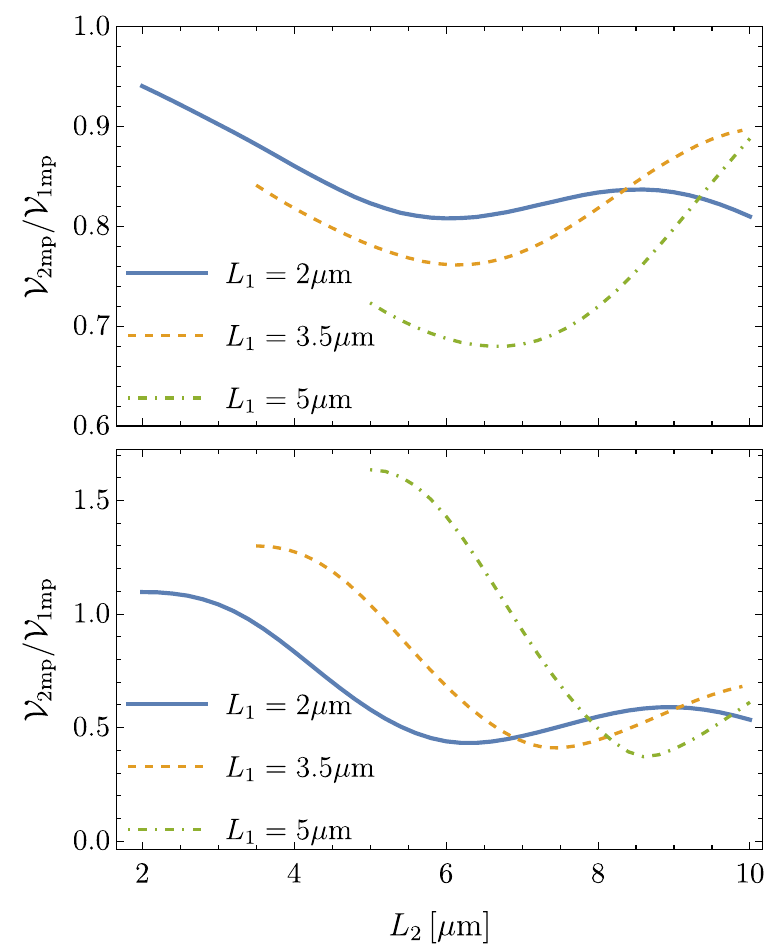}
    \put(75,12.5){(b)}
    \put(75,57){(a)}
    \end{overpic}
    \caption{Comparison of the visibility of the HOM dip in the case with two ($\mathcal{V}_\text{2mp}$) or a single mixing point ($\mathcal{V}_\text{1mp}$). (a) Weak mixing regime, with $\ta=T_1=T_2=0.9$. (b) Strong mixing regime, with $\ta=T_1=T_2=0.5$. In all plots, we have chosen $T=0.5$ and $L_\text{A}=L_1$, depending on the different curves, as specified in the legend. The visibility ratio $\mathcal{V}_\text{2mp}/\mathcal{V}_\text{1mp}$ is plotted as a function of $L_2$, for $L_2>L_1$. All other parameters are the same as in Fig.~\ref{fig:HOM_weak-vs-strong-mixing}.}
    \label{fig:visibility_ratio}
\end{figure}
We conclude this Section by comparing the exact result~\eqref{eq:HOM-full_2vs0} in the presence of two mixing points on a given edge, Fig.~\ref{fig:HOM-full}(b), with a configuration with a single mixing point on the same edge, {namely} Fig.~\ref{fig:HOM-full}(a) {with $S_\text{B}=\mathbbm{1}$}. Furthermore, we assume the position $L_\text{A}$ of the mixing point in the single-point setup to be equal to the position $L_1$ of the first mixing point in the two-point configuration. Likewise, we take $\ta=T_1$; in this way, the two settings simply differ by the addition of the second mixing point, located at $L_2$. With these assumptions, we investigate how the visibility of the HOM dip changes due to the addition of the second mixing point.

The result is shown in Fig.~\ref{fig:visibility_ratio}, where we plot the ratio $\mathcal{V}_\text{2mp}/\mathcal{V}_\text{1mp}$ between the visibility in the two- ($\mathcal{V}_\text{2mp}$) and single-point configurations ($\mathcal{V}_\text{1mp}$). This ratio is shown as a function of the position $L_2>L_1$ of the second mixing point, for different positions $L_1$ of the first one. Because of the larger number of terms in Eq.~\eqref{eq:HOM-full_2vs0} compared to Eq.~\eqref{eq:HOM-full_1vs0}, it is natural to expect a reduction of the visibility when two mixing points are present. Indeed, more terms containing the function $\mathcal{S}_0$ need to be synchronized in order to retain a good visibility and this is more difficult if more time delays are present. This expectation is confirmed for a generic choice of the parameters in the case of weak mixing, shown in Fig.~\ref{fig:visibility_ratio}(a). Interestingly enough, however, in the strong mixing case it is possible to have a better visibility with two mixing points than with one only, at least when $L_2$ is close to $L_1$, see Fig.~\ref{fig:visibility_ratio}(b). This is because the terms $\mathcal{S}_\text{b}$ and $\mathcal{S}_\text{d}$, unlike those in Eq.~\eqref{eq:HOM_2vs0_direct}, can become negative. Apparently, in the strong mixing regime and for $L_2$ close to $L_1$ the negativity of the interference terms is enough to compensate the otherwise larger positive value of the contribution $\mathcal{S}_\text{a}$, eventually leading to a better visibility of the HOM dip compared with the setup with a single mixing point.

\section{Conclusions and outlook}\label{sec:conclusion}

We have investigated the influence of channel mixing in fermionic Hong-Ou-Mandel experiments in the integer quantum Hall regime at filling factor $\nu=2$. We have shown that tunneling events between the copropagating edge channels where the injected electronic excitations propagate are responsible for an incomplete suppression of the current cross correlations detected at the output of the HOM interferometer. In other words, the presence of mixing reduces the visibility of the so-called Pauli dip (or HOM dip) occurring when the two sources at the input arms of the interferometer are synchronized. This is because inter-channel tunneling events mix excitations injected on different channels and propagating at possibly different velocities in such a way that there is always a delay in the arrival times of the excitations at the beamsplitter, which results in a finite noise. On the contrary, Coulomb interactions cannot account for a reduced visibility of the HOM dip if the injected states are generated via voltage pulses.

Possible directions for further investigations, well beyond the purpose of this work, include the analysis of the combined effect of interactions and mixing, as well as the challenging task of considering a large number of mixing points and studying the crossover towards the incoherent regime, where the oscillations in the HOM signal are expected to fade out.
{It would also be interesting to extend the analysis of HOM interferometry to more complex edge structures including counterpropagating modes and understand how this affects the conclusions presented in this work.}

Our results highlight the importance of mixing in the interpretation of fermionic HOM experiments and are particularly relevant for the analysis of future measurements, in particular in the fractional quantum Hall regime \cite{Taktak2021}. There, a nonzero HOM dip is expected to be directly related to the anyonic statistics of the fractionally charged quasiparticles, allowing one to extract their statistical angle. However, we have shown in this paper that a reduction of the HOM dip can also be attributed to mixing, which should be then properly taken into account for a correct interpretation of the measurement outcome.
Finally, our results could find applications in graphene which is, thanks to its robustness to decoherence, a very promising test bed for Hong-Ou-Mandel experiments. Indeed, it has recently been shown that any defect of the crystalline structure can lead to mixing of copropagating edge states \cite{PhysRevLett.126.146803}. Therefore, our approach should be carefully considered in any graphene Hong-Ou-Mandel interferometer in the quantum Hall regime.

\acknowledgements
This work has received funding from the European Union's H2020 research and innovation program under grant agreement No.~862683.
Funding from the Knut and Alice Wallenberg Foundation through the Academy Fellow program (J.S. and M.A.) is also gratefully acknowledged. M.A. would like to thank C. Sp\r{a}nsl\"{a}tt for useful discussions. D.C.G., I.T. and P.R. thank funding by the French ANR contract FullyQuantum (ANR-16-CE30-0015).

\appendix

\section{Detailed expressions for the HOM noise in Sec.~\ref{sec:mix2}}
\label{app:2mixing-points}
In this Appendix, we provide further details concerning the case of two mixing points in the same edge, discussed in Sec.~\ref{sec:mix2}. In particular, we give the explicit expressions of the functions $\mathcal{X}$, $\mathcal{Y}$ and $\mathcal{Z}$ contained in Eqs.~\eqref{eq:HOM_2vs0_dir-int}, \eqref{eq:HOM_2vs0_1pi} and \eqref{eq:HOM_2vs0_tpi}, respectively. They can all be expressed in terms of the following integral:
\begin{equation}
    \mathcal{I}(\Delta,{l})=\int d\omega f(\omega)f(-\omega-\hbar\Omega{l})e^{i\omega\Delta}\,,
\end{equation}
given by
\begin{equation}
    \mathcal{I}(\Delta,{l})=\frac{\hbar}{\Delta}\frac{e^{{l}/\bar\theta}e^{-i{l}\bar\Delta/2}}{e^{{l}/\bar\theta}-1}\frac{2\pi\bar\Delta\bar\theta}{\sinh(\pi\bar\Delta\bar\theta)}\sin({l}\bar\Delta/2)\,,
\end{equation}
where $\bar\Delta=\Omega\Delta$ and $\bar\theta=k_\text{B}\theta/\hbar\Omega$.
In special cases, it reduces to: $\mathcal{I}(0,0)=k_\text{B}\theta$ and
\begin{equation}
    \begin{split}
        \mathcal{I}(0,{l})&=\frac{\hbar\Omega{l}\,e^{\hbar\Omega{l}/k_\text{B}\theta}}{e^{\hbar\Omega{l}/k_\text{B}\theta}-1}\,,\\
        \mathcal{I}(\Delta,0)&=\frac{\pi\Delta (k_\text{B}\theta)^2}{\hbar}\,\text{csch}\left(\frac{\pi k_\text{B}\theta\Delta}{\hbar}\right)\,.
    \end{split}
\end{equation}
In terms of this integral, we have
\begin{equation}
    \mathcal{X}(\Delta)=4\mathcal{I}(\Delta,0)\text{Re}\Big(e^{iq\Omega\Delta}\sum_{l\in\mathbb{Z}}|p_{l}|^2e^{i{l}\Omega\Delta}-1\Big),
\end{equation}
where
\begin{equation}
    p_{l}=\frac{1}{\mathcal{T}}\int_{-\mathcal{T}/2}^{\mathcal{T}/2}dt\,e^{i{l}\Omega t}e^{\frac{ie}{\hbar}\int_{-\infty}^t dt' V_\text{ac}(t')}\,,
\end{equation}
are the photoassisted amplitudes associated with the voltage $V(t)=V_\text{dc}+V_\text{ac}(t)$ and
\begin{equation}
    q=\frac{e V_\text{dc}}{\hbar\Omega}
\end{equation}
represents the average number of particles injected in one period of the driving.
Next, the function $\mathcal{Y}$ is given by
\begin{equation}
\begin{split}
    \mathcal{Y}(\Delta)&=2\,\text{Re}\Big[e^{2iq\Omega\Delta}\sum_{l\in\mathbb{Z}} e^{2i\Omega{l}}\tilde{p}_{l}(\Delta)\tilde{p}_{l}^*(-\Delta)\mathcal{I}(2\Delta,{l})\Big]\\
    &\quad-2\mathcal{I}(2\Delta,0)\,,
\end{split}
\end{equation}
with the modified photoassisted amplitudes defined in Eq.~\eqref{eq:ptilde}, namely
\begin{equation}
    \tilde{p}_l(\Delta)=\sum_{m\in\mathbb{Z}} p_m p_{m-l}^* e^{im\Omega\Delta}\,.
\end{equation}
In the limit of zero time delay, one finds
\begin{equation}
    \mathcal{X}(0)=\mathcal{Y}(0)=0\,.
\label{eq:xy_zero-lim}
\end{equation}
Finally, we have
\begin{equation}
\begin{split}
    \mathcal{Z}(\Delta_a,\Delta_b)&= 2\,\text{Re}\Big\{e^{iq\Omega\Delta_{ba}}\sum_{l\in\mathbb{Z}} e^{i{l}\Omega\Delta_{ba}}\mathcal{I}(\Delta_{ba},{l})\\
    &\qquad\times[\tilde{p}_{l}^*(\Delta_a)\tilde{p}_{l}(\Delta_b)+\tilde{p}_{l}^*(-\Delta_b)\tilde{p}_{l}(-\Delta_a)]\Big\}\\
    &\quad-4\mathcal{I}(\Delta_{ba},0),
\end{split}
\end{equation}
where $\Delta_{ba}=\Delta_b-\Delta_a$. This function is symmetric in its arguments and for $\Delta_a=\Delta_b=\Delta$ reduces to
\begin{equation}
    \mathcal{Z}(\Delta,\Delta)=2\mathcal{S}_0(\Delta)\,.
\label{eq:z_special-case}
\end{equation}
By using Eqs.~\eqref{eq:xy_zero-lim} and~\eqref{eq:z_special-case}, one can easily show that in the limit $L_2\to L_1$, where the two mixing points merge into a single one [see Fig.~\ref{fig:HOM-full}(b)], the HOM noise~\eqref{eq:HOM-full_2vs0} reduces to a form that is structurally identical to~\eqref{eq:HOM-full_1vs0}, but with renormalized prefactors. Furthermore, Eq.~\eqref{eq:HOM-full_1vs0} is recovered when $T_2=1$, as it should be.


\bibliography{Refs.bib}

\begin{thebibliography}{104}%
\makeatletter
\providecommand \@ifxundefined [1]{%
 \@ifx{#1\undefined}
}%
\providecommand \@ifnum [1]{%
 \ifnum #1\expandafter \@firstoftwo
 \else \expandafter \@secondoftwo
 \fi
}%
\providecommand \@ifx [1]{%
 \ifx #1\expandafter \@firstoftwo
 \else \expandafter \@secondoftwo
 \fi
}%
\providecommand \natexlab [1]{#1}%
\providecommand \enquote  [1]{``#1''}%
\providecommand \bibnamefont  [1]{#1}%
\providecommand \bibfnamefont [1]{#1}%
\providecommand \citenamefont [1]{#1}%
\providecommand \href@noop [0]{\@secondoftwo}%
\providecommand \href [0]{\begingroup \@sanitize@url \@href}%
\providecommand \@href[1]{\@@startlink{#1}\@@href}%
\providecommand \@@href[1]{\endgroup#1\@@endlink}%
\providecommand \@sanitize@url [0]{\catcode `\\12\catcode `\$12\catcode
  `\&12\catcode `\#12\catcode `\^12\catcode `\_12\catcode `\%12\relax}%
\providecommand \@@startlink[1]{}%
\providecommand \@@endlink[0]{}%
\providecommand \url  [0]{\begingroup\@sanitize@url \@url }%
\providecommand \@url [1]{\endgroup\@href {#1}{\urlprefix }}%
\providecommand \urlprefix  [0]{URL }%
\providecommand \Eprint [0]{\href }%
\providecommand \doibase [0]{http://dx.doi.org/}%
\providecommand \selectlanguage [0]{\@gobble}%
\providecommand \bibinfo  [0]{\@secondoftwo}%
\providecommand \bibfield  [0]{\@secondoftwo}%
\providecommand \translation [1]{[#1]}%
\providecommand \BibitemOpen [0]{}%
\providecommand \bibitemStop [0]{}%
\providecommand \bibitemNoStop [0]{.\EOS\space}%
\providecommand \EOS [0]{\spacefactor3000\relax}%
\providecommand \BibitemShut  [1]{\csname bibitem#1\endcsname}%
\let\auto@bib@innerbib\@empty
\bibitem [{\citenamefont {F\`eve}\ \emph {et~al.}(2007)\citenamefont {F\`eve},
  \citenamefont {Mah\'e}, \citenamefont {Berroir}, \citenamefont {Kontos},
  \citenamefont {Pla\c{c}ais}, \citenamefont {Glattli}, \citenamefont
  {Cavanna}, \citenamefont {Etienne},\ and\ \citenamefont {Jin}}]{Feve2007}%
  \BibitemOpen
  \bibfield  {author} {\bibinfo {author} {\bibfnamefont {G.}~\bibnamefont
  {F\`eve}}, \bibinfo {author} {\bibfnamefont {A.}~\bibnamefont {Mah\'e}},
  \bibinfo {author} {\bibfnamefont {J.-M.}\ \bibnamefont {Berroir}}, \bibinfo
  {author} {\bibfnamefont {T.}~\bibnamefont {Kontos}}, \bibinfo {author}
  {\bibfnamefont {B.}~\bibnamefont {Pla\c{c}ais}}, \bibinfo {author}
  {\bibfnamefont {D.~C.}\ \bibnamefont {Glattli}}, \bibinfo {author}
  {\bibfnamefont {A.}~\bibnamefont {Cavanna}}, \bibinfo {author} {\bibfnamefont
  {B.}~\bibnamefont {Etienne}}, \ and\ \bibinfo {author} {\bibfnamefont
  {Y.}~\bibnamefont {Jin}},\ }\bibfield  {title} {\enquote {\bibinfo {title}
  {{An On-Demand Coherent Single-Electron Source}},}\ }\href {\doibase
  10.1126/science.1141243} {\bibfield  {journal} {\bibinfo  {journal}
  {Science}\ }\textbf {\bibinfo {volume} {316}},\ \bibinfo {pages} {1169--1172}
  (\bibinfo {year} {2007})}\BibitemShut {NoStop}%
\bibitem [{\citenamefont {Hermelin}\ \emph {et~al.}(2011)\citenamefont
  {Hermelin}, \citenamefont {Takada}, \citenamefont {Yamamoto}, \citenamefont
  {Tarucha}, \citenamefont {Wieck}, \citenamefont {Saminadayar}, \citenamefont
  {B{\"a}uerle},\ and\ \citenamefont {Meunier}}]{Hermelin2011}%
  \BibitemOpen
  \bibfield  {author} {\bibinfo {author} {\bibfnamefont {S.}~\bibnamefont
  {Hermelin}}, \bibinfo {author} {\bibfnamefont {S.}~\bibnamefont {Takada}},
  \bibinfo {author} {\bibfnamefont {M.}~\bibnamefont {Yamamoto}}, \bibinfo
  {author} {\bibfnamefont {S.}~\bibnamefont {Tarucha}}, \bibinfo {author}
  {\bibfnamefont {A.~D.}\ \bibnamefont {Wieck}}, \bibinfo {author}
  {\bibfnamefont {L.}~\bibnamefont {Saminadayar}}, \bibinfo {author}
  {\bibfnamefont {C.}~\bibnamefont {B{\"a}uerle}}, \ and\ \bibinfo {author}
  {\bibfnamefont {T.}~\bibnamefont {Meunier}},\ }\bibfield  {title} {\enquote
  {\bibinfo {title} {Electrons surfing on a sound wave as a platform for
  quantum optics with flying electrons},}\ }\href {\doibase
  10.1038/nature10416} {\bibfield  {journal} {\bibinfo  {journal} {Nature}\
  }\textbf {\bibinfo {volume} {477}},\ \bibinfo {pages} {435--438} (\bibinfo
  {year} {2011})}\BibitemShut {NoStop}%
\bibitem [{\citenamefont {Dubois}\ \emph
  {et~al.}(2013{\natexlab{a}})\citenamefont {Dubois}, \citenamefont {Jullien},
  \citenamefont {Portier}, \citenamefont {Roche}, \citenamefont {Cavanna},
  \citenamefont {Jin}, \citenamefont {Wegscheider}, \citenamefont {Roulleau},\
  and\ \citenamefont {Glattli}}]{Dubois2013}%
  \BibitemOpen
  \bibfield  {author} {\bibinfo {author} {\bibfnamefont {J.}~\bibnamefont
  {Dubois}}, \bibinfo {author} {\bibfnamefont {T.}~\bibnamefont {Jullien}},
  \bibinfo {author} {\bibfnamefont {F.}~\bibnamefont {Portier}}, \bibinfo
  {author} {\bibfnamefont {P.}~\bibnamefont {Roche}}, \bibinfo {author}
  {\bibfnamefont {A.}~\bibnamefont {Cavanna}}, \bibinfo {author} {\bibfnamefont
  {Y.}~\bibnamefont {Jin}}, \bibinfo {author} {\bibfnamefont {W.}~\bibnamefont
  {Wegscheider}}, \bibinfo {author} {\bibfnamefont {P.}~\bibnamefont
  {Roulleau}}, \ and\ \bibinfo {author} {\bibfnamefont {D.~C.}\ \bibnamefont
  {Glattli}},\ }\bibfield  {title} {\enquote {\bibinfo {title}
  {Minimal-excitation states for electron quantum optics using levitons},}\
  }\href {\doibase 10.1038/nature12713} {\bibfield  {journal} {\bibinfo
  {journal} {Nature}\ }\textbf {\bibinfo {volume} {502}},\ \bibinfo {pages}
  {659--663} (\bibinfo {year} {2013}{\natexlab{a}})}\BibitemShut {NoStop}%
\bibitem [{\citenamefont {Fletcher}\ \emph {et~al.}(2013)\citenamefont
  {Fletcher}, \citenamefont {See}, \citenamefont {Howe}, \citenamefont
  {Pepper}, \citenamefont {Giblin}, \citenamefont {Griffiths}, \citenamefont
  {Jones}, \citenamefont {Farrer}, \citenamefont {Ritchie}, \citenamefont
  {Janssen},\ and\ \citenamefont {Kataoka}}]{Fletcher2013}%
  \BibitemOpen
  \bibfield  {author} {\bibinfo {author} {\bibfnamefont {J.~D.}\ \bibnamefont
  {Fletcher}}, \bibinfo {author} {\bibfnamefont {P.}~\bibnamefont {See}},
  \bibinfo {author} {\bibfnamefont {H.}~\bibnamefont {Howe}}, \bibinfo {author}
  {\bibfnamefont {M.}~\bibnamefont {Pepper}}, \bibinfo {author} {\bibfnamefont
  {S.~P.}\ \bibnamefont {Giblin}}, \bibinfo {author} {\bibfnamefont {J.~P.}\
  \bibnamefont {Griffiths}}, \bibinfo {author} {\bibfnamefont {G.~A.~C.}\
  \bibnamefont {Jones}}, \bibinfo {author} {\bibfnamefont {I.}~\bibnamefont
  {Farrer}}, \bibinfo {author} {\bibfnamefont {D.~A.}\ \bibnamefont {Ritchie}},
  \bibinfo {author} {\bibfnamefont {T.~J. B.~M.}\ \bibnamefont {Janssen}}, \
  and\ \bibinfo {author} {\bibfnamefont {M.}~\bibnamefont {Kataoka}},\
  }\bibfield  {title} {\enquote {\bibinfo {title} {{Clock-Controlled Emission
  of Single-Electron Wave Packets in a Solid-State Circuit}},}\ }\href
  {\doibase 10.1103/PhysRevLett.111.216807} {\bibfield  {journal} {\bibinfo
  {journal} {Phys. Rev. Lett.}\ }\textbf {\bibinfo {volume} {111}},\ \bibinfo
  {pages} {216807} (\bibinfo {year} {2013})}\BibitemShut {NoStop}%
\bibitem [{\citenamefont {Grenier}\ \emph {et~al.}(2011)\citenamefont
  {Grenier}, \citenamefont {Herv\'e}, \citenamefont {F\`eve},\ and\
  \citenamefont {Degiovanni}}]{Grenier2011}%
  \BibitemOpen
  \bibfield  {author} {\bibinfo {author} {\bibfnamefont {C.}~\bibnamefont
  {Grenier}}, \bibinfo {author} {\bibfnamefont {R.}~\bibnamefont {Herv\'e}},
  \bibinfo {author} {\bibfnamefont {G.}~\bibnamefont {F\`eve}}, \ and\ \bibinfo
  {author} {\bibfnamefont {P.}~\bibnamefont {Degiovanni}},\ }\bibfield  {title}
  {\enquote {\bibinfo {title} {{Electron Quantum Optics In Quantum Hall Edge
  Channels}},}\ }\href {\doibase 10.1142/S0217984911026772} {\bibfield
  {journal} {\bibinfo  {journal} {Modern Physics Letters B}\ }\textbf {\bibinfo
  {volume} {25}},\ \bibinfo {pages} {1053--1073} (\bibinfo {year}
  {2011})}\BibitemShut {NoStop}%
\bibitem [{\citenamefont {Bocquillon}\ \emph {et~al.}(2014)\citenamefont
  {Bocquillon}, \citenamefont {Freulon}, \citenamefont {Parmentier},
  \citenamefont {Berroir}, \citenamefont {Pla\c{c}ais}, \citenamefont {Wahl},
  \citenamefont {Rech}, \citenamefont {Jonckheere}, \citenamefont {Martin},
  \citenamefont {Grenier}, \citenamefont {Ferraro}, \citenamefont
  {Degiovanni},\ and\ \citenamefont {F\`eve}}]{Bocquillon2013review}%
  \BibitemOpen
  \bibfield  {author} {\bibinfo {author} {\bibfnamefont {E.}~\bibnamefont
  {Bocquillon}}, \bibinfo {author} {\bibfnamefont {V.}~\bibnamefont {Freulon}},
  \bibinfo {author} {\bibfnamefont {F.~D.}\ \bibnamefont {Parmentier}},
  \bibinfo {author} {\bibfnamefont {J.-M.}\ \bibnamefont {Berroir}}, \bibinfo
  {author} {\bibfnamefont {B.}~\bibnamefont {Pla\c{c}ais}}, \bibinfo {author}
  {\bibfnamefont {C.}~\bibnamefont {Wahl}}, \bibinfo {author} {\bibfnamefont
  {J.}~\bibnamefont {Rech}}, \bibinfo {author} {\bibfnamefont {T.}~\bibnamefont
  {Jonckheere}}, \bibinfo {author} {\bibfnamefont {T.}~\bibnamefont {Martin}},
  \bibinfo {author} {\bibfnamefont {C.}~\bibnamefont {Grenier}}, \bibinfo
  {author} {\bibfnamefont {D.}~\bibnamefont {Ferraro}}, \bibinfo {author}
  {\bibfnamefont {P.}~\bibnamefont {Degiovanni}}, \ and\ \bibinfo {author}
  {\bibfnamefont {G.}~\bibnamefont {F\`eve}},\ }\bibfield  {title} {\enquote
  {\bibinfo {title} {Electron quantum optics in ballistic chiral conductors},}\
  }\href {\doibase 10.1002/andp.201300181} {\bibfield  {journal} {\bibinfo
  {journal} {Ann. Phys. (Berlin)}\ }\textbf {\bibinfo {volume} {526}},\
  \bibinfo {pages} {1} (\bibinfo {year} {2014})}\BibitemShut {NoStop}%
\bibitem [{\citenamefont {Bäuerle}\ \emph {et~al.}(2018)\citenamefont
  {Bäuerle}, \citenamefont {Glattli}, \citenamefont {Meunier}, \citenamefont
  {Portier}, \citenamefont {Roche}, \citenamefont {Roulleau}, \citenamefont
  {Takada},\ and\ \citenamefont {Waintal}}]{Bauerle2018}%
  \BibitemOpen
  \bibfield  {author} {\bibinfo {author} {\bibfnamefont {C.}~\bibnamefont
  {Bäuerle}}, \bibinfo {author} {\bibfnamefont {D.~C.}\ \bibnamefont
  {Glattli}}, \bibinfo {author} {\bibfnamefont {T.}~\bibnamefont {Meunier}},
  \bibinfo {author} {\bibfnamefont {F.}~\bibnamefont {Portier}}, \bibinfo
  {author} {\bibfnamefont {P.}~\bibnamefont {Roche}}, \bibinfo {author}
  {\bibfnamefont {P.}~\bibnamefont {Roulleau}}, \bibinfo {author}
  {\bibfnamefont {S.}~\bibnamefont {Takada}}, \ and\ \bibinfo {author}
  {\bibfnamefont {X.}~\bibnamefont {Waintal}},\ }\bibfield  {title} {\enquote
  {\bibinfo {title} {Coherent control of single electrons: a review of current
  progress},}\ }\href {\doibase 10.1088/1361-6633/aaa98a} {\bibfield  {journal}
  {\bibinfo  {journal} {Rep. Progr. Phys.}\ }\textbf {\bibinfo {volume} {81}},\
  \bibinfo {pages} {056503} (\bibinfo {year} {2018})}\BibitemShut {NoStop}%
\bibitem [{\citenamefont {Ol'khovskaya}\ \emph {et~al.}(2008)\citenamefont
  {Ol'khovskaya}, \citenamefont {Splettstoesser}, \citenamefont {Moskalets},\
  and\ \citenamefont {B\"uttiker}}]{Olkhovskaya2008}%
  \BibitemOpen
  \bibfield  {author} {\bibinfo {author} {\bibfnamefont {S.}~\bibnamefont
  {Ol'khovskaya}}, \bibinfo {author} {\bibfnamefont {J.}~\bibnamefont
  {Splettstoesser}}, \bibinfo {author} {\bibfnamefont {M.}~\bibnamefont
  {Moskalets}}, \ and\ \bibinfo {author} {\bibfnamefont {M.}~\bibnamefont
  {B\"uttiker}},\ }\bibfield  {title} {\enquote {\bibinfo {title} {{Shot Noise
  of a Mesoscopic Two-Particle Collider}},}\ }\href {\doibase
  10.1103/PhysRevLett.101.166802} {\bibfield  {journal} {\bibinfo  {journal}
  {Phys. Rev. Lett.}\ }\textbf {\bibinfo {volume} {101}},\ \bibinfo {pages}
  {166802} (\bibinfo {year} {2008})}\BibitemShut {NoStop}%
\bibitem [{\citenamefont {Jonckheere}\ \emph {et~al.}(2012)\citenamefont
  {Jonckheere}, \citenamefont {Rech}, \citenamefont {Wahl},\ and\ \citenamefont
  {Martin}}]{Jonckheere2012}%
  \BibitemOpen
  \bibfield  {author} {\bibinfo {author} {\bibfnamefont {T.}~\bibnamefont
  {Jonckheere}}, \bibinfo {author} {\bibfnamefont {J.}~\bibnamefont {Rech}},
  \bibinfo {author} {\bibfnamefont {C.}~\bibnamefont {Wahl}}, \ and\ \bibinfo
  {author} {\bibfnamefont {T.}~\bibnamefont {Martin}},\ }\bibfield  {title}
  {\enquote {\bibinfo {title} {{Electron and hole Hong-Ou-Mandel
  interferometry}},}\ }\href {\doibase 10.1103/PhysRevB.86.125425} {\bibfield
  {journal} {\bibinfo  {journal} {Phys. Rev. B}\ }\textbf {\bibinfo {volume}
  {86}},\ \bibinfo {pages} {125425} (\bibinfo {year} {2012})}\BibitemShut
  {NoStop}%
\bibitem [{\citenamefont {Dubois}\ \emph
  {et~al.}(2013{\natexlab{b}})\citenamefont {Dubois}, \citenamefont {Jullien},
  \citenamefont {Grenier}, \citenamefont {Degiovanni}, \citenamefont
  {Roulleau},\ and\ \citenamefont {Glattli}}]{Dubois2013PRB}%
  \BibitemOpen
  \bibfield  {author} {\bibinfo {author} {\bibfnamefont {J.}~\bibnamefont
  {Dubois}}, \bibinfo {author} {\bibfnamefont {T.}~\bibnamefont {Jullien}},
  \bibinfo {author} {\bibfnamefont {C.}~\bibnamefont {Grenier}}, \bibinfo
  {author} {\bibfnamefont {P.}~\bibnamefont {Degiovanni}}, \bibinfo {author}
  {\bibfnamefont {P.}~\bibnamefont {Roulleau}}, \ and\ \bibinfo {author}
  {\bibfnamefont {D.~C.}\ \bibnamefont {Glattli}},\ }\bibfield  {title}
  {\enquote {\bibinfo {title} {{Integer and fractional charge Lorentzian
  voltage pulses analyzed in the framework of photon-assisted shot noise}},}\
  }\href {\doibase 10.1103/PhysRevB.88.085301} {\bibfield  {journal} {\bibinfo
  {journal} {Phys. Rev. B}\ }\textbf {\bibinfo {volume} {88}},\ \bibinfo
  {pages} {085301} (\bibinfo {year} {2013}{\natexlab{b}})}\BibitemShut
  {NoStop}%
\bibitem [{\citenamefont {Haack}\ \emph {et~al.}(2013)\citenamefont {Haack},
  \citenamefont {Moskalets},\ and\ \citenamefont {B\"uttiker}}]{Haack2013}%
  \BibitemOpen
  \bibfield  {author} {\bibinfo {author} {\bibfnamefont {G.}~\bibnamefont
  {Haack}}, \bibinfo {author} {\bibfnamefont {M.}~\bibnamefont {Moskalets}}, \
  and\ \bibinfo {author} {\bibfnamefont {M.}~\bibnamefont {B\"uttiker}},\
  }\bibfield  {title} {\enquote {\bibinfo {title} {Glauber coherence of
  single-electron sources},}\ }\href {\doibase 10.1103/PhysRevB.87.201302}
  {\bibfield  {journal} {\bibinfo  {journal} {Phys. Rev. B}\ }\textbf {\bibinfo
  {volume} {87}},\ \bibinfo {pages} {201302} (\bibinfo {year}
  {2013})}\BibitemShut {NoStop}%
\bibitem [{\citenamefont {Moskalets}\ \emph {et~al.}(2013)\citenamefont
  {Moskalets}, \citenamefont {Haack},\ and\ \citenamefont
  {B\"uttiker}}]{Moskalets2013}%
  \BibitemOpen
  \bibfield  {author} {\bibinfo {author} {\bibfnamefont {M.}~\bibnamefont
  {Moskalets}}, \bibinfo {author} {\bibfnamefont {G.}~\bibnamefont {Haack}}, \
  and\ \bibinfo {author} {\bibfnamefont {M.}~\bibnamefont {B\"uttiker}},\
  }\bibfield  {title} {\enquote {\bibinfo {title} {Single-electron source:
  Adiabatic versus nonadiabatic emission},}\ }\href {\doibase
  10.1103/PhysRevB.87.125429} {\bibfield  {journal} {\bibinfo  {journal} {Phys.
  Rev. B}\ }\textbf {\bibinfo {volume} {87}},\ \bibinfo {pages} {125429}
  (\bibinfo {year} {2013})}\BibitemShut {NoStop}%
\bibitem [{\citenamefont {Grenier}\ \emph {et~al.}(2013)\citenamefont
  {Grenier}, \citenamefont {Dubois}, \citenamefont {Jullien}, \citenamefont
  {Roulleau}, \citenamefont {Glattli},\ and\ \citenamefont
  {Degiovanni}}]{Grenier2013}%
  \BibitemOpen
  \bibfield  {author} {\bibinfo {author} {\bibfnamefont {C.}~\bibnamefont
  {Grenier}}, \bibinfo {author} {\bibfnamefont {J.}~\bibnamefont {Dubois}},
  \bibinfo {author} {\bibfnamefont {T.}~\bibnamefont {Jullien}}, \bibinfo
  {author} {\bibfnamefont {P.}~\bibnamefont {Roulleau}}, \bibinfo {author}
  {\bibfnamefont {D.~C.}\ \bibnamefont {Glattli}}, \ and\ \bibinfo {author}
  {\bibfnamefont {P.}~\bibnamefont {Degiovanni}},\ }\bibfield  {title}
  {\enquote {\bibinfo {title} {Fractionalization of minimal excitations in
  integer quantum hall edge channels},}\ }\href {\doibase
  10.1103/PhysRevB.88.085302} {\bibfield  {journal} {\bibinfo  {journal} {Phys.
  Rev. B}\ }\textbf {\bibinfo {volume} {88}},\ \bibinfo {pages} {085302}
  (\bibinfo {year} {2013})}\BibitemShut {NoStop}%
\bibitem [{\citenamefont {Ferraro}\ \emph {et~al.}(2013)\citenamefont
  {Ferraro}, \citenamefont {Feller}, \citenamefont {Ghibaudo}, \citenamefont
  {Thibierge}, \citenamefont {Bocquillon}, \citenamefont {F\`eve},
  \citenamefont {Grenier},\ and\ \citenamefont {Degiovanni}}]{Ferraro2013}%
  \BibitemOpen
  \bibfield  {author} {\bibinfo {author} {\bibfnamefont {D.}~\bibnamefont
  {Ferraro}}, \bibinfo {author} {\bibfnamefont {A.}~\bibnamefont {Feller}},
  \bibinfo {author} {\bibfnamefont {A.}~\bibnamefont {Ghibaudo}}, \bibinfo
  {author} {\bibfnamefont {E.}~\bibnamefont {Thibierge}}, \bibinfo {author}
  {\bibfnamefont {E.}~\bibnamefont {Bocquillon}}, \bibinfo {author}
  {\bibfnamefont {G.}~\bibnamefont {F\`eve}}, \bibinfo {author} {\bibfnamefont
  {C.}~\bibnamefont {Grenier}}, \ and\ \bibinfo {author} {\bibfnamefont
  {P.}~\bibnamefont {Degiovanni}},\ }\bibfield  {title} {\enquote {\bibinfo
  {title} {{Wigner function approach to single electron coherence in quantum
  Hall edge channels}},}\ }\href {\doibase 10.1103/PhysRevB.88.205303}
  {\bibfield  {journal} {\bibinfo  {journal} {Phys. Rev. B}\ }\textbf {\bibinfo
  {volume} {88}},\ \bibinfo {pages} {205303} (\bibinfo {year}
  {2013})}\BibitemShut {NoStop}%
\bibitem [{\citenamefont {Wahl}\ \emph {et~al.}(2014)\citenamefont {Wahl},
  \citenamefont {Rech}, \citenamefont {Jonckheere},\ and\ \citenamefont
  {Martin}}]{Wahl2014}%
  \BibitemOpen
  \bibfield  {author} {\bibinfo {author} {\bibfnamefont {C.}~\bibnamefont
  {Wahl}}, \bibinfo {author} {\bibfnamefont {J.}~\bibnamefont {Rech}}, \bibinfo
  {author} {\bibfnamefont {T.}~\bibnamefont {Jonckheere}}, \ and\ \bibinfo
  {author} {\bibfnamefont {T.}~\bibnamefont {Martin}},\ }\bibfield  {title}
  {\enquote {\bibinfo {title} {{Interactions and Charge Fractionalization in an
  Electronic Hong-Ou-Mandel Interferometer}},}\ }\href {\doibase
  10.1103/PhysRevLett.112.046802} {\bibfield  {journal} {\bibinfo  {journal}
  {Phys. Rev. Lett.}\ }\textbf {\bibinfo {volume} {112}},\ \bibinfo {pages}
  {046802} (\bibinfo {year} {2014})}\BibitemShut {NoStop}%
\bibitem [{\citenamefont {Ferraro}\ \emph {et~al.}(2014)\citenamefont
  {Ferraro}, \citenamefont {Roussel}, \citenamefont {Cabart}, \citenamefont
  {Thibierge}, \citenamefont {F\`eve}, \citenamefont {Grenier},\ and\
  \citenamefont {Degiovanni}}]{Ferraro2014}%
  \BibitemOpen
  \bibfield  {author} {\bibinfo {author} {\bibfnamefont {D.}~\bibnamefont
  {Ferraro}}, \bibinfo {author} {\bibfnamefont {B.}~\bibnamefont {Roussel}},
  \bibinfo {author} {\bibfnamefont {C.}~\bibnamefont {Cabart}}, \bibinfo
  {author} {\bibfnamefont {E.}~\bibnamefont {Thibierge}}, \bibinfo {author}
  {\bibfnamefont {G.}~\bibnamefont {F\`eve}}, \bibinfo {author} {\bibfnamefont
  {C.}~\bibnamefont {Grenier}}, \ and\ \bibinfo {author} {\bibfnamefont
  {P.}~\bibnamefont {Degiovanni}},\ }\bibfield  {title} {\enquote {\bibinfo
  {title} {{Real-Time Decoherence of Landau and Levitov Quasiparticles in
  Quantum Hall Edge Channels}},}\ }\href {\doibase
  10.1103/PhysRevLett.113.166403} {\bibfield  {journal} {\bibinfo  {journal}
  {Phys. Rev. Lett.}\ }\textbf {\bibinfo {volume} {113}},\ \bibinfo {pages}
  {166403} (\bibinfo {year} {2014})}\BibitemShut {NoStop}%
\bibitem [{\citenamefont {Moskalets}(2015)}]{Moskalets2015}%
  \BibitemOpen
  \bibfield  {author} {\bibinfo {author} {\bibfnamefont {M.}~\bibnamefont
  {Moskalets}},\ }\bibfield  {title} {\enquote {\bibinfo {title} {First-order
  correlation function of a stream of single-electron wave packets},}\ }\href
  {\doibase 10.1103/PhysRevB.91.195431} {\bibfield  {journal} {\bibinfo
  {journal} {Phys. Rev. B}\ }\textbf {\bibinfo {volume} {91}},\ \bibinfo
  {pages} {195431} (\bibinfo {year} {2015})}\BibitemShut {NoStop}%
\bibitem [{\citenamefont {Moskalets}(2016)}]{Moskalets2016}%
  \BibitemOpen
  \bibfield  {author} {\bibinfo {author} {\bibfnamefont {M.}~\bibnamefont
  {Moskalets}},\ }\bibfield  {title} {\enquote {\bibinfo {title} {{Fractionally
  Charged Zero-Energy Single-Particle Excitations in a Driven Fermi Sea}},}\
  }\href {\doibase 10.1103/PhysRevLett.117.046801} {\bibfield  {journal}
  {\bibinfo  {journal} {Phys. Rev. Lett.}\ }\textbf {\bibinfo {volume} {117}},\
  \bibinfo {pages} {046801} (\bibinfo {year} {2016})}\BibitemShut {NoStop}%
\bibitem [{\citenamefont {Moskalets}\ and\ \citenamefont
  {Haack}(2016)}]{Moskalets2016b}%
  \BibitemOpen
  \bibfield  {author} {\bibinfo {author} {\bibfnamefont {M.}~\bibnamefont
  {Moskalets}}\ and\ \bibinfo {author} {\bibfnamefont {G.}~\bibnamefont
  {Haack}},\ }\bibfield  {title} {\enquote {\bibinfo {title} {Single-electron
  coherence: Finite temperature versus pure dephasing},}\ }\href {\doibase
  https://doi.org/10.1016/j.physe.2015.09.002} {\bibfield  {journal} {\bibinfo
  {journal} {Physica E: Low-dimensional Systems and Nanostructures}\ }\textbf
  {\bibinfo {volume} {75}},\ \bibinfo {pages} {358--369} (\bibinfo {year}
  {2016})}\BibitemShut {NoStop}%
\bibitem [{\citenamefont {Hofer}\ \emph {et~al.}(2017)\citenamefont {Hofer},
  \citenamefont {Dasenbrook},\ and\ \citenamefont {Flindt}}]{Hofer2017}%
  \BibitemOpen
  \bibfield  {author} {\bibinfo {author} {\bibfnamefont {P.~P.}\ \bibnamefont
  {Hofer}}, \bibinfo {author} {\bibfnamefont {D.}~\bibnamefont {Dasenbrook}}, \
  and\ \bibinfo {author} {\bibfnamefont {C.}~\bibnamefont {Flindt}},\
  }\bibfield  {title} {\enquote {\bibinfo {title} {On-demand entanglement
  generation using dynamic single-electron sources},}\ }\href {\doibase
  10.1002/pssb.201600582} {\bibfield  {journal} {\bibinfo  {journal} {Phys.
  Status Solidi B}\ }\textbf {\bibinfo {volume} {254}},\ \bibinfo {pages}
  {1600582} (\bibinfo {year} {2017})}\BibitemShut {NoStop}%
\bibitem [{\citenamefont {Glattli}\ and\ \citenamefont
  {Roulleau}(2018)}]{Glattli2018}%
  \BibitemOpen
  \bibfield  {author} {\bibinfo {author} {\bibfnamefont {D.~C.}\ \bibnamefont
  {Glattli}}\ and\ \bibinfo {author} {\bibfnamefont {P.}~\bibnamefont
  {Roulleau}},\ }\bibfield  {title} {\enquote {\bibinfo {title} {Pseudorandom
  binary injection of levitons for electron quantum optics},}\ }\href {\doibase
  10.1103/PhysRevB.97.125407} {\bibfield  {journal} {\bibinfo  {journal} {Phys.
  Rev. B}\ }\textbf {\bibinfo {volume} {97}},\ \bibinfo {pages} {125407}
  (\bibinfo {year} {2018})}\BibitemShut {NoStop}%
\bibitem [{\citenamefont {Cabart}\ \emph {et~al.}(2018)\citenamefont {Cabart},
  \citenamefont {Roussel}, \citenamefont {F\`eve},\ and\ \citenamefont
  {Degiovanni}}]{Cabart2018}%
  \BibitemOpen
  \bibfield  {author} {\bibinfo {author} {\bibfnamefont {C.}~\bibnamefont
  {Cabart}}, \bibinfo {author} {\bibfnamefont {B.}~\bibnamefont {Roussel}},
  \bibinfo {author} {\bibfnamefont {G.}~\bibnamefont {F\`eve}}, \ and\ \bibinfo
  {author} {\bibfnamefont {P.}~\bibnamefont {Degiovanni}},\ }\bibfield  {title}
  {\enquote {\bibinfo {title} {Taming electronic decoherence in one-dimensional
  chiral ballistic quantum conductors},}\ }\href {\doibase
  10.1103/PhysRevB.98.155302} {\bibfield  {journal} {\bibinfo  {journal} {Phys.
  Rev. B}\ }\textbf {\bibinfo {volume} {98}},\ \bibinfo {pages} {155302}
  (\bibinfo {year} {2018})}\BibitemShut {NoStop}%
\bibitem [{\citenamefont {Misiorny}\ \emph {et~al.}(2018)\citenamefont
  {Misiorny}, \citenamefont {F\`eve},\ and\ \citenamefont
  {Splettstoesser}}]{Misiorny2018}%
  \BibitemOpen
  \bibfield  {author} {\bibinfo {author} {\bibfnamefont {M.}~\bibnamefont
  {Misiorny}}, \bibinfo {author} {\bibfnamefont {G.}~\bibnamefont {F\`eve}}, \
  and\ \bibinfo {author} {\bibfnamefont {J.}~\bibnamefont {Splettstoesser}},\
  }\bibfield  {title} {\enquote {\bibinfo {title} {{Shaping charge excitations
  in chiral edge states with a time-dependent gate voltage}},}\ }\href
  {\doibase 10.1103/PhysRevB.97.075426} {\bibfield  {journal} {\bibinfo
  {journal} {Phys. Rev. B}\ }\textbf {\bibinfo {volume} {97}},\ \bibinfo
  {pages} {075426} (\bibinfo {year} {2018})}\BibitemShut {NoStop}%
\bibitem [{\citenamefont {Dashti}\ \emph {et~al.}(2019)\citenamefont {Dashti},
  \citenamefont {Misiorny}, \citenamefont {Kheradsoud}, \citenamefont
  {Samuelsson},\ and\ \citenamefont {Splettstoesser}}]{Dashti2019}%
  \BibitemOpen
  \bibfield  {author} {\bibinfo {author} {\bibfnamefont {N.}~\bibnamefont
  {Dashti}}, \bibinfo {author} {\bibfnamefont {M.}~\bibnamefont {Misiorny}},
  \bibinfo {author} {\bibfnamefont {S.}~\bibnamefont {Kheradsoud}}, \bibinfo
  {author} {\bibfnamefont {P.}~\bibnamefont {Samuelsson}}, \ and\ \bibinfo
  {author} {\bibfnamefont {J.}~\bibnamefont {Splettstoesser}},\ }\bibfield
  {title} {\enquote {\bibinfo {title} {{Minimal excitation single-particle
  emitters: Comparison of charge-transport and energy-transport properties}},}\
  }\href {\doibase 10.1103/PhysRevB.100.035405} {\bibfield  {journal} {\bibinfo
   {journal} {Phys. Rev. B}\ }\textbf {\bibinfo {volume} {100}},\ \bibinfo
  {pages} {035405} (\bibinfo {year} {2019})}\BibitemShut {NoStop}%
\bibitem [{\citenamefont {Yin}(2019)}]{Yin_2019}%
  \BibitemOpen
  \bibfield  {author} {\bibinfo {author} {\bibfnamefont {Y.}~\bibnamefont
  {Yin}},\ }\bibfield  {title} {\enquote {\bibinfo {title} {Quasiparticle
  states of on-demand coherent electron sources},}\ }\href {\doibase
  10.1088/1361-648x/ab0fc4} {\bibfield  {journal} {\bibinfo  {journal} {Journal
  of Physics: Condensed Matter}\ }\textbf {\bibinfo {volume} {31}},\ \bibinfo
  {pages} {245301} (\bibinfo {year} {2019})}\BibitemShut {NoStop}%
\bibitem [{\citenamefont {Yue}\ and\ \citenamefont {Yin}(2021)}]{Yue2021}%
  \BibitemOpen
  \bibfield  {author} {\bibinfo {author} {\bibfnamefont {X.~K.}\ \bibnamefont
  {Yue}}\ and\ \bibinfo {author} {\bibfnamefont {Y.}~\bibnamefont {Yin}},\
  }\bibfield  {title} {\enquote {\bibinfo {title} {Quasiparticle states for
  integer- and fractional-charged electron wave packets},}\ }\href {\doibase
  10.1103/PhysRevB.103.245429} {\bibfield  {journal} {\bibinfo  {journal}
  {Phys. Rev. B}\ }\textbf {\bibinfo {volume} {103}},\ \bibinfo {pages}
  {245429} (\bibinfo {year} {2021})}\BibitemShut {NoStop}%
\bibitem [{\citenamefont {Bocquillon}\ \emph {et~al.}(2012)\citenamefont
  {Bocquillon}, \citenamefont {Parmentier}, \citenamefont {Grenier},
  \citenamefont {Berroir}, \citenamefont {Degiovanni}, \citenamefont {Glattli},
  \citenamefont {Pla\ifmmode~\mbox{\c{c}}\else \c{c}\fi{}ais}, \citenamefont
  {Cavanna}, \citenamefont {Jin},\ and\ \citenamefont
  {F\`eve}}]{Bocquillon2012}%
  \BibitemOpen
  \bibfield  {author} {\bibinfo {author} {\bibfnamefont {E.}~\bibnamefont
  {Bocquillon}}, \bibinfo {author} {\bibfnamefont {F.~D.}\ \bibnamefont
  {Parmentier}}, \bibinfo {author} {\bibfnamefont {C.}~\bibnamefont {Grenier}},
  \bibinfo {author} {\bibfnamefont {J.-M.}\ \bibnamefont {Berroir}}, \bibinfo
  {author} {\bibfnamefont {P.}~\bibnamefont {Degiovanni}}, \bibinfo {author}
  {\bibfnamefont {D.~C.}\ \bibnamefont {Glattli}}, \bibinfo {author}
  {\bibfnamefont {B.}~\bibnamefont {Pla\ifmmode~\mbox{\c{c}}\else
  \c{c}\fi{}ais}}, \bibinfo {author} {\bibfnamefont {A.}~\bibnamefont
  {Cavanna}}, \bibinfo {author} {\bibfnamefont {Y.}~\bibnamefont {Jin}}, \ and\
  \bibinfo {author} {\bibfnamefont {G.}~\bibnamefont {F\`eve}},\ }\bibfield
  {title} {\enquote {\bibinfo {title} {{Electron Quantum Optics: Partitioning
  Electrons One by One}},}\ }\href {\doibase 10.1103/PhysRevLett.108.196803}
  {\bibfield  {journal} {\bibinfo  {journal} {Phys. Rev. Lett.}\ }\textbf
  {\bibinfo {volume} {108}},\ \bibinfo {pages} {196803} (\bibinfo {year}
  {2012})}\BibitemShut {NoStop}%
\bibitem [{\citenamefont {Parmentier}\ \emph {et~al.}(2012)\citenamefont
  {Parmentier}, \citenamefont {Bocquillon}, \citenamefont {Berroir},
  \citenamefont {Glattli}, \citenamefont {Pla\ifmmode~\mbox{\c{c}}\else
  \c{c}\fi{}ais}, \citenamefont {F\`eve}, \citenamefont {Albert}, \citenamefont
  {Flindt},\ and\ \citenamefont {B\"uttiker}}]{Parmentier2012}%
  \BibitemOpen
  \bibfield  {author} {\bibinfo {author} {\bibfnamefont {F.~D.}\ \bibnamefont
  {Parmentier}}, \bibinfo {author} {\bibfnamefont {E.}~\bibnamefont
  {Bocquillon}}, \bibinfo {author} {\bibfnamefont {J.-M.}\ \bibnamefont
  {Berroir}}, \bibinfo {author} {\bibfnamefont {D.~C.}\ \bibnamefont
  {Glattli}}, \bibinfo {author} {\bibfnamefont {B.}~\bibnamefont
  {Pla\ifmmode~\mbox{\c{c}}\else \c{c}\fi{}ais}}, \bibinfo {author}
  {\bibfnamefont {G.}~\bibnamefont {F\`eve}}, \bibinfo {author} {\bibfnamefont
  {M.}~\bibnamefont {Albert}}, \bibinfo {author} {\bibfnamefont
  {C.}~\bibnamefont {Flindt}}, \ and\ \bibinfo {author} {\bibfnamefont
  {M.}~\bibnamefont {B\"uttiker}},\ }\bibfield  {title} {\enquote {\bibinfo
  {title} {Current noise spectrum of a single-particle emitter: Theory and
  experiment},}\ }\href {\doibase 10.1103/PhysRevB.85.165438} {\bibfield
  {journal} {\bibinfo  {journal} {Phys. Rev. B}\ }\textbf {\bibinfo {volume}
  {85}},\ \bibinfo {pages} {165438} (\bibinfo {year} {2012})}\BibitemShut
  {NoStop}%
\bibitem [{\citenamefont {Bocquillon}\ \emph
  {et~al.}(2013{\natexlab{a}})\citenamefont {Bocquillon}, \citenamefont
  {Freulon}, \citenamefont {Berroir}, \citenamefont {Degiovanni}, \citenamefont
  {Pla{\c{c}}ais}, \citenamefont {Cavanna}, \citenamefont {Jin},\ and\
  \citenamefont {F{\`e}ve}}]{Bocquillon2013}%
  \BibitemOpen
  \bibfield  {author} {\bibinfo {author} {\bibfnamefont {E.}~\bibnamefont
  {Bocquillon}}, \bibinfo {author} {\bibfnamefont {V.}~\bibnamefont {Freulon}},
  \bibinfo {author} {\bibfnamefont {J.-.~M.}\ \bibnamefont {Berroir}}, \bibinfo
  {author} {\bibfnamefont {P.}~\bibnamefont {Degiovanni}}, \bibinfo {author}
  {\bibfnamefont {B.}~\bibnamefont {Pla{\c{c}}ais}}, \bibinfo {author}
  {\bibfnamefont {A.}~\bibnamefont {Cavanna}}, \bibinfo {author} {\bibfnamefont
  {Y.}~\bibnamefont {Jin}}, \ and\ \bibinfo {author} {\bibfnamefont
  {G.}~\bibnamefont {F{\`e}ve}},\ }\bibfield  {title} {\enquote {\bibinfo
  {title} {{Separation of neutral and charge modes in one-dimensional chiral
  edge channels}},}\ }\href {\doibase 10.1038/ncomms2788} {\bibfield  {journal}
  {\bibinfo  {journal} {Nature Communications}\ }\textbf {\bibinfo {volume}
  {4}},\ \bibinfo {pages} {1839} (\bibinfo {year}
  {2013}{\natexlab{a}})}\BibitemShut {NoStop}%
\bibitem [{\citenamefont {Bocquillon}\ \emph
  {et~al.}(2013{\natexlab{b}})\citenamefont {Bocquillon}, \citenamefont
  {Freulon}, \citenamefont {Berroir}, \citenamefont {Degiovanni}, \citenamefont
  {Pla\c{c}ais}, \citenamefont {Cavanna}, \citenamefont {Jin},\ and\
  \citenamefont {F\`eve}}]{Bocuillon2013Science}%
  \BibitemOpen
  \bibfield  {author} {\bibinfo {author} {\bibfnamefont {E.}~\bibnamefont
  {Bocquillon}}, \bibinfo {author} {\bibfnamefont {V.}~\bibnamefont {Freulon}},
  \bibinfo {author} {\bibfnamefont {J.-M.}\ \bibnamefont {Berroir}}, \bibinfo
  {author} {\bibfnamefont {P.}~\bibnamefont {Degiovanni}}, \bibinfo {author}
  {\bibfnamefont {B.}~\bibnamefont {Pla\c{c}ais}}, \bibinfo {author}
  {\bibfnamefont {A.}~\bibnamefont {Cavanna}}, \bibinfo {author} {\bibfnamefont
  {Y.}~\bibnamefont {Jin}}, \ and\ \bibinfo {author} {\bibfnamefont
  {G.}~\bibnamefont {F\`eve}},\ }\bibfield  {title} {\enquote {\bibinfo {title}
  {{Coherence and Indistinguishability of Single Electrons Emitted by
  Independent Sources}},}\ }\href {\doibase 10.1126/science.1232572} {\bibfield
   {journal} {\bibinfo  {journal} {Science}\ }\textbf {\bibinfo {volume}
  {339}},\ \bibinfo {pages} {1054--1057} (\bibinfo {year}
  {2013}{\natexlab{b}})}\BibitemShut {NoStop}%
\bibitem [{\citenamefont {Jullien}\ \emph {et~al.}(2014)\citenamefont
  {Jullien}, \citenamefont {Roulleau}, \citenamefont {Roche}, \citenamefont
  {Cavanna}, \citenamefont {Jin},\ and\ \citenamefont {Glattli}}]{Jullien2014}%
  \BibitemOpen
  \bibfield  {author} {\bibinfo {author} {\bibfnamefont {T.}~\bibnamefont
  {Jullien}}, \bibinfo {author} {\bibfnamefont {P.}~\bibnamefont {Roulleau}},
  \bibinfo {author} {\bibfnamefont {B.}~\bibnamefont {Roche}}, \bibinfo
  {author} {\bibfnamefont {A.}~\bibnamefont {Cavanna}}, \bibinfo {author}
  {\bibfnamefont {Y.}~\bibnamefont {Jin}}, \ and\ \bibinfo {author}
  {\bibfnamefont {D.~C.}\ \bibnamefont {Glattli}},\ }\bibfield  {title}
  {\enquote {\bibinfo {title} {Quantum tomography of an electron},}\ }\href
  {\doibase 10.1038/nature13821} {\bibfield  {journal} {\bibinfo  {journal}
  {Nature}\ }\textbf {\bibinfo {volume} {514}},\ \bibinfo {pages} {603--607}
  (\bibinfo {year} {2014})}\BibitemShut {NoStop}%
\bibitem [{\citenamefont {Battista}\ \emph {et~al.}(2014)\citenamefont
  {Battista}, \citenamefont {Haupt},\ and\ \citenamefont
  {Splettstoesser}}]{Battista2014}%
  \BibitemOpen
  \bibfield  {author} {\bibinfo {author} {\bibfnamefont {F.}~\bibnamefont
  {Battista}}, \bibinfo {author} {\bibfnamefont {F.}~\bibnamefont {Haupt}}, \
  and\ \bibinfo {author} {\bibfnamefont {J.}~\bibnamefont {Splettstoesser}},\
  }\bibfield  {title} {\enquote {\bibinfo {title} {{Energy and power
  fluctuations in ac-driven coherent conductors}},}\ }\href {\doibase
  10.1103/PhysRevB.90.085418} {\bibfield  {journal} {\bibinfo  {journal} {Phys.
  Rev. B}\ }\textbf {\bibinfo {volume} {90}},\ \bibinfo {pages} {085418}
  (\bibinfo {year} {2014})}\BibitemShut {NoStop}%
\bibitem [{\citenamefont {Freulon}\ \emph {et~al.}(2015)\citenamefont
  {Freulon}, \citenamefont {Marguerite}, \citenamefont {Berroir}, \citenamefont
  {Pla{\c{c}}ais}, \citenamefont {Cavanna}, \citenamefont {Jin},\ and\
  \citenamefont {F{\`e}ve}}]{Freulon2015}%
  \BibitemOpen
  \bibfield  {author} {\bibinfo {author} {\bibfnamefont {V.}~\bibnamefont
  {Freulon}}, \bibinfo {author} {\bibfnamefont {A.}~\bibnamefont {Marguerite}},
  \bibinfo {author} {\bibfnamefont {J.-M.}\ \bibnamefont {Berroir}}, \bibinfo
  {author} {\bibfnamefont {B.}~\bibnamefont {Pla{\c{c}}ais}}, \bibinfo {author}
  {\bibfnamefont {A.}~\bibnamefont {Cavanna}}, \bibinfo {author} {\bibfnamefont
  {Y.}~\bibnamefont {Jin}}, \ and\ \bibinfo {author} {\bibfnamefont
  {G.}~\bibnamefont {F{\`e}ve}},\ }\bibfield  {title} {\enquote {\bibinfo
  {title} {{Hong-Ou-Mandel experiment for temporal investigation of
  single-electron fractionalization}},}\ }\href {\doibase 10.1038/ncomms7854}
  {\bibfield  {journal} {\bibinfo  {journal} {Nature Communications}\ }\textbf
  {\bibinfo {volume} {6}},\ \bibinfo {pages} {6854} (\bibinfo {year}
  {2015})}\BibitemShut {NoStop}%
\bibitem [{\citenamefont {Ubbelohde}\ \emph {et~al.}(2015)\citenamefont
  {Ubbelohde}, \citenamefont {Hohls}, \citenamefont {Kashcheyevs},
  \citenamefont {Wagner}, \citenamefont {Fricke}, \citenamefont {K{\"a}stner},
  \citenamefont {Pierz}, \citenamefont {Schumacher},\ and\ \citenamefont
  {Haug}}]{Ubbelohde2015}%
  \BibitemOpen
  \bibfield  {author} {\bibinfo {author} {\bibfnamefont {N.}~\bibnamefont
  {Ubbelohde}}, \bibinfo {author} {\bibfnamefont {F.}~\bibnamefont {Hohls}},
  \bibinfo {author} {\bibfnamefont {V.}~\bibnamefont {Kashcheyevs}}, \bibinfo
  {author} {\bibfnamefont {T.}~\bibnamefont {Wagner}}, \bibinfo {author}
  {\bibfnamefont {L.}~\bibnamefont {Fricke}}, \bibinfo {author} {\bibfnamefont
  {B.}~\bibnamefont {K{\"a}stner}}, \bibinfo {author} {\bibfnamefont
  {K.}~\bibnamefont {Pierz}}, \bibinfo {author} {\bibfnamefont {H.~W.}\
  \bibnamefont {Schumacher}}, \ and\ \bibinfo {author} {\bibfnamefont {R.~J.}\
  \bibnamefont {Haug}},\ }\bibfield  {title} {\enquote {\bibinfo {title}
  {Partitioning of on-demand electron pairs},}\ }\href {\doibase
  10.1038/nnano.2014.275} {\bibfield  {journal} {\bibinfo  {journal} {Nature
  Nanotechnology}\ }\textbf {\bibinfo {volume} {10}},\ \bibinfo {pages}
  {46--49} (\bibinfo {year} {2015})}\BibitemShut {NoStop}%
\bibitem [{\citenamefont {Vanevi\ifmmode~\acute{c}\else \'{c}\fi{}}\ \emph
  {et~al.}(2016)\citenamefont {Vanevi\ifmmode~\acute{c}\else \'{c}\fi{}},
  \citenamefont {Gabelli}, \citenamefont {Belzig},\ and\ \citenamefont
  {Reulet}}]{Vanevic2016}%
  \BibitemOpen
  \bibfield  {author} {\bibinfo {author} {\bibfnamefont {M.}~\bibnamefont
  {Vanevi\ifmmode~\acute{c}\else \'{c}\fi{}}}, \bibinfo {author} {\bibfnamefont
  {J.}~\bibnamefont {Gabelli}}, \bibinfo {author} {\bibfnamefont
  {W.}~\bibnamefont {Belzig}}, \ and\ \bibinfo {author} {\bibfnamefont
  {B.}~\bibnamefont {Reulet}},\ }\bibfield  {title} {\enquote {\bibinfo {title}
  {Electron and electron-hole quasiparticle states in a driven quantum
  contact},}\ }\href {\doibase 10.1103/PhysRevB.93.041416} {\bibfield
  {journal} {\bibinfo  {journal} {Phys. Rev. B}\ }\textbf {\bibinfo {volume}
  {93}},\ \bibinfo {pages} {041416} (\bibinfo {year} {2016})}\BibitemShut
  {NoStop}%
\bibitem [{\citenamefont {Kataoka}\ \emph {et~al.}(2016)\citenamefont
  {Kataoka}, \citenamefont {Johnson}, \citenamefont {Emary}, \citenamefont
  {See}, \citenamefont {Griffiths}, \citenamefont {Jones}, \citenamefont
  {Farrer}, \citenamefont {Ritchie}, \citenamefont {Pepper},\ and\
  \citenamefont {Janssen}}]{Kataoka2016}%
  \BibitemOpen
  \bibfield  {author} {\bibinfo {author} {\bibfnamefont {M.}~\bibnamefont
  {Kataoka}}, \bibinfo {author} {\bibfnamefont {N.}~\bibnamefont {Johnson}},
  \bibinfo {author} {\bibfnamefont {C.}~\bibnamefont {Emary}}, \bibinfo
  {author} {\bibfnamefont {P.}~\bibnamefont {See}}, \bibinfo {author}
  {\bibfnamefont {J.~P.}\ \bibnamefont {Griffiths}}, \bibinfo {author}
  {\bibfnamefont {G.~A.~C.}\ \bibnamefont {Jones}}, \bibinfo {author}
  {\bibfnamefont {I.}~\bibnamefont {Farrer}}, \bibinfo {author} {\bibfnamefont
  {D.~A.}\ \bibnamefont {Ritchie}}, \bibinfo {author} {\bibfnamefont
  {M.}~\bibnamefont {Pepper}}, \ and\ \bibinfo {author} {\bibfnamefont {T.~J.
  B.~M.}\ \bibnamefont {Janssen}},\ }\bibfield  {title} {\enquote {\bibinfo
  {title} {{Time-of-Flight Measurements of Single-Electron Wave Packets in
  Quantum Hall Edge States}},}\ }\href {\doibase
  10.1103/PhysRevLett.116.126803} {\bibfield  {journal} {\bibinfo  {journal}
  {Phys. Rev. Lett.}\ }\textbf {\bibinfo {volume} {116}},\ \bibinfo {pages}
  {126803} (\bibinfo {year} {2016})}\BibitemShut {NoStop}%
\bibitem [{\citenamefont {Glattli}\ and\ \citenamefont
  {Roulleau}(2016)}]{Glattli2016}%
  \BibitemOpen
  \bibfield  {author} {\bibinfo {author} {\bibfnamefont {D.}~\bibnamefont
  {Glattli}}\ and\ \bibinfo {author} {\bibfnamefont {P.}~\bibnamefont
  {Roulleau}},\ }\bibfield  {title} {\enquote {\bibinfo {title} {{Hanbury-Brown
  Twiss noise correlation with time controlled quasi-particles in ballistic
  quantum conductors}},}\ }\href {\doibase
  https://doi.org/10.1016/j.physe.2015.10.034} {\bibfield  {journal} {\bibinfo
  {journal} {Physica E: Low-dimensional Systems and Nanostructures}\ }\textbf
  {\bibinfo {volume} {76}},\ \bibinfo {pages} {216--222} (\bibinfo {year}
  {2016})}\BibitemShut {NoStop}%
\bibitem [{\citenamefont {Bisognin}\ \emph {et~al.}(2019)\citenamefont
  {Bisognin}, \citenamefont {Marguerite}, \citenamefont {Roussel},
  \citenamefont {Kumar}, \citenamefont {Cabart}, \citenamefont {Chapdelaine},
  \citenamefont {Mohammad-Djafari}, \citenamefont {Berroir}, \citenamefont
  {Bocquillon}, \citenamefont {Pla{\c{c}}ais}, \citenamefont {Cavanna},
  \citenamefont {Gennser}, \citenamefont {Jin}, \citenamefont {Degiovanni},\
  and\ \citenamefont {F{\`e}ve}}]{Bisognin2019}%
  \BibitemOpen
  \bibfield  {author} {\bibinfo {author} {\bibfnamefont {R.}~\bibnamefont
  {Bisognin}}, \bibinfo {author} {\bibfnamefont {A.}~\bibnamefont
  {Marguerite}}, \bibinfo {author} {\bibfnamefont {B.}~\bibnamefont {Roussel}},
  \bibinfo {author} {\bibfnamefont {M.}~\bibnamefont {Kumar}}, \bibinfo
  {author} {\bibfnamefont {C.}~\bibnamefont {Cabart}}, \bibinfo {author}
  {\bibfnamefont {C.}~\bibnamefont {Chapdelaine}}, \bibinfo {author}
  {\bibfnamefont {A.}~\bibnamefont {Mohammad-Djafari}}, \bibinfo {author}
  {\bibfnamefont {J.-M.}\ \bibnamefont {Berroir}}, \bibinfo {author}
  {\bibfnamefont {E.}~\bibnamefont {Bocquillon}}, \bibinfo {author}
  {\bibfnamefont {B.}~\bibnamefont {Pla{\c{c}}ais}}, \bibinfo {author}
  {\bibfnamefont {A.}~\bibnamefont {Cavanna}}, \bibinfo {author} {\bibfnamefont
  {U.}~\bibnamefont {Gennser}}, \bibinfo {author} {\bibfnamefont
  {Y.}~\bibnamefont {Jin}}, \bibinfo {author} {\bibfnamefont {P.}~\bibnamefont
  {Degiovanni}}, \ and\ \bibinfo {author} {\bibfnamefont {G.}~\bibnamefont
  {F{\`e}ve}},\ }\bibfield  {title} {\enquote {\bibinfo {title} {Quantum
  tomography of electrical currents},}\ }\href {\doibase
  10.1038/s41467-019-11369-5} {\bibfield  {journal} {\bibinfo  {journal}
  {Nature Communications}\ }\textbf {\bibinfo {volume} {10}},\ \bibinfo {pages}
  {3379} (\bibinfo {year} {2019})}\BibitemShut {NoStop}%
\bibitem [{\citenamefont {Ferraro}\ \emph {et~al.}(2015)\citenamefont
  {Ferraro}, \citenamefont {Rech}, \citenamefont {Jonckheere},\ and\
  \citenamefont {Martin}}]{Ferraro2015}%
  \BibitemOpen
  \bibfield  {author} {\bibinfo {author} {\bibfnamefont {D.}~\bibnamefont
  {Ferraro}}, \bibinfo {author} {\bibfnamefont {J.}~\bibnamefont {Rech}},
  \bibinfo {author} {\bibfnamefont {T.}~\bibnamefont {Jonckheere}}, \ and\
  \bibinfo {author} {\bibfnamefont {T.}~\bibnamefont {Martin}},\ }\bibfield
  {title} {\enquote {\bibinfo {title} {{Nonlocal interference and
  Hong-Ou-Mandel collisions of single Bogoliubov quasiparticles}},}\ }\href
  {\doibase 10.1103/PhysRevB.91.075406} {\bibfield  {journal} {\bibinfo
  {journal} {Phys. Rev. B}\ }\textbf {\bibinfo {volume} {91}},\ \bibinfo
  {pages} {075406} (\bibinfo {year} {2015})}\BibitemShut {NoStop}%
\bibitem [{\citenamefont {Ferraro}\ \emph {et~al.}(2017)\citenamefont
  {Ferraro}, \citenamefont {Jonckheere}, \citenamefont {Rech},\ and\
  \citenamefont {Martin}}]{Ferraro2016}%
  \BibitemOpen
  \bibfield  {author} {\bibinfo {author} {\bibfnamefont {D.}~\bibnamefont
  {Ferraro}}, \bibinfo {author} {\bibfnamefont {T.}~\bibnamefont {Jonckheere}},
  \bibinfo {author} {\bibfnamefont {J.}~\bibnamefont {Rech}}, \ and\ \bibinfo
  {author} {\bibfnamefont {T.}~\bibnamefont {Martin}},\ }\bibfield  {title}
  {\enquote {\bibinfo {title} {{Electronic quantum optics beyond the integer
  quantum Hall effect}},}\ }\href {\doibase 10.1002/pssb.201600531} {\bibfield
  {journal} {\bibinfo  {journal} {Phys. Status Solidi B}\ }\textbf {\bibinfo
  {volume} {254}},\ \bibinfo {pages} {1600531} (\bibinfo {year}
  {2017})}\BibitemShut {NoStop}%
\bibitem [{\citenamefont {Slobodeniuk}\ \emph {et~al.}(2016)\citenamefont
  {Slobodeniuk}, \citenamefont {Idrisov},\ and\ \citenamefont
  {Sukhorukov}}]{Slobodeniuk2016}%
  \BibitemOpen
  \bibfield  {author} {\bibinfo {author} {\bibfnamefont {A.~O.}\ \bibnamefont
  {Slobodeniuk}}, \bibinfo {author} {\bibfnamefont {E.~G.}\ \bibnamefont
  {Idrisov}}, \ and\ \bibinfo {author} {\bibfnamefont {E.~V.}\ \bibnamefont
  {Sukhorukov}},\ }\bibfield  {title} {\enquote {\bibinfo {title} {{Relaxation
  of an electron wave packet at the quantum Hall edge at filling factor
  $\ensuremath{\nu}=2$}},}\ }\href {\doibase 10.1103/PhysRevB.93.035421}
  {\bibfield  {journal} {\bibinfo  {journal} {Phys. Rev. B}\ }\textbf {\bibinfo
  {volume} {93}},\ \bibinfo {pages} {035421} (\bibinfo {year}
  {2016})}\BibitemShut {NoStop}%
\bibitem [{\citenamefont {Glattli}\ and\ \citenamefont
  {Roulleau}(2017)}]{Glattli2017}%
  \BibitemOpen
  \bibfield  {author} {\bibinfo {author} {\bibfnamefont {C.~C.}\ \bibnamefont
  {Glattli}}\ and\ \bibinfo {author} {\bibfnamefont {P.}~\bibnamefont
  {Roulleau}},\ }\bibfield  {title} {\enquote {\bibinfo {title} {Levitons for
  electron quantum optics},}\ }\href {\doibase 10.1002/pssb.201600650}
  {\bibfield  {journal} {\bibinfo  {journal} {Phys. Status Solidi B}\ }\textbf
  {\bibinfo {volume} {254}},\ \bibinfo {pages} {1600650} (\bibinfo {year}
  {2017})}\BibitemShut {NoStop}%
\bibitem [{\citenamefont {Rech}\ \emph {et~al.}(2017)\citenamefont {Rech},
  \citenamefont {Ferraro}, \citenamefont {Jonckheere}, \citenamefont
  {Vannucci}, \citenamefont {Sassetti},\ and\ \citenamefont
  {Martin}}]{Rech2017}%
  \BibitemOpen
  \bibfield  {author} {\bibinfo {author} {\bibfnamefont {J.}~\bibnamefont
  {Rech}}, \bibinfo {author} {\bibfnamefont {D.}~\bibnamefont {Ferraro}},
  \bibinfo {author} {\bibfnamefont {T.}~\bibnamefont {Jonckheere}}, \bibinfo
  {author} {\bibfnamefont {L.}~\bibnamefont {Vannucci}}, \bibinfo {author}
  {\bibfnamefont {M.}~\bibnamefont {Sassetti}}, \ and\ \bibinfo {author}
  {\bibfnamefont {T.}~\bibnamefont {Martin}},\ }\bibfield  {title} {\enquote
  {\bibinfo {title} {{Minimal Excitations in the Fractional Quantum Hall
  Regime}},}\ }\href {\doibase 10.1103/PhysRevLett.118.076801} {\bibfield
  {journal} {\bibinfo  {journal} {Phys. Rev. Lett.}\ }\textbf {\bibinfo
  {volume} {118}},\ \bibinfo {pages} {076801} (\bibinfo {year}
  {2017})}\BibitemShut {NoStop}%
\bibitem [{\citenamefont {Litinski}\ \emph {et~al.}(2017)\citenamefont
  {Litinski}, \citenamefont {Brouwer},\ and\ \citenamefont
  {Filippone}}]{Litinski2017}%
  \BibitemOpen
  \bibfield  {author} {\bibinfo {author} {\bibfnamefont {D.}~\bibnamefont
  {Litinski}}, \bibinfo {author} {\bibfnamefont {P.~W.}\ \bibnamefont
  {Brouwer}}, \ and\ \bibinfo {author} {\bibfnamefont {M.}~\bibnamefont
  {Filippone}},\ }\bibfield  {title} {\enquote {\bibinfo {title} {{Interacting
  mesoscopic capacitor out of equilibrium}},}\ }\href {\doibase
  10.1103/PhysRevB.96.085429} {\bibfield  {journal} {\bibinfo  {journal} {Phys.
  Rev. B}\ }\textbf {\bibinfo {volume} {96}},\ \bibinfo {pages} {085429}
  (\bibinfo {year} {2017})}\BibitemShut {NoStop}%
\bibitem [{\citenamefont {Vannucci}\ \emph {et~al.}(2017)\citenamefont
  {Vannucci}, \citenamefont {Ronetti}, \citenamefont {Rech}, \citenamefont
  {Ferraro}, \citenamefont {Jonckheere}, \citenamefont {Martin},\ and\
  \citenamefont {Sassetti}}]{Vannucci2017}%
  \BibitemOpen
  \bibfield  {author} {\bibinfo {author} {\bibfnamefont {L.}~\bibnamefont
  {Vannucci}}, \bibinfo {author} {\bibfnamefont {F.}~\bibnamefont {Ronetti}},
  \bibinfo {author} {\bibfnamefont {J.}~\bibnamefont {Rech}}, \bibinfo {author}
  {\bibfnamefont {D.}~\bibnamefont {Ferraro}}, \bibinfo {author} {\bibfnamefont
  {T.}~\bibnamefont {Jonckheere}}, \bibinfo {author} {\bibfnamefont
  {T.}~\bibnamefont {Martin}}, \ and\ \bibinfo {author} {\bibfnamefont
  {M.}~\bibnamefont {Sassetti}},\ }\bibfield  {title} {\enquote {\bibinfo
  {title} {{Minimal excitation states for heat transport in driven quantum Hall
  systems}},}\ }\href {\doibase 10.1103/PhysRevB.95.245415} {\bibfield
  {journal} {\bibinfo  {journal} {Phys. Rev. B}\ }\textbf {\bibinfo {volume}
  {95}},\ \bibinfo {pages} {245415} (\bibinfo {year} {2017})}\BibitemShut
  {NoStop}%
\bibitem [{\citenamefont {Ferraro}\ \emph {et~al.}(2018)\citenamefont
  {Ferraro}, \citenamefont {Ronetti}, \citenamefont {Vannucci}, \citenamefont
  {Acciai}, \citenamefont {Rech}, \citenamefont {Jockheere}, \citenamefont
  {Martin},\ and\ \citenamefont {Sassetti}}]{Ferraro2018}%
  \BibitemOpen
  \bibfield  {author} {\bibinfo {author} {\bibfnamefont {D.}~\bibnamefont
  {Ferraro}}, \bibinfo {author} {\bibfnamefont {F.}~\bibnamefont {Ronetti}},
  \bibinfo {author} {\bibfnamefont {L.}~\bibnamefont {Vannucci}}, \bibinfo
  {author} {\bibfnamefont {M.}~\bibnamefont {Acciai}}, \bibinfo {author}
  {\bibfnamefont {J.}~\bibnamefont {Rech}}, \bibinfo {author} {\bibfnamefont
  {T.}~\bibnamefont {Jockheere}}, \bibinfo {author} {\bibfnamefont
  {T.}~\bibnamefont {Martin}}, \ and\ \bibinfo {author} {\bibfnamefont
  {M.}~\bibnamefont {Sassetti}},\ }\bibfield  {title} {\enquote {\bibinfo
  {title} {{Hong-Ou-Mandel characterization of multiply charged Levitons}},}\
  }\href {\doibase 10.1140/epjst/e2018-800074-1} {\bibfield  {journal}
  {\bibinfo  {journal} {The European Physical Journal Special Topics}\ }\textbf
  {\bibinfo {volume} {227}},\ \bibinfo {pages} {1345--1359} (\bibinfo {year}
  {2018})}\BibitemShut {NoStop}%
\bibitem [{\citenamefont {Ronetti}\ \emph {et~al.}(2018)\citenamefont
  {Ronetti}, \citenamefont {Vannucci}, \citenamefont {Ferraro}, \citenamefont
  {Jonckheere}, \citenamefont {Rech}, \citenamefont {Martin},\ and\
  \citenamefont {Sassetti}}]{Ronetti2018a}%
  \BibitemOpen
  \bibfield  {author} {\bibinfo {author} {\bibfnamefont {F.}~\bibnamefont
  {Ronetti}}, \bibinfo {author} {\bibfnamefont {L.}~\bibnamefont {Vannucci}},
  \bibinfo {author} {\bibfnamefont {D.}~\bibnamefont {Ferraro}}, \bibinfo
  {author} {\bibfnamefont {T.}~\bibnamefont {Jonckheere}}, \bibinfo {author}
  {\bibfnamefont {J.}~\bibnamefont {Rech}}, \bibinfo {author} {\bibfnamefont
  {T.}~\bibnamefont {Martin}}, \ and\ \bibinfo {author} {\bibfnamefont
  {M.}~\bibnamefont {Sassetti}},\ }\bibfield  {title} {\enquote {\bibinfo
  {title} {{Crystallization of levitons in the fractional quantum Hall
  regime}},}\ }\href {\doibase 10.1103/PhysRevB.98.075401} {\bibfield
  {journal} {\bibinfo  {journal} {Phys. Rev. B}\ }\textbf {\bibinfo {volume}
  {98}},\ \bibinfo {pages} {075401} (\bibinfo {year} {2018})}\BibitemShut
  {NoStop}%
\bibitem [{\citenamefont {Acciai}\ \emph
  {et~al.}(2019{\natexlab{a}})\citenamefont {Acciai}, \citenamefont {Ronetti},
  \citenamefont {Ferraro}, \citenamefont {Rech}, \citenamefont {Jonckheere},
  \citenamefont {Sassetti},\ and\ \citenamefont {Martin}}]{Acciai2019}%
  \BibitemOpen
  \bibfield  {author} {\bibinfo {author} {\bibfnamefont {M.}~\bibnamefont
  {Acciai}}, \bibinfo {author} {\bibfnamefont {F.}~\bibnamefont {Ronetti}},
  \bibinfo {author} {\bibfnamefont {D.}~\bibnamefont {Ferraro}}, \bibinfo
  {author} {\bibfnamefont {J.}~\bibnamefont {Rech}}, \bibinfo {author}
  {\bibfnamefont {T.}~\bibnamefont {Jonckheere}}, \bibinfo {author}
  {\bibfnamefont {M.}~\bibnamefont {Sassetti}}, \ and\ \bibinfo {author}
  {\bibfnamefont {T.}~\bibnamefont {Martin}},\ }\bibfield  {title} {\enquote
  {\bibinfo {title} {Levitons in superconducting point contacts},}\ }\href
  {\doibase 10.1103/PhysRevB.100.085418} {\bibfield  {journal} {\bibinfo
  {journal} {Phys. Rev. B}\ }\textbf {\bibinfo {volume} {100}},\ \bibinfo
  {pages} {085418} (\bibinfo {year} {2019}{\natexlab{a}})}\BibitemShut
  {NoStop}%
\bibitem [{\citenamefont {Acciai}\ \emph
  {et~al.}(2019{\natexlab{b}})\citenamefont {Acciai}, \citenamefont {Calzona},
  \citenamefont {Carrega}, \citenamefont {Martin},\ and\ \citenamefont
  {Sassetti}}]{Acciai_2019b}%
  \BibitemOpen
  \bibfield  {author} {\bibinfo {author} {\bibfnamefont {M.}~\bibnamefont
  {Acciai}}, \bibinfo {author} {\bibfnamefont {A.}~\bibnamefont {Calzona}},
  \bibinfo {author} {\bibfnamefont {M.}~\bibnamefont {Carrega}}, \bibinfo
  {author} {\bibfnamefont {T.}~\bibnamefont {Martin}}, \ and\ \bibinfo {author}
  {\bibfnamefont {M.}~\bibnamefont {Sassetti}},\ }\bibfield  {title} {\enquote
  {\bibinfo {title} {Spectral properties of interacting helical channels driven
  by lorentzian pulses},}\ }\href {\doibase 10.1088/1367-2630/ab494b}
  {\bibfield  {journal} {\bibinfo  {journal} {New Journal of Physics}\ }\textbf
  {\bibinfo {volume} {21}},\ \bibinfo {pages} {103031} (\bibinfo {year}
  {2019}{\natexlab{b}})}\BibitemShut {NoStop}%
\bibitem [{\citenamefont {Ronetti}\ \emph {et~al.}(2019)\citenamefont
  {Ronetti}, \citenamefont {Vannucci}, \citenamefont {Ferraro}, \citenamefont
  {Jonckheere}, \citenamefont {Rech}, \citenamefont {Martin},\ and\
  \citenamefont {Sassetti}}]{Ronetti2019}%
  \BibitemOpen
  \bibfield  {author} {\bibinfo {author} {\bibfnamefont {F.}~\bibnamefont
  {Ronetti}}, \bibinfo {author} {\bibfnamefont {L.}~\bibnamefont {Vannucci}},
  \bibinfo {author} {\bibfnamefont {D.}~\bibnamefont {Ferraro}}, \bibinfo
  {author} {\bibfnamefont {T.}~\bibnamefont {Jonckheere}}, \bibinfo {author}
  {\bibfnamefont {J.}~\bibnamefont {Rech}}, \bibinfo {author} {\bibfnamefont
  {T.}~\bibnamefont {Martin}}, \ and\ \bibinfo {author} {\bibfnamefont
  {M.}~\bibnamefont {Sassetti}},\ }\bibfield  {title} {\enquote {\bibinfo
  {title} {{Hong-Ou-Mandel heat noise in the quantum Hall regime}},}\ }\href
  {\doibase 10.1103/PhysRevB.99.205406} {\bibfield  {journal} {\bibinfo
  {journal} {Phys. Rev. B}\ }\textbf {\bibinfo {volume} {99}},\ \bibinfo
  {pages} {205406} (\bibinfo {year} {2019})}\BibitemShut {NoStop}%
\bibitem [{\citenamefont {Ronetti}\ \emph {et~al.}(2020)\citenamefont
  {Ronetti}, \citenamefont {Carrega},\ and\ \citenamefont
  {Sassetti}}]{Ronetti2020}%
  \BibitemOpen
  \bibfield  {author} {\bibinfo {author} {\bibfnamefont {F.}~\bibnamefont
  {Ronetti}}, \bibinfo {author} {\bibfnamefont {M.}~\bibnamefont {Carrega}}, \
  and\ \bibinfo {author} {\bibfnamefont {M.}~\bibnamefont {Sassetti}},\
  }\bibfield  {title} {\enquote {\bibinfo {title} {{Levitons in helical liquids
  with Rashba spin-orbit coupling probed by a superconducting contact}},}\
  }\href {\doibase 10.1103/PhysRevResearch.2.013203} {\bibfield  {journal}
  {\bibinfo  {journal} {Phys. Rev. Research}\ }\textbf {\bibinfo {volume}
  {2}},\ \bibinfo {pages} {013203} (\bibinfo {year} {2020})}\BibitemShut
  {NoStop}%
\bibitem [{\citenamefont {Carrega}\ \emph {et~al.}(2021)\citenamefont
  {Carrega}, \citenamefont {Chirolli}, \citenamefont {Heun},\ and\
  \citenamefont {Sorba}}]{Carrega2021}%
  \BibitemOpen
  \bibfield  {author} {\bibinfo {author} {\bibfnamefont {M.}~\bibnamefont
  {Carrega}}, \bibinfo {author} {\bibfnamefont {L.}~\bibnamefont {Chirolli}},
  \bibinfo {author} {\bibfnamefont {S.}~\bibnamefont {Heun}}, \ and\ \bibinfo
  {author} {\bibfnamefont {L.}~\bibnamefont {Sorba}},\ }\bibfield  {title}
  {\enquote {\bibinfo {title} {{Anyons in quantum Hall interferometry}},}\
  }\href {\doibase 10.1038/s42254-021-00351-0} {\bibfield  {journal} {\bibinfo
  {journal} {Nature Reviews Physics}\ }\textbf {\bibinfo {volume} {3}},\
  \bibinfo {pages} {698--711} (\bibinfo {year} {2021})}\BibitemShut {NoStop}%
\bibitem [{\citenamefont {Kapfer}\ \emph {et~al.}(2019)\citenamefont {Kapfer},
  \citenamefont {Roulleau}, \citenamefont {Santin}, \citenamefont {Farrer},
  \citenamefont {Ritchie},\ and\ \citenamefont {Glattli}}]{Kapfer2019}%
  \BibitemOpen
  \bibfield  {author} {\bibinfo {author} {\bibfnamefont {M.}~\bibnamefont
  {Kapfer}}, \bibinfo {author} {\bibfnamefont {P.}~\bibnamefont {Roulleau}},
  \bibinfo {author} {\bibfnamefont {M.}~\bibnamefont {Santin}}, \bibinfo
  {author} {\bibfnamefont {I.}~\bibnamefont {Farrer}}, \bibinfo {author}
  {\bibfnamefont {D.~A.}\ \bibnamefont {Ritchie}}, \ and\ \bibinfo {author}
  {\bibfnamefont {D.~C.}\ \bibnamefont {Glattli}},\ }\bibfield  {title}
  {\enquote {\bibinfo {title} {{A Josephson relation for fractionally charged
  anyons}},}\ }\href {\doibase 10.1126/science.aau3539} {\bibfield  {journal}
  {\bibinfo  {journal} {Science}\ }\textbf {\bibinfo {volume} {363}},\ \bibinfo
  {pages} {846--849} (\bibinfo {year} {2019})}\BibitemShut {NoStop}%
\bibitem [{\citenamefont {Bartolomei}\ \emph {et~al.}(2020)\citenamefont
  {Bartolomei}, \citenamefont {Kumar}, \citenamefont {Bisognin}, \citenamefont
  {Marguerite}, \citenamefont {Berroir}, \citenamefont {Bocquillon},
  \citenamefont {Pla\c{c}ais}, \citenamefont {Cavanna}, \citenamefont {Dong},
  \citenamefont {Gennser}, \citenamefont {Jin},\ and\ \citenamefont
  {F\`eve}}]{Bartolomei2020}%
  \BibitemOpen
  \bibfield  {author} {\bibinfo {author} {\bibfnamefont {H.}~\bibnamefont
  {Bartolomei}}, \bibinfo {author} {\bibfnamefont {M.}~\bibnamefont {Kumar}},
  \bibinfo {author} {\bibfnamefont {R.}~\bibnamefont {Bisognin}}, \bibinfo
  {author} {\bibfnamefont {A.}~\bibnamefont {Marguerite}}, \bibinfo {author}
  {\bibfnamefont {J.-M.}\ \bibnamefont {Berroir}}, \bibinfo {author}
  {\bibfnamefont {E.}~\bibnamefont {Bocquillon}}, \bibinfo {author}
  {\bibfnamefont {B.}~\bibnamefont {Pla\c{c}ais}}, \bibinfo {author}
  {\bibfnamefont {A.}~\bibnamefont {Cavanna}}, \bibinfo {author} {\bibfnamefont
  {Q.}~\bibnamefont {Dong}}, \bibinfo {author} {\bibfnamefont {U.}~\bibnamefont
  {Gennser}}, \bibinfo {author} {\bibfnamefont {Y.}~\bibnamefont {Jin}}, \ and\
  \bibinfo {author} {\bibfnamefont {G.}~\bibnamefont {F\`eve}},\ }\bibfield
  {title} {\enquote {\bibinfo {title} {Fractional statistics in anyon
  collisions},}\ }\href {\doibase 10.1126/science.aaz5601} {\bibfield
  {journal} {\bibinfo  {journal} {Science}\ }\textbf {\bibinfo {volume}
  {368}},\ \bibinfo {pages} {173--177} (\bibinfo {year} {2020})}\BibitemShut
  {NoStop}%
\bibitem [{\citenamefont {Hong}\ \emph {et~al.}(1987)\citenamefont {Hong},
  \citenamefont {Ou},\ and\ \citenamefont {Mandel}}]{Hong1987}%
  \BibitemOpen
  \bibfield  {author} {\bibinfo {author} {\bibfnamefont {C.~K.}\ \bibnamefont
  {Hong}}, \bibinfo {author} {\bibfnamefont {Z.~Y.}\ \bibnamefont {Ou}}, \ and\
  \bibinfo {author} {\bibfnamefont {L.}~\bibnamefont {Mandel}},\ }\bibfield
  {title} {\enquote {\bibinfo {title} {Measurement of subpicosecond time
  intervals between two photons by interference},}\ }\href {\doibase
  10.1103/PhysRevLett.59.2044} {\bibfield  {journal} {\bibinfo  {journal}
  {Phys. Rev. Lett.}\ }\textbf {\bibinfo {volume} {59}},\ \bibinfo {pages}
  {2044--2046} (\bibinfo {year} {1987})}\BibitemShut {NoStop}%
\bibitem [{\citenamefont {Taktak}\ \emph {et~al.}(2022)\citenamefont {Taktak},
  \citenamefont {Kapfer}, \citenamefont {Nath}, \citenamefont {Roulleau},
  \citenamefont {Acciai}, \citenamefont {Splettstoesser}, \citenamefont
  {Farrer}, \citenamefont {Ritchie},\ and\ \citenamefont
  {Glattli}}]{Taktak2021}%
  \BibitemOpen
  \bibfield  {author} {\bibinfo {author} {\bibfnamefont {I.}~\bibnamefont
  {Taktak}}, \bibinfo {author} {\bibfnamefont {M.}~\bibnamefont {Kapfer}},
  \bibinfo {author} {\bibfnamefont {J.}~\bibnamefont {Nath}}, \bibinfo {author}
  {\bibfnamefont {P.}~\bibnamefont {Roulleau}}, \bibinfo {author}
  {\bibfnamefont {M.}~\bibnamefont {Acciai}}, \bibinfo {author} {\bibfnamefont
  {J.}~\bibnamefont {Splettstoesser}}, \bibinfo {author} {\bibfnamefont
  {I.}~\bibnamefont {Farrer}}, \bibinfo {author} {\bibfnamefont {D.~A.}\
  \bibnamefont {Ritchie}}, \ and\ \bibinfo {author} {\bibfnamefont {D.~C.}\
  \bibnamefont {Glattli}},\ }\href@noop {} {\enquote {\bibinfo {title}
  {{Two-particle time-domain interferometry in the Fractional Quantum Hall
  Effect regime}},}\ } (\bibinfo {year} {2022}),\ \Eprint
  {http://arxiv.org/abs/2201.09553} {arXiv:2201.09553 [cond-mat.mes-hall]}
  \BibitemShut {NoStop}%
\bibitem [{\citenamefont {Safi}(2014)}]{safi2014}%
  \BibitemOpen
  \bibfield  {author} {\bibinfo {author} {\bibfnamefont {I.}~\bibnamefont
  {Safi}},\ }\href@noop {} {\enquote {\bibinfo {title} {{Time-dependent
  Transport in arbitrary extended driven tunnel junctions}},}\ } (\bibinfo
  {year} {2014}),\ \Eprint {http://arxiv.org/abs/1401.5950} {arXiv:1401.5950
  [cond-mat.mes-hall]} \BibitemShut {NoStop}%
\bibitem [{\citenamefont {Safi}(2019)}]{Safi2019}%
  \BibitemOpen
  \bibfield  {author} {\bibinfo {author} {\bibfnamefont {I.}~\bibnamefont
  {Safi}},\ }\bibfield  {title} {\enquote {\bibinfo {title} {{Driven quantum
  circuits and conductors: A unifying perturbative approach}},}\ }\href
  {\doibase 10.1103/PhysRevB.99.045101} {\bibfield  {journal} {\bibinfo
  {journal} {Phys. Rev. B}\ }\textbf {\bibinfo {volume} {99}},\ \bibinfo
  {pages} {045101} (\bibinfo {year} {2019})}\BibitemShut {NoStop}%
\bibitem [{\citenamefont {Rebora}\ \emph {et~al.}(2020)\citenamefont {Rebora},
  \citenamefont {Acciai}, \citenamefont {Ferraro},\ and\ \citenamefont
  {Sassetti}}]{Rebora2020}%
  \BibitemOpen
  \bibfield  {author} {\bibinfo {author} {\bibfnamefont {G.}~\bibnamefont
  {Rebora}}, \bibinfo {author} {\bibfnamefont {M.}~\bibnamefont {Acciai}},
  \bibinfo {author} {\bibfnamefont {D.}~\bibnamefont {Ferraro}}, \ and\
  \bibinfo {author} {\bibfnamefont {M.}~\bibnamefont {Sassetti}},\ }\bibfield
  {title} {\enquote {\bibinfo {title} {Collisional interferometry of levitons
  in quantum hall edge channels at $\ensuremath{\nu}=2$},}\ }\href {\doibase
  10.1103/PhysRevB.101.245310} {\bibfield  {journal} {\bibinfo  {journal}
  {Phys. Rev. B}\ }\textbf {\bibinfo {volume} {101}},\ \bibinfo {pages}
  {245310} (\bibinfo {year} {2020})}\BibitemShut {NoStop}%
\bibitem [{\citenamefont {Alphenaar}\ \emph {et~al.}(1990)\citenamefont
  {Alphenaar}, \citenamefont {McEuen}, \citenamefont {Wheeler},\ and\
  \citenamefont {Sacks}}]{Alphenaar_1990}%
  \BibitemOpen
  \bibfield  {author} {\bibinfo {author} {\bibfnamefont {B.~W.}\ \bibnamefont
  {Alphenaar}}, \bibinfo {author} {\bibfnamefont {P.~L.}\ \bibnamefont
  {McEuen}}, \bibinfo {author} {\bibfnamefont {R.~G.}\ \bibnamefont {Wheeler}},
  \ and\ \bibinfo {author} {\bibfnamefont {R.~N.}\ \bibnamefont {Sacks}},\
  }\bibfield  {title} {\enquote {\bibinfo {title} {{Selective equilibration
  among the current-carrying states in the quantum Hall regime}},}\ }\href
  {\doibase 10.1103/PhysRevLett.64.677} {\bibfield  {journal} {\bibinfo
  {journal} {Phys. Rev. Lett.}\ }\textbf {\bibinfo {volume} {64}},\ \bibinfo
  {pages} {677--680} (\bibinfo {year} {1990})}\BibitemShut {NoStop}%
\bibitem [{\citenamefont {Hirai}\ \emph {et~al.}(1995)\citenamefont {Hirai},
  \citenamefont {Komiyama}, \citenamefont {Fukatsu}, \citenamefont {Osada},
  \citenamefont {Shiraki},\ and\ \citenamefont {Toyoshima}}]{Hirai_1995}%
  \BibitemOpen
  \bibfield  {author} {\bibinfo {author} {\bibfnamefont {H.}~\bibnamefont
  {Hirai}}, \bibinfo {author} {\bibfnamefont {S.}~\bibnamefont {Komiyama}},
  \bibinfo {author} {\bibfnamefont {S.}~\bibnamefont {Fukatsu}}, \bibinfo
  {author} {\bibfnamefont {T.}~\bibnamefont {Osada}}, \bibinfo {author}
  {\bibfnamefont {Y.}~\bibnamefont {Shiraki}}, \ and\ \bibinfo {author}
  {\bibfnamefont {H.}~\bibnamefont {Toyoshima}},\ }\bibfield  {title} {\enquote
  {\bibinfo {title} {{Dependence of inter-edge-channel scattering on
  temperature and magnetic field: Insight into the edge-confining
  potential}},}\ }\href {\doibase 10.1103/PhysRevB.52.11159} {\bibfield
  {journal} {\bibinfo  {journal} {Phys. Rev. B}\ }\textbf {\bibinfo {volume}
  {52}},\ \bibinfo {pages} {11159--11164} (\bibinfo {year} {1995})}\BibitemShut
  {NoStop}%
\bibitem [{\citenamefont {Palacios}\ and\ \citenamefont
  {Tejedor}(1992)}]{Palacios_1992}%
  \BibitemOpen
  \bibfield  {author} {\bibinfo {author} {\bibfnamefont {J.~J.}\ \bibnamefont
  {Palacios}}\ and\ \bibinfo {author} {\bibfnamefont {C.}~\bibnamefont
  {Tejedor}},\ }\bibfield  {title} {\enquote {\bibinfo {title} {{Effects of
  geometry on edge states in magnetic fields: Adiabatic and nonadiabatic
  behavior}},}\ }\href {\doibase 10.1103/PhysRevB.45.9059} {\bibfield
  {journal} {\bibinfo  {journal} {Phys. Rev. B}\ }\textbf {\bibinfo {volume}
  {45}},\ \bibinfo {pages} {9059--9064} (\bibinfo {year} {1992})}\BibitemShut
  {NoStop}%
\bibitem [{\citenamefont {Takagaki}\ \emph {et~al.}(1994)\citenamefont
  {Takagaki}, \citenamefont {Friedland}, \citenamefont {Herfort}, \citenamefont
  {Kostial},\ and\ \citenamefont {Ploog}}]{Takagaki_1994}%
  \BibitemOpen
  \bibfield  {author} {\bibinfo {author} {\bibfnamefont {Y.}~\bibnamefont
  {Takagaki}}, \bibinfo {author} {\bibfnamefont {K.~J.}\ \bibnamefont
  {Friedland}}, \bibinfo {author} {\bibfnamefont {J.}~\bibnamefont {Herfort}},
  \bibinfo {author} {\bibfnamefont {H.}~\bibnamefont {Kostial}}, \ and\
  \bibinfo {author} {\bibfnamefont {K.}~\bibnamefont {Ploog}},\ }\bibfield
  {title} {\enquote {\bibinfo {title} {{Inter-edge-state scattering in the
  spin-polarized quantum Hall regime with current injection into inner
  states}},}\ }\href {\doibase 10.1103/PhysRevB.50.4456} {\bibfield  {journal}
  {\bibinfo  {journal} {Phys. Rev. B}\ }\textbf {\bibinfo {volume} {50}},\
  \bibinfo {pages} {4456--4462} (\bibinfo {year} {1994})}\BibitemShut {NoStop}%
\bibitem [{\citenamefont {Acremann}\ \emph {et~al.}(1999)\citenamefont
  {Acremann}, \citenamefont {Heinzel}, \citenamefont {Ensslin}, \citenamefont
  {Gini}, \citenamefont {Melchior},\ and\ \citenamefont
  {Holland}}]{Acremann_1999}%
  \BibitemOpen
  \bibfield  {author} {\bibinfo {author} {\bibfnamefont {Y.}~\bibnamefont
  {Acremann}}, \bibinfo {author} {\bibfnamefont {T.}~\bibnamefont {Heinzel}},
  \bibinfo {author} {\bibfnamefont {K.}~\bibnamefont {Ensslin}}, \bibinfo
  {author} {\bibfnamefont {E.}~\bibnamefont {Gini}}, \bibinfo {author}
  {\bibfnamefont {H.}~\bibnamefont {Melchior}}, \ and\ \bibinfo {author}
  {\bibfnamefont {M.}~\bibnamefont {Holland}},\ }\bibfield  {title} {\enquote
  {\bibinfo {title} {{Individual scatterers as microscopic origin of
  equilibration between spin-polarized edge channels in the quantum Hall
  regime}},}\ }\href {\doibase 10.1103/PhysRevB.59.2116} {\bibfield  {journal}
  {\bibinfo  {journal} {Phys. Rev. B}\ }\textbf {\bibinfo {volume} {59}},\
  \bibinfo {pages} {2116--2119} (\bibinfo {year} {1999})}\BibitemShut {NoStop}%
\bibitem [{\citenamefont {M\"uller}\ \emph {et~al.}(1990)\citenamefont
  {M\"uller}, \citenamefont {Weiss}, \citenamefont {Koch}, \citenamefont {von
  Klitzing}, \citenamefont {Nickel}, \citenamefont {Schlapp},\ and\
  \citenamefont {L\"osch}}]{Muller_1990}%
  \BibitemOpen
  \bibfield  {author} {\bibinfo {author} {\bibfnamefont {G.}~\bibnamefont
  {M\"uller}}, \bibinfo {author} {\bibfnamefont {D.}~\bibnamefont {Weiss}},
  \bibinfo {author} {\bibfnamefont {S.}~\bibnamefont {Koch}}, \bibinfo {author}
  {\bibfnamefont {K.}~\bibnamefont {von Klitzing}}, \bibinfo {author}
  {\bibfnamefont {H.}~\bibnamefont {Nickel}}, \bibinfo {author} {\bibfnamefont
  {W.}~\bibnamefont {Schlapp}}, \ and\ \bibinfo {author} {\bibfnamefont
  {R.}~\bibnamefont {L\"osch}},\ }\bibfield  {title} {\enquote {\bibinfo
  {title} {{Edge channels and the role of contacts in the quantum Hall
  regime}},}\ }\href {\doibase 10.1103/PhysRevB.42.7633} {\bibfield  {journal}
  {\bibinfo  {journal} {Phys. Rev. B}\ }\textbf {\bibinfo {volume} {42}},\
  \bibinfo {pages} {7633--7636} (\bibinfo {year} {1990})}\BibitemShut {NoStop}%
\bibitem [{\citenamefont {Khaetskii}(1992)}]{Khaetskii_1992}%
  \BibitemOpen
  \bibfield  {author} {\bibinfo {author} {\bibfnamefont {A.~V.}\ \bibnamefont
  {Khaetskii}},\ }\bibfield  {title} {\enquote {\bibinfo {title} {{Transitions
  between spin-split edge channels in the quantum-Hall-effect regime}},}\
  }\href {\doibase 10.1103/PhysRevB.45.13777} {\bibfield  {journal} {\bibinfo
  {journal} {Phys. Rev. B}\ }\textbf {\bibinfo {volume} {45}},\ \bibinfo
  {pages} {13777--13780} (\bibinfo {year} {1992})}\BibitemShut {NoStop}%
\bibitem [{\citenamefont {Polyakov}(1996)}]{Polyakov_1996}%
  \BibitemOpen
  \bibfield  {author} {\bibinfo {author} {\bibfnamefont {D.~G.}\ \bibnamefont
  {Polyakov}},\ }\bibfield  {title} {\enquote {\bibinfo {title} {{Spin-flip
  scattering in the quantum Hall regime}},}\ }\href {\doibase
  10.1103/PhysRevB.53.15777} {\bibfield  {journal} {\bibinfo  {journal} {Phys.
  Rev. B}\ }\textbf {\bibinfo {volume} {53}},\ \bibinfo {pages} {15777--15788}
  (\bibinfo {year} {1996})}\BibitemShut {NoStop}%
\bibitem [{\citenamefont {Pala}\ \emph {et~al.}(2005)\citenamefont {Pala},
  \citenamefont {Governale}, \citenamefont {Z\"ulicke},\ and\ \citenamefont
  {Iannaccone}}]{Pala_2005}%
  \BibitemOpen
  \bibfield  {author} {\bibinfo {author} {\bibfnamefont {M.~G.}\ \bibnamefont
  {Pala}}, \bibinfo {author} {\bibfnamefont {M.}~\bibnamefont {Governale}},
  \bibinfo {author} {\bibfnamefont {U.}~\bibnamefont {Z\"ulicke}}, \ and\
  \bibinfo {author} {\bibfnamefont {G.}~\bibnamefont {Iannaccone}},\ }\bibfield
   {title} {\enquote {\bibinfo {title} {{Rashba spin precession in quantum-Hall
  edge channels}},}\ }\href {\doibase 10.1103/PhysRevB.71.115306} {\bibfield
  {journal} {\bibinfo  {journal} {Phys. Rev. B}\ }\textbf {\bibinfo {volume}
  {71}},\ \bibinfo {pages} {115306} (\bibinfo {year} {2005})}\BibitemShut
  {NoStop}%
\bibitem [{\citenamefont {Johnson}\ and\ \citenamefont
  {MacDonald}(1991)}]{Johnson1991}%
  \BibitemOpen
  \bibfield  {author} {\bibinfo {author} {\bibfnamefont {M.~D.}\ \bibnamefont
  {Johnson}}\ and\ \bibinfo {author} {\bibfnamefont {A.~H.}\ \bibnamefont
  {MacDonald}},\ }\bibfield  {title} {\enquote {\bibinfo {title} {Composite
  edges in the \ensuremath{\nu}=2/3 fractional quantum hall effect},}\ }\href
  {\doibase 10.1103/PhysRevLett.67.2060} {\bibfield  {journal} {\bibinfo
  {journal} {Phys. Rev. Lett.}\ }\textbf {\bibinfo {volume} {67}},\ \bibinfo
  {pages} {2060--2063} (\bibinfo {year} {1991})}\BibitemShut {NoStop}%
\bibitem [{\citenamefont {Kane}\ \emph {et~al.}(1994)\citenamefont {Kane},
  \citenamefont {Fisher},\ and\ \citenamefont {Polchinski}}]{Kane1994}%
  \BibitemOpen
  \bibfield  {author} {\bibinfo {author} {\bibfnamefont {C.~L.}\ \bibnamefont
  {Kane}}, \bibinfo {author} {\bibfnamefont {M.~P.~A.}\ \bibnamefont {Fisher}},
  \ and\ \bibinfo {author} {\bibfnamefont {J.}~\bibnamefont {Polchinski}},\
  }\bibfield  {title} {\enquote {\bibinfo {title} {Randomness at the edge:
  Theory of quantum hall transport at filling \ensuremath{\nu}=2/3},}\ }\href
  {\doibase 10.1103/PhysRevLett.72.4129} {\bibfield  {journal} {\bibinfo
  {journal} {Phys. Rev. Lett.}\ }\textbf {\bibinfo {volume} {72}},\ \bibinfo
  {pages} {4129--4132} (\bibinfo {year} {1994})}\BibitemShut {NoStop}%
\bibitem [{\citenamefont {Meir}(1994)}]{Meir1994}%
  \BibitemOpen
  \bibfield  {author} {\bibinfo {author} {\bibfnamefont {Y.}~\bibnamefont
  {Meir}},\ }\bibfield  {title} {\enquote {\bibinfo {title} {Composite edge
  states in the \ensuremath{\nu}=2/3 fractional quantum hall regime},}\ }\href
  {\doibase 10.1103/PhysRevLett.72.2624} {\bibfield  {journal} {\bibinfo
  {journal} {Phys. Rev. Lett.}\ }\textbf {\bibinfo {volume} {72}},\ \bibinfo
  {pages} {2624--2627} (\bibinfo {year} {1994})}\BibitemShut {NoStop}%
\bibitem [{\citenamefont {Sp\aa{}nsl\"att}\ \emph {et~al.}(2019)\citenamefont
  {Sp\aa{}nsl\"att}, \citenamefont {Park}, \citenamefont {Gefen},\ and\
  \citenamefont {Mirlin}}]{Spaanslaett2019}%
  \BibitemOpen
  \bibfield  {author} {\bibinfo {author} {\bibfnamefont {C.}~\bibnamefont
  {Sp\aa{}nsl\"att}}, \bibinfo {author} {\bibfnamefont {J.}~\bibnamefont
  {Park}}, \bibinfo {author} {\bibfnamefont {Y.}~\bibnamefont {Gefen}}, \ and\
  \bibinfo {author} {\bibfnamefont {A.~D.}\ \bibnamefont {Mirlin}},\ }\bibfield
   {title} {\enquote {\bibinfo {title} {Topological classification of shot
  noise on fractional quantum hall edges},}\ }\href {\doibase
  10.1103/PhysRevLett.123.137701} {\bibfield  {journal} {\bibinfo  {journal}
  {Phys. Rev. Lett.}\ }\textbf {\bibinfo {volume} {123}},\ \bibinfo {pages}
  {137701} (\bibinfo {year} {2019})}\BibitemShut {NoStop}%
\bibitem [{\citenamefont {Splettstoesser}\ \emph {et~al.}(2009)\citenamefont
  {Splettstoesser}, \citenamefont {Moskalets},\ and\ \citenamefont
  {B\"uttiker}}]{Splettstoesser2009}%
  \BibitemOpen
  \bibfield  {author} {\bibinfo {author} {\bibfnamefont {J.}~\bibnamefont
  {Splettstoesser}}, \bibinfo {author} {\bibfnamefont {M.}~\bibnamefont
  {Moskalets}}, \ and\ \bibinfo {author} {\bibfnamefont {M.}~\bibnamefont
  {B\"uttiker}},\ }\bibfield  {title} {\enquote {\bibinfo {title}
  {{Two-Particle Nonlocal Aharonov-Bohm Effect from Two Single-Particle
  Emitters}},}\ }\href {\doibase 10.1103/PhysRevLett.103.076804} {\bibfield
  {journal} {\bibinfo  {journal} {Phys. Rev. Lett.}\ }\textbf {\bibinfo
  {volume} {103}},\ \bibinfo {pages} {076804} (\bibinfo {year}
  {2009})}\BibitemShut {NoStop}%
\bibitem [{\citenamefont {Haack}\ \emph {et~al.}(2011)\citenamefont {Haack},
  \citenamefont {Moskalets}, \citenamefont {Splettstoesser},\ and\
  \citenamefont {B\"uttiker}}]{Haack2011}%
  \BibitemOpen
  \bibfield  {author} {\bibinfo {author} {\bibfnamefont {G.}~\bibnamefont
  {Haack}}, \bibinfo {author} {\bibfnamefont {M.}~\bibnamefont {Moskalets}},
  \bibinfo {author} {\bibfnamefont {J.}~\bibnamefont {Splettstoesser}}, \ and\
  \bibinfo {author} {\bibfnamefont {M.}~\bibnamefont {B\"uttiker}},\ }\bibfield
   {title} {\enquote {\bibinfo {title} {{Coherence of single-electron sources
  from Mach-Zehnder interferometry}},}\ }\href {\doibase
  10.1103/PhysRevB.84.081303} {\bibfield  {journal} {\bibinfo  {journal} {Phys.
  Rev. B}\ }\textbf {\bibinfo {volume} {84}},\ \bibinfo {pages} {081303}
  (\bibinfo {year} {2011})}\BibitemShut {NoStop}%
\bibitem [{\citenamefont {Juergens}\ \emph {et~al.}(2011)\citenamefont
  {Juergens}, \citenamefont {Splettstoesser},\ and\ \citenamefont
  {Moskalets}}]{Juergens_2011}%
  \BibitemOpen
  \bibfield  {author} {\bibinfo {author} {\bibfnamefont {S.}~\bibnamefont
  {Juergens}}, \bibinfo {author} {\bibfnamefont {J.}~\bibnamefont
  {Splettstoesser}}, \ and\ \bibinfo {author} {\bibfnamefont {M.}~\bibnamefont
  {Moskalets}},\ }\bibfield  {title} {\enquote {\bibinfo {title}
  {Single-particle interference versus two-particle collisions},}\ }\href
  {\doibase 10.1209/0295-5075/96/37011} {\bibfield  {journal} {\bibinfo
  {journal} {{EPL} (Europhysics Letters)}\ }\textbf {\bibinfo {volume} {96}},\
  \bibinfo {pages} {37011} (\bibinfo {year} {2011})}\BibitemShut {NoStop}%
\bibitem [{\citenamefont {Hofer}\ and\ \citenamefont
  {Flindt}(2014)}]{Hofer2014}%
  \BibitemOpen
  \bibfield  {author} {\bibinfo {author} {\bibfnamefont {P.~P.}\ \bibnamefont
  {Hofer}}\ and\ \bibinfo {author} {\bibfnamefont {C.}~\bibnamefont {Flindt}},\
  }\bibfield  {title} {\enquote {\bibinfo {title} {{Mach-Zehnder interferometry
  with periodic voltage pulses}},}\ }\href {\doibase
  10.1103/PhysRevB.90.235416} {\bibfield  {journal} {\bibinfo  {journal} {Phys.
  Rev. B}\ }\textbf {\bibinfo {volume} {90}},\ \bibinfo {pages} {235416}
  (\bibinfo {year} {2014})}\BibitemShut {NoStop}%
\bibitem [{\citenamefont {Rossell\'o}\ \emph {et~al.}(2015)\citenamefont
  {Rossell\'o}, \citenamefont {Battista}, \citenamefont {Moskalets},\ and\
  \citenamefont {Splettstoesser}}]{Rossello2015}%
  \BibitemOpen
  \bibfield  {author} {\bibinfo {author} {\bibfnamefont {G.}~\bibnamefont
  {Rossell\'o}}, \bibinfo {author} {\bibfnamefont {F.}~\bibnamefont
  {Battista}}, \bibinfo {author} {\bibfnamefont {M.}~\bibnamefont {Moskalets}},
  \ and\ \bibinfo {author} {\bibfnamefont {J.}~\bibnamefont {Splettstoesser}},\
  }\bibfield  {title} {\enquote {\bibinfo {title} {{Interference and
  multiparticle effects in a Mach-Zehnder interferometer with single-particle
  sources}},}\ }\href {\doibase 10.1103/PhysRevB.91.115438} {\bibfield
  {journal} {\bibinfo  {journal} {Phys. Rev. B}\ }\textbf {\bibinfo {volume}
  {91}},\ \bibinfo {pages} {115438} (\bibinfo {year} {2015})}\BibitemShut
  {NoStop}%
\bibitem [{\citenamefont {Kotilahti}\ \emph {et~al.}(2021)\citenamefont
  {Kotilahti}, \citenamefont {Burset}, \citenamefont {Moskalets},\ and\
  \citenamefont {Flindt}}]{Kotilahti2021}%
  \BibitemOpen
  \bibfield  {author} {\bibinfo {author} {\bibfnamefont {J.}~\bibnamefont
  {Kotilahti}}, \bibinfo {author} {\bibfnamefont {P.}~\bibnamefont {Burset}},
  \bibinfo {author} {\bibfnamefont {M.}~\bibnamefont {Moskalets}}, \ and\
  \bibinfo {author} {\bibfnamefont {C.}~\bibnamefont {Flindt}},\ }\bibfield
  {title} {\enquote {\bibinfo {title} {{Multi-Particle Interference in an
  Electronic Mach–Zehnder Interferometer}},}\ }\href {\doibase
  10.3390/e23060736} {\bibfield  {journal} {\bibinfo  {journal} {Entropy}\
  }\textbf {\bibinfo {volume} {23}} (\bibinfo {year} {2021}),\
  10.3390/e23060736}\BibitemShut {NoStop}%
\bibitem [{\citenamefont {Jain}(1989)}]{Jain1989}%
  \BibitemOpen
  \bibfield  {author} {\bibinfo {author} {\bibfnamefont {J.~K.}\ \bibnamefont
  {Jain}},\ }\bibfield  {title} {\enquote {\bibinfo {title} {{Composite-fermion
  approach for the fractional quantum Hall effect}},}\ }\href {\doibase
  10.1103/PhysRevLett.63.199} {\bibfield  {journal} {\bibinfo  {journal} {Phys.
  Rev. Lett.}\ }\textbf {\bibinfo {volume} {63}},\ \bibinfo {pages} {199--202}
  (\bibinfo {year} {1989})}\BibitemShut {NoStop}%
\bibitem [{\citenamefont {Marguerite}\ \emph {et~al.}(2016)\citenamefont
  {Marguerite}, \citenamefont {Cabart}, \citenamefont {Wahl}, \citenamefont
  {Roussel}, \citenamefont {Freulon}, \citenamefont {Ferraro}, \citenamefont
  {Grenier}, \citenamefont {Berroir}, \citenamefont
  {Pla\ifmmode~\mbox{\c{c}}\else \c{c}\fi{}ais}, \citenamefont {Jonckheere},
  \citenamefont {Rech}, \citenamefont {Martin}, \citenamefont {Degiovanni},
  \citenamefont {Cavanna}, \citenamefont {Jin},\ and\ \citenamefont
  {F\`eve}}]{Marguerite2016}%
  \BibitemOpen
  \bibfield  {author} {\bibinfo {author} {\bibfnamefont {A.}~\bibnamefont
  {Marguerite}}, \bibinfo {author} {\bibfnamefont {C.}~\bibnamefont {Cabart}},
  \bibinfo {author} {\bibfnamefont {C.}~\bibnamefont {Wahl}}, \bibinfo {author}
  {\bibfnamefont {B.}~\bibnamefont {Roussel}}, \bibinfo {author} {\bibfnamefont
  {V.}~\bibnamefont {Freulon}}, \bibinfo {author} {\bibfnamefont
  {D.}~\bibnamefont {Ferraro}}, \bibinfo {author} {\bibfnamefont
  {C.}~\bibnamefont {Grenier}}, \bibinfo {author} {\bibfnamefont {J.-M.}\
  \bibnamefont {Berroir}}, \bibinfo {author} {\bibfnamefont {B.}~\bibnamefont
  {Pla\ifmmode~\mbox{\c{c}}\else \c{c}\fi{}ais}}, \bibinfo {author}
  {\bibfnamefont {T.}~\bibnamefont {Jonckheere}}, \bibinfo {author}
  {\bibfnamefont {J.}~\bibnamefont {Rech}}, \bibinfo {author} {\bibfnamefont
  {T.}~\bibnamefont {Martin}}, \bibinfo {author} {\bibfnamefont
  {P.}~\bibnamefont {Degiovanni}}, \bibinfo {author} {\bibfnamefont
  {A.}~\bibnamefont {Cavanna}}, \bibinfo {author} {\bibfnamefont
  {Y.}~\bibnamefont {Jin}}, \ and\ \bibinfo {author} {\bibfnamefont
  {G.}~\bibnamefont {F\`eve}},\ }\bibfield  {title} {\enquote {\bibinfo {title}
  {Decoherence and relaxation of a single electron in a one-dimensional
  conductor},}\ }\href {\doibase 10.1103/PhysRevB.94.115311} {\bibfield
  {journal} {\bibinfo  {journal} {Phys. Rev. B}\ }\textbf {\bibinfo {volume}
  {94}},\ \bibinfo {pages} {115311} (\bibinfo {year} {2016})}\BibitemShut
  {NoStop}%
\bibitem [{\citenamefont {Moskalets}(2011)}]{moskalets-book}%
  \BibitemOpen
  \bibfield  {author} {\bibinfo {author} {\bibfnamefont {M.~V.}\ \bibnamefont
  {Moskalets}},\ }\href {\doibase 10.1142/p822} {\emph {\bibinfo {title}
  {{Scattering Matrix Approach to Non-Stationary Quantum Transport}}}}\
  (\bibinfo  {publisher} {Imperial College Press},\ \bibinfo {year}
  {2011})\BibitemShut {NoStop}%
\bibitem [{\citenamefont {Levitov}\ \emph {et~al.}(1996)\citenamefont
  {Levitov}, \citenamefont {Lee},\ and\ \citenamefont {Lesovik}}]{Levitov1996}%
  \BibitemOpen
  \bibfield  {author} {\bibinfo {author} {\bibfnamefont {L.~S.}\ \bibnamefont
  {Levitov}}, \bibinfo {author} {\bibfnamefont {H.}~\bibnamefont {Lee}}, \ and\
  \bibinfo {author} {\bibfnamefont {G.~B.}\ \bibnamefont {Lesovik}},\
  }\bibfield  {title} {\enquote {\bibinfo {title} {Electron counting statistics
  and coherent states of electric current},}\ }\href {\doibase
  10.1063/1.531672} {\bibfield  {journal} {\bibinfo  {journal} {Journal of
  Mathematical Physics}\ }\textbf {\bibinfo {volume} {37}},\ \bibinfo {pages}
  {4845--4866} (\bibinfo {year} {1996})}\BibitemShut {NoStop}%
\bibitem [{\citenamefont {Ivanov}\ \emph {et~al.}(1997)\citenamefont {Ivanov},
  \citenamefont {Lee},\ and\ \citenamefont {Levitov}}]{Ivanov1997}%
  \BibitemOpen
  \bibfield  {author} {\bibinfo {author} {\bibfnamefont {D.~A.}\ \bibnamefont
  {Ivanov}}, \bibinfo {author} {\bibfnamefont {H.~W.}\ \bibnamefont {Lee}}, \
  and\ \bibinfo {author} {\bibfnamefont {L.~S.}\ \bibnamefont {Levitov}},\
  }\bibfield  {title} {\enquote {\bibinfo {title} {Coherent states of
  alternating current},}\ }\href {\doibase 10.1103/PhysRevB.56.6839} {\bibfield
   {journal} {\bibinfo  {journal} {Phys. Rev. B}\ }\textbf {\bibinfo {volume}
  {56}},\ \bibinfo {pages} {6839--6850} (\bibinfo {year} {1997})}\BibitemShut
  {NoStop}%
\bibitem [{\citenamefont {Keeling}\ \emph {et~al.}(2006)\citenamefont
  {Keeling}, \citenamefont {Klich},\ and\ \citenamefont
  {Levitov}}]{Keeling2006}%
  \BibitemOpen
  \bibfield  {author} {\bibinfo {author} {\bibfnamefont {J.}~\bibnamefont
  {Keeling}}, \bibinfo {author} {\bibfnamefont {I.}~\bibnamefont {Klich}}, \
  and\ \bibinfo {author} {\bibfnamefont {L.~S.}\ \bibnamefont {Levitov}},\
  }\bibfield  {title} {\enquote {\bibinfo {title} {{Minimal Excitation States
  of Electrons in One-Dimensional Wires}},}\ }\href {\doibase
  10.1103/PhysRevLett.97.116403} {\bibfield  {journal} {\bibinfo  {journal}
  {Phys. Rev. Lett.}\ }\textbf {\bibinfo {volume} {97}},\ \bibinfo {pages}
  {116403} (\bibinfo {year} {2006})}\BibitemShut {NoStop}%
\bibitem [{\citenamefont {Inoue}\ \emph {et~al.}(2014)\citenamefont {Inoue},
  \citenamefont {Grivnin}, \citenamefont {Ofek}, \citenamefont {Neder},
  \citenamefont {Heiblum}, \citenamefont {Umansky},\ and\ \citenamefont
  {Mahalu}}]{Inoue2014}%
  \BibitemOpen
  \bibfield  {author} {\bibinfo {author} {\bibfnamefont {H.}~\bibnamefont
  {Inoue}}, \bibinfo {author} {\bibfnamefont {A.}~\bibnamefont {Grivnin}},
  \bibinfo {author} {\bibfnamefont {N.}~\bibnamefont {Ofek}}, \bibinfo {author}
  {\bibfnamefont {I.}~\bibnamefont {Neder}}, \bibinfo {author} {\bibfnamefont
  {M.}~\bibnamefont {Heiblum}}, \bibinfo {author} {\bibfnamefont
  {V.}~\bibnamefont {Umansky}}, \ and\ \bibinfo {author} {\bibfnamefont
  {D.}~\bibnamefont {Mahalu}},\ }\bibfield  {title} {\enquote {\bibinfo {title}
  {{Charge Fractionalization in the Integer Quantum Hall Effect}},}\ }\href
  {\doibase 10.1103/PhysRevLett.112.166801} {\bibfield  {journal} {\bibinfo
  {journal} {Phys. Rev. Lett.}\ }\textbf {\bibinfo {volume} {112}},\ \bibinfo
  {pages} {166801} (\bibinfo {year} {2014})}\BibitemShut {NoStop}%
\bibitem [{\citenamefont {Hashisaka}\ \emph {et~al.}(2017)\citenamefont
  {Hashisaka}, \citenamefont {Hiyama}, \citenamefont {Akiho}, \citenamefont
  {Muraki},\ and\ \citenamefont {Fujisawa}}]{Hashisaka2017}%
  \BibitemOpen
  \bibfield  {author} {\bibinfo {author} {\bibfnamefont {M.}~\bibnamefont
  {Hashisaka}}, \bibinfo {author} {\bibfnamefont {N.}~\bibnamefont {Hiyama}},
  \bibinfo {author} {\bibfnamefont {T.}~\bibnamefont {Akiho}}, \bibinfo
  {author} {\bibfnamefont {K.}~\bibnamefont {Muraki}}, \ and\ \bibinfo {author}
  {\bibfnamefont {T.}~\bibnamefont {Fujisawa}},\ }\bibfield  {title} {\enquote
  {\bibinfo {title} {{Waveform measurement of charge- and spin-density
  wavepackets in a chiral Tomonaga--Luttinger liquid}},}\ }\href {\doibase
  10.1038/nphys4062} {\bibfield  {journal} {\bibinfo  {journal} {Nature
  Physics}\ }\textbf {\bibinfo {volume} {13}},\ \bibinfo {pages} {559--562}
  (\bibinfo {year} {2017})}\BibitemShut {NoStop}%
\bibitem [{\citenamefont {Acciai}\ \emph {et~al.}(2018)\citenamefont {Acciai},
  \citenamefont {Carrega}, \citenamefont {Rech}, \citenamefont {Jonckheere},
  \citenamefont {Martin},\ and\ \citenamefont {Sassetti}}]{Acciai2018}%
  \BibitemOpen
  \bibfield  {author} {\bibinfo {author} {\bibfnamefont {M.}~\bibnamefont
  {Acciai}}, \bibinfo {author} {\bibfnamefont {M.}~\bibnamefont {Carrega}},
  \bibinfo {author} {\bibfnamefont {J.}~\bibnamefont {Rech}}, \bibinfo {author}
  {\bibfnamefont {T.}~\bibnamefont {Jonckheere}}, \bibinfo {author}
  {\bibfnamefont {T.}~\bibnamefont {Martin}}, \ and\ \bibinfo {author}
  {\bibfnamefont {M.}~\bibnamefont {Sassetti}},\ }\bibfield  {title} {\enquote
  {\bibinfo {title} {Probing interactions via nonequilibrium momentum
  distribution and noise in integer quantum hall systems at
  $\ensuremath{\nu}=2$},}\ }\href {\doibase 10.1103/PhysRevB.98.035426}
  {\bibfield  {journal} {\bibinfo  {journal} {Phys. Rev. B}\ }\textbf {\bibinfo
  {volume} {98}},\ \bibinfo {pages} {035426} (\bibinfo {year}
  {2018})}\BibitemShut {NoStop}%
\bibitem [{\citenamefont {Safi}\ and\ \citenamefont {Schulz}(1995)}]{Safi1995}%
  \BibitemOpen
  \bibfield  {author} {\bibinfo {author} {\bibfnamefont {I.}~\bibnamefont
  {Safi}}\ and\ \bibinfo {author} {\bibfnamefont {H.~J.}\ \bibnamefont
  {Schulz}},\ }\bibfield  {title} {\enquote {\bibinfo {title} {Transport in an
  inhomogeneous interacting one-dimensional system},}\ }\href {\doibase
  10.1103/PhysRevB.52.R17040} {\bibfield  {journal} {\bibinfo  {journal} {Phys.
  Rev. B}\ }\textbf {\bibinfo {volume} {52}},\ \bibinfo {pages}
  {R17040--R17043} (\bibinfo {year} {1995})}\BibitemShut {NoStop}%
\bibitem [{\citenamefont {Pham}\ \emph {et~al.}(2000)\citenamefont {Pham},
  \citenamefont {Gabay},\ and\ \citenamefont {Lederer}}]{Pham2000}%
  \BibitemOpen
  \bibfield  {author} {\bibinfo {author} {\bibfnamefont {K.-V.}\ \bibnamefont
  {Pham}}, \bibinfo {author} {\bibfnamefont {M.}~\bibnamefont {Gabay}}, \ and\
  \bibinfo {author} {\bibfnamefont {P.}~\bibnamefont {Lederer}},\ }\bibfield
  {title} {\enquote {\bibinfo {title} {{Fractional excitations in the Luttinger
  liquid}},}\ }\href {\doibase 10.1103/PhysRevB.61.16397} {\bibfield  {journal}
  {\bibinfo  {journal} {Phys. Rev. B}\ }\textbf {\bibinfo {volume} {61}},\
  \bibinfo {pages} {16397--16422} (\bibinfo {year} {2000})}\BibitemShut
  {NoStop}%
\bibitem [{\citenamefont {Steinberg}\ \emph {et~al.}(2008)\citenamefont
  {Steinberg}, \citenamefont {Barak}, \citenamefont {Yacoby}, \citenamefont
  {Pfeiffer}, \citenamefont {West}, \citenamefont {Halperin},\ and\
  \citenamefont {Le~Hur}}]{Steinberg2008}%
  \BibitemOpen
  \bibfield  {author} {\bibinfo {author} {\bibfnamefont {H.}~\bibnamefont
  {Steinberg}}, \bibinfo {author} {\bibfnamefont {G.}~\bibnamefont {Barak}},
  \bibinfo {author} {\bibfnamefont {A.}~\bibnamefont {Yacoby}}, \bibinfo
  {author} {\bibfnamefont {L.~N.}\ \bibnamefont {Pfeiffer}}, \bibinfo {author}
  {\bibfnamefont {K.~W.}\ \bibnamefont {West}}, \bibinfo {author}
  {\bibfnamefont {B.~I.}\ \bibnamefont {Halperin}}, \ and\ \bibinfo {author}
  {\bibfnamefont {K.}~\bibnamefont {Le~Hur}},\ }\bibfield  {title} {\enquote
  {\bibinfo {title} {Charge fractionalization in quantum wires},}\ }\href
  {\doibase 10.1038/nphys810} {\bibfield  {journal} {\bibinfo  {journal}
  {Nature Physics}\ }\textbf {\bibinfo {volume} {4}},\ \bibinfo {pages}
  {116--119} (\bibinfo {year} {2008})}\BibitemShut {NoStop}%
\bibitem [{\citenamefont {Kamata}\ \emph {et~al.}(2014)\citenamefont {Kamata},
  \citenamefont {Kumada}, \citenamefont {Hashisaka}, \citenamefont {Muraki},\
  and\ \citenamefont {Fujisawa}}]{Kamata2014}%
  \BibitemOpen
  \bibfield  {author} {\bibinfo {author} {\bibfnamefont {H.}~\bibnamefont
  {Kamata}}, \bibinfo {author} {\bibfnamefont {N.}~\bibnamefont {Kumada}},
  \bibinfo {author} {\bibfnamefont {M.}~\bibnamefont {Hashisaka}}, \bibinfo
  {author} {\bibfnamefont {K.}~\bibnamefont {Muraki}}, \ and\ \bibinfo {author}
  {\bibfnamefont {T.}~\bibnamefont {Fujisawa}},\ }\bibfield  {title} {\enquote
  {\bibinfo {title} {{Fractionalized wave packets from an artificial
  Tomonaga--Luttinger liquid}},}\ }\href {\doibase 10.1038/nnano.2013.312}
  {\bibfield  {journal} {\bibinfo  {journal} {Nature Nanotechnology}\ }\textbf
  {\bibinfo {volume} {9}},\ \bibinfo {pages} {177--181} (\bibinfo {year}
  {2014})}\BibitemShut {NoStop}%
\bibitem [{\citenamefont {Acciai}\ \emph {et~al.}(2017)\citenamefont {Acciai},
  \citenamefont {Calzona}, \citenamefont {Dolcetto}, \citenamefont {Schmidt},\
  and\ \citenamefont {Sassetti}}]{Acciai2017}%
  \BibitemOpen
  \bibfield  {author} {\bibinfo {author} {\bibfnamefont {M.}~\bibnamefont
  {Acciai}}, \bibinfo {author} {\bibfnamefont {A.}~\bibnamefont {Calzona}},
  \bibinfo {author} {\bibfnamefont {G.}~\bibnamefont {Dolcetto}}, \bibinfo
  {author} {\bibfnamefont {T.~L.}\ \bibnamefont {Schmidt}}, \ and\ \bibinfo
  {author} {\bibfnamefont {M.}~\bibnamefont {Sassetti}},\ }\bibfield  {title}
  {\enquote {\bibinfo {title} {{Charge and energy fractionalization mechanism
  in one-dimensional channels}},}\ }\href {\doibase 10.1103/PhysRevB.96.075144}
  {\bibfield  {journal} {\bibinfo  {journal} {Phys. Rev. B}\ }\textbf {\bibinfo
  {volume} {96}},\ \bibinfo {pages} {075144} (\bibinfo {year}
  {2017})}\BibitemShut {NoStop}%
\bibitem [{\citenamefont {Lin}\ \emph {et~al.}(2021)\citenamefont {Lin},
  \citenamefont {Hashisaka}, \citenamefont {Akiho}, \citenamefont {Muraki},\
  and\ \citenamefont {Fujisawa}}]{Lin2021}%
  \BibitemOpen
  \bibfield  {author} {\bibinfo {author} {\bibfnamefont {C.}~\bibnamefont
  {Lin}}, \bibinfo {author} {\bibfnamefont {M.}~\bibnamefont {Hashisaka}},
  \bibinfo {author} {\bibfnamefont {T.}~\bibnamefont {Akiho}}, \bibinfo
  {author} {\bibfnamefont {K.}~\bibnamefont {Muraki}}, \ and\ \bibinfo {author}
  {\bibfnamefont {T.}~\bibnamefont {Fujisawa}},\ }\bibfield  {title} {\enquote
  {\bibinfo {title} {{Quantized charge fractionalization at quantum Hall Y
  junctions in the disorder dominated regime}},}\ }\href {\doibase
  10.1038/s41467-020-20395-7} {\bibfield  {journal} {\bibinfo  {journal}
  {Nature Communications}\ }\textbf {\bibinfo {volume} {12}},\ \bibinfo {pages}
  {131} (\bibinfo {year} {2021})}\BibitemShut {NoStop}%
\bibitem [{\citenamefont {Levkivskyi}\ and\ \citenamefont
  {Sukhorukov}(2008)}]{Levkivskyi2008}%
  \BibitemOpen
  \bibfield  {author} {\bibinfo {author} {\bibfnamefont {I.~P.}\ \bibnamefont
  {Levkivskyi}}\ and\ \bibinfo {author} {\bibfnamefont {E.~V.}\ \bibnamefont
  {Sukhorukov}},\ }\bibfield  {title} {\enquote {\bibinfo {title} {{Dephasing
  in the electronic Mach-Zehnder interferometer at filling factor
  $\ensuremath{\nu}=2$}},}\ }\href {\doibase 10.1103/PhysRevB.78.045322}
  {\bibfield  {journal} {\bibinfo  {journal} {Phys. Rev. B}\ }\textbf {\bibinfo
  {volume} {78}},\ \bibinfo {pages} {045322} (\bibinfo {year}
  {2008})}\BibitemShut {NoStop}%
\bibitem [{\citenamefont {Degiovanni}\ \emph {et~al.}(2010)\citenamefont
  {Degiovanni}, \citenamefont {Grenier}, \citenamefont {F\`eve}, \citenamefont
  {Altimiras}, \citenamefont {le~Sueur},\ and\ \citenamefont
  {Pierre}}]{Degiovanni2010}%
  \BibitemOpen
  \bibfield  {author} {\bibinfo {author} {\bibfnamefont {P.}~\bibnamefont
  {Degiovanni}}, \bibinfo {author} {\bibfnamefont {C.}~\bibnamefont {Grenier}},
  \bibinfo {author} {\bibfnamefont {G.}~\bibnamefont {F\`eve}}, \bibinfo
  {author} {\bibfnamefont {C.}~\bibnamefont {Altimiras}}, \bibinfo {author}
  {\bibfnamefont {H.}~\bibnamefont {le~Sueur}}, \ and\ \bibinfo {author}
  {\bibfnamefont {F.}~\bibnamefont {Pierre}},\ }\bibfield  {title} {\enquote
  {\bibinfo {title} {{Plasmon scattering approach to energy exchange and
  high-frequency noise in $\ensuremath{\nu}=2$ quantum Hall edge channels}},}\
  }\href {\doibase 10.1103/PhysRevB.81.121302} {\bibfield  {journal} {\bibinfo
  {journal} {Phys. Rev. B}\ }\textbf {\bibinfo {volume} {81}},\ \bibinfo
  {pages} {121302} (\bibinfo {year} {2010})}\BibitemShut {NoStop}%
\bibitem [{\citenamefont {le~Sueur}\ \emph {et~al.}(2010)\citenamefont
  {le~Sueur}, \citenamefont {Altimiras}, \citenamefont {Gennser}, \citenamefont
  {Cavanna}, \citenamefont {Mailly},\ and\ \citenamefont {Pierre}}]{Sueur2010}%
  \BibitemOpen
  \bibfield  {author} {\bibinfo {author} {\bibfnamefont {H.}~\bibnamefont
  {le~Sueur}}, \bibinfo {author} {\bibfnamefont {C.}~\bibnamefont {Altimiras}},
  \bibinfo {author} {\bibfnamefont {U.}~\bibnamefont {Gennser}}, \bibinfo
  {author} {\bibfnamefont {A.}~\bibnamefont {Cavanna}}, \bibinfo {author}
  {\bibfnamefont {D.}~\bibnamefont {Mailly}}, \ and\ \bibinfo {author}
  {\bibfnamefont {F.}~\bibnamefont {Pierre}},\ }\bibfield  {title} {\enquote
  {\bibinfo {title} {{Energy Relaxation in the Integer Quantum Hall Regime}},}\
  }\href {\doibase 10.1103/PhysRevLett.105.056803} {\bibfield  {journal}
  {\bibinfo  {journal} {Phys. Rev. Lett.}\ }\textbf {\bibinfo {volume} {105}},\
  \bibinfo {pages} {056803} (\bibinfo {year} {2010})}\BibitemShut {NoStop}%
\bibitem [{\citenamefont {Kr{\"a}henmann}\ \emph {et~al.}(2019)\citenamefont
  {Kr{\"a}henmann}, \citenamefont {Fischer}, \citenamefont {R{\"o}{\"o}sli},
  \citenamefont {Ihn}, \citenamefont {Reichl}, \citenamefont {Wegscheider},
  \citenamefont {Ensslin}, \citenamefont {Gefen},\ and\ \citenamefont
  {Meir}}]{Krahenmann2019}%
  \BibitemOpen
  \bibfield  {author} {\bibinfo {author} {\bibfnamefont {T.}~\bibnamefont
  {Kr{\"a}henmann}}, \bibinfo {author} {\bibfnamefont {S.~G.}\ \bibnamefont
  {Fischer}}, \bibinfo {author} {\bibfnamefont {M.}~\bibnamefont
  {R{\"o}{\"o}sli}}, \bibinfo {author} {\bibfnamefont {T.}~\bibnamefont {Ihn}},
  \bibinfo {author} {\bibfnamefont {C.}~\bibnamefont {Reichl}}, \bibinfo
  {author} {\bibfnamefont {W.}~\bibnamefont {Wegscheider}}, \bibinfo {author}
  {\bibfnamefont {K.}~\bibnamefont {Ensslin}}, \bibinfo {author} {\bibfnamefont
  {Y.}~\bibnamefont {Gefen}}, \ and\ \bibinfo {author} {\bibfnamefont
  {Y.}~\bibnamefont {Meir}},\ }\bibfield  {title} {\enquote {\bibinfo {title}
  {{Auger-spectroscopy in quantum Hall edge channels and the missing energy
  problem}},}\ }\href {\doibase 10.1038/s41467-019-11888-1} {\bibfield
  {journal} {\bibinfo  {journal} {Nature Communications}\ }\textbf {\bibinfo
  {volume} {10}},\ \bibinfo {pages} {3915} (\bibinfo {year}
  {2019})}\BibitemShut {NoStop}%
\bibitem [{\citenamefont {Rodriguez}\ \emph {et~al.}(2020)\citenamefont
  {Rodriguez}, \citenamefont {Parmentier}, \citenamefont {Ferraro},
  \citenamefont {Roulleau}, \citenamefont {Gennser}, \citenamefont {Cavanna},
  \citenamefont {Sassetti}, \citenamefont {Portier}, \citenamefont {Mailly},\
  and\ \citenamefont {Roche}}]{Rodriguez2020}%
  \BibitemOpen
  \bibfield  {author} {\bibinfo {author} {\bibfnamefont {R.~H.}\ \bibnamefont
  {Rodriguez}}, \bibinfo {author} {\bibfnamefont {F.~D.}\ \bibnamefont
  {Parmentier}}, \bibinfo {author} {\bibfnamefont {D.}~\bibnamefont {Ferraro}},
  \bibinfo {author} {\bibfnamefont {P.}~\bibnamefont {Roulleau}}, \bibinfo
  {author} {\bibfnamefont {U.}~\bibnamefont {Gennser}}, \bibinfo {author}
  {\bibfnamefont {A.}~\bibnamefont {Cavanna}}, \bibinfo {author} {\bibfnamefont
  {M.}~\bibnamefont {Sassetti}}, \bibinfo {author} {\bibfnamefont
  {F.}~\bibnamefont {Portier}}, \bibinfo {author} {\bibfnamefont
  {D.}~\bibnamefont {Mailly}}, \ and\ \bibinfo {author} {\bibfnamefont
  {P.}~\bibnamefont {Roche}},\ }\bibfield  {title} {\enquote {\bibinfo {title}
  {{Relaxation and revival of quasiparticles injected in an interacting quantum
  Hall liquid}},}\ }\href {\doibase 10.1038/s41467-020-16331-4} {\bibfield
  {journal} {\bibinfo  {journal} {Nature Communications}\ }\textbf {\bibinfo
  {volume} {11}},\ \bibinfo {pages} {2426} (\bibinfo {year}
  {2020})}\BibitemShut {NoStop}%
\bibitem [{\citenamefont {Rebora}\ \emph {et~al.}(2021)\citenamefont {Rebora},
  \citenamefont {Ferraro}, \citenamefont {Rodriguez}, \citenamefont
  {Parmentier}, \citenamefont {Roche},\ and\ \citenamefont
  {Sassetti}}]{Rebora2021}%
  \BibitemOpen
  \bibfield  {author} {\bibinfo {author} {\bibfnamefont {G.}~\bibnamefont
  {Rebora}}, \bibinfo {author} {\bibfnamefont {D.}~\bibnamefont {Ferraro}},
  \bibinfo {author} {\bibfnamefont {R.~H.}\ \bibnamefont {Rodriguez}}, \bibinfo
  {author} {\bibfnamefont {F.~D.}\ \bibnamefont {Parmentier}}, \bibinfo
  {author} {\bibfnamefont {P.}~\bibnamefont {Roche}}, \ and\ \bibinfo {author}
  {\bibfnamefont {M.}~\bibnamefont {Sassetti}},\ }\bibfield  {title} {\enquote
  {\bibinfo {title} {{Electronic Wave-Packets in Integer Quantum Hall Edge
  Channels: Relaxation and Dissipative Effects}},}\ }\href {\doibase
  10.3390/e23020138} {\bibfield  {journal} {\bibinfo  {journal} {Entropy}\
  }\textbf {\bibinfo {volume} {23}} (\bibinfo {year} {2021}),\
  10.3390/e23020138}\BibitemShut {NoStop}%
\bibitem [{\citenamefont {Goremykina}\ and\ \citenamefont
  {Sukhorukov}(2018)}]{Goremykina2018}%
  \BibitemOpen
  \bibfield  {author} {\bibinfo {author} {\bibfnamefont {A.~S.}\ \bibnamefont
  {Goremykina}}\ and\ \bibinfo {author} {\bibfnamefont {E.~V.}\ \bibnamefont
  {Sukhorukov}},\ }\bibfield  {title} {\enquote {\bibinfo {title} {{Coherence
  recovery mechanisms of quantum Hall edge states}},}\ }\href {\doibase
  10.1103/PhysRevB.97.115418} {\bibfield  {journal} {\bibinfo  {journal} {Phys.
  Rev. B}\ }\textbf {\bibinfo {volume} {97}},\ \bibinfo {pages} {115418}
  (\bibinfo {year} {2018})}\BibitemShut {NoStop}%
\bibitem [{\citenamefont {Lesovik}\ and\ \citenamefont
  {Levitov}(1994)}]{Lesovik1994}%
  \BibitemOpen
  \bibfield  {author} {\bibinfo {author} {\bibfnamefont {G.~B.}\ \bibnamefont
  {Lesovik}}\ and\ \bibinfo {author} {\bibfnamefont {L.~S.}\ \bibnamefont
  {Levitov}},\ }\bibfield  {title} {\enquote {\bibinfo {title} {{Noise in an ac
  biased junction: Nonstationary Aharonov-Bohm effec}},}\ }\href {\doibase
  10.1103/PhysRevLett.72.538} {\bibfield  {journal} {\bibinfo  {journal} {Phys.
  Rev. Lett.}\ }\textbf {\bibinfo {volume} {72}},\ \bibinfo {pages} {538--541}
  (\bibinfo {year} {1994})}\BibitemShut {NoStop}%
\bibitem [{\citenamefont {Pedersen}\ and\ \citenamefont
  {B\"uttiker}(1998)}]{Pedersen1998}%
  \BibitemOpen
  \bibfield  {author} {\bibinfo {author} {\bibfnamefont {M.~H.}\ \bibnamefont
  {Pedersen}}\ and\ \bibinfo {author} {\bibfnamefont {M.}~\bibnamefont
  {B\"uttiker}},\ }\bibfield  {title} {\enquote {\bibinfo {title} {Scattering
  theory of photon-assisted electron transport},}\ }\href {\doibase
  10.1103/PhysRevB.58.12993} {\bibfield  {journal} {\bibinfo  {journal} {Phys.
  Rev. B}\ }\textbf {\bibinfo {volume} {58}},\ \bibinfo {pages} {12993--13006}
  (\bibinfo {year} {1998})}\BibitemShut {NoStop}%
\bibitem [{\citenamefont {Cr\'epieux}\ \emph {et~al.}(2004)\citenamefont
  {Cr\'epieux}, \citenamefont {Devillard},\ and\ \citenamefont
  {Martin}}]{Crepieux2004}%
  \BibitemOpen
  \bibfield  {author} {\bibinfo {author} {\bibfnamefont {A.}~\bibnamefont
  {Cr\'epieux}}, \bibinfo {author} {\bibfnamefont {P.}~\bibnamefont
  {Devillard}}, \ and\ \bibinfo {author} {\bibfnamefont {T.}~\bibnamefont
  {Martin}},\ }\bibfield  {title} {\enquote {\bibinfo {title} {{Photoassisted
  current and shot noise in the fractional quantum Hall effect}},}\ }\href
  {\doibase 10.1103/PhysRevB.69.205302} {\bibfield  {journal} {\bibinfo
  {journal} {Phys. Rev. B}\ }\textbf {\bibinfo {volume} {69}},\ \bibinfo
  {pages} {205302} (\bibinfo {year} {2004})}\BibitemShut {NoStop}%
\bibitem [{\citenamefont {Jo}\ \emph {et~al.}(2021)\citenamefont {Jo},
  \citenamefont {Brasseur}, \citenamefont {Assouline}, \citenamefont {Fleury},
  \citenamefont {Sim}, \citenamefont {Watanabe}, \citenamefont {Taniguchi},
  \citenamefont {Dumnernpanich}, \citenamefont {Roche}, \citenamefont
  {Glattli}, \citenamefont {Kumada}, \citenamefont {Parmentier},\ and\
  \citenamefont {Roulleau}}]{PhysRevLett.126.146803}%
  \BibitemOpen
  \bibfield  {author} {\bibinfo {author} {\bibfnamefont {M.}~\bibnamefont
  {Jo}}, \bibinfo {author} {\bibfnamefont {P.}~\bibnamefont {Brasseur}},
  \bibinfo {author} {\bibfnamefont {A.}~\bibnamefont {Assouline}}, \bibinfo
  {author} {\bibfnamefont {G.}~\bibnamefont {Fleury}}, \bibinfo {author}
  {\bibfnamefont {H.-S.}\ \bibnamefont {Sim}}, \bibinfo {author} {\bibfnamefont
  {K.}~\bibnamefont {Watanabe}}, \bibinfo {author} {\bibfnamefont
  {T.}~\bibnamefont {Taniguchi}}, \bibinfo {author} {\bibfnamefont
  {W.}~\bibnamefont {Dumnernpanich}}, \bibinfo {author} {\bibfnamefont
  {P.}~\bibnamefont {Roche}}, \bibinfo {author} {\bibfnamefont {D.~C.}\
  \bibnamefont {Glattli}}, \bibinfo {author} {\bibfnamefont {N.}~\bibnamefont
  {Kumada}}, \bibinfo {author} {\bibfnamefont {F.~D.}\ \bibnamefont
  {Parmentier}}, \ and\ \bibinfo {author} {\bibfnamefont {P.}~\bibnamefont
  {Roulleau}},\ }\bibfield  {title} {\enquote {\bibinfo {title} {Quantum hall
  valley splitters and a tunable mach-zehnder interferometer in graphene},}\
  }\href {\doibase 10.1103/PhysRevLett.126.146803} {\bibfield  {journal}
  {\bibinfo  {journal} {Phys. Rev. Lett.}\ }\textbf {\bibinfo {volume} {126}},\
  \bibinfo {pages} {146803} (\bibinfo {year} {2021})}\BibitemShut {NoStop}%
\end{thebibliography}%

\end{document}